\documentclass[10pt]{article}
\usepackage[latin1]{inputenc}
\usepackage{amsmath}
\usepackage{amsfonts}
\usepackage{amssymb}
\usepackage{mathrsfs}

\usepackage[cal=boondox,scr=boondoxo]{mathalfa}

\usepackage{float}
\usepackage{subfig}
\usepackage{fancybox,graphicx}
\usepackage{subfig}
\usepackage{caption}
\usepackage{color}
\usepackage{authblk}

\usepackage[colorlinks]{hyperref}
\usepackage{hyperref}

\newcommand{\doi}[1]{{doi:~\href{https://doi.org/#1}{\nolinkurl{#1}}}\rmFullStop}

\newcommand*{\rmFullStop}{\rmifnextchar{.}{}{}}

\makeatletter
\newcommand{\rmifnextchar}[3]{%
  \begingroup
  \ltx@LocToksA{\endgroup#2}%
  \ltx@LocToksB{\endgroup#3}%
  \ltx@ifnextchar{#1}{%
    \def\next{\the\ltx@LocToksA}%
    \afterassignment\next
    \let\scratch= %
  }{%
    \the\ltx@LocToksB
  }%
}
\makeatother %doi command

\usepackage{accents}
\usepackage[titletoc,title]{appendix}
\usepackage{cite}

\usepackage{mathtools}

\DeclarePairedDelimiter\floor{\lfloor}{\rfloor}

\usepackage[top=2in, bottom=1.5in, left=1in, right=1in]{geometry}

\def\componentsII{
\left(
\begin{matrix}
    z\cos\gamma_n \vspace{0.3cm} \\
    z\sin\gamma_n\vspace{0.3cm} \\
    \mathscr{R}_n\cos(\gamma_n+\beta_n)-x\cos\gamma_n+y\sin\gamma_n
\end{matrix}
\right)
}

\def\change{
\scriptsize
\begin{matrix}
    
    y \rightarrow \Lambda_{n}(\boldsymbol{r})
    \vspace{0.05cm} \\
    r \rightarrow \lambda_{n}(\boldsymbol{r})
\end{matrix}
\normalsize
}

\title{Electrostatic Field of Angular-Dependent Surface Electrodes} 
 
\author[1,2]{Robert Salazar}
\author[1,3]{Camilo Bayona-Roa}
\author[1]{J. S.~Sol{\'\i}s-Chaves}
\affil[1]{Universidad ECCI, Cra. 19 No. 49-20, Bogot\'a, Colombia, C\'odigo Postal 111311}
\affil[2]{Departamento de F\'isica, Universidad de los Andes - Bogot\'a, Colombia}
\affil[3]{Centro de Ingenier\'ia Avanzada Investigaci\'on y Desarrollo, CIAID- Bogot\'a, Colombia}
\begin{document}

    \maketitle

\begin{abstract}
We present an analytic strategy to find the electric field generated by surface electrode SE with angular dependent potential. This system is 
a planar region $\mathcal{A}$ kept at a fixed but non-uniform electric potential $V(\phi)$ with an arbitrary angular dependence.  We show that the generated electric field is due to the contribution of two fields: one that depends on the circulation on the contour of the planar region ---in a Biot-Savart-Like (BSL) term---, and another one that accounts for the angular variations of the potential in $\mathcal{A}$. This approach can be used to find exact solutions of the BSL electric field for circular or polygonal contours of the planar region with periodic distributions of the electric potential. 
Analytic results are validated with numerical computations and the Finite Element Method.
\end{abstract}

\section{Introduction}
The computation of the electric field generated by a stationary charge density distribution is a standard problem of physics and engineering. 
Conceptually, the problem is simple:
the electric potential due to a density charge distribution can be found by solving the Laplace's equation. 
However, obtaining an exact solution can be difficult depending on the charge density's distribution in space. 
In principle, it is possible to find analytic solutions to this problem in very different settings \cite{hummer1996electrostatic,lekner2010analytical,ciftja2013calculation,polyakov2015new,mccreery2018electric}. 
A particular case has to do with a fixed but not uniform potential distribution over a boundary surface. 
This type of distribution of charges on the boundary surfaces implies to solve a complex Dirichlet problem.
At best, this system is integrable, as it occurs with some axially symmetric distribution of charges on a sphere or a disk \cite{jackson1999classical,griffiths2005introduction}.
In those cases, standard strategies of separation of variables can be used to find the electric potential. 
However, the computation of the electric potential for more complicated geometries often requires numerical approximations as the only possible approach.      

\begin{figure}[h]
  \begin{minipage}[b]{0.5\linewidth}
    
   \includegraphics[width=0.9\textwidth]{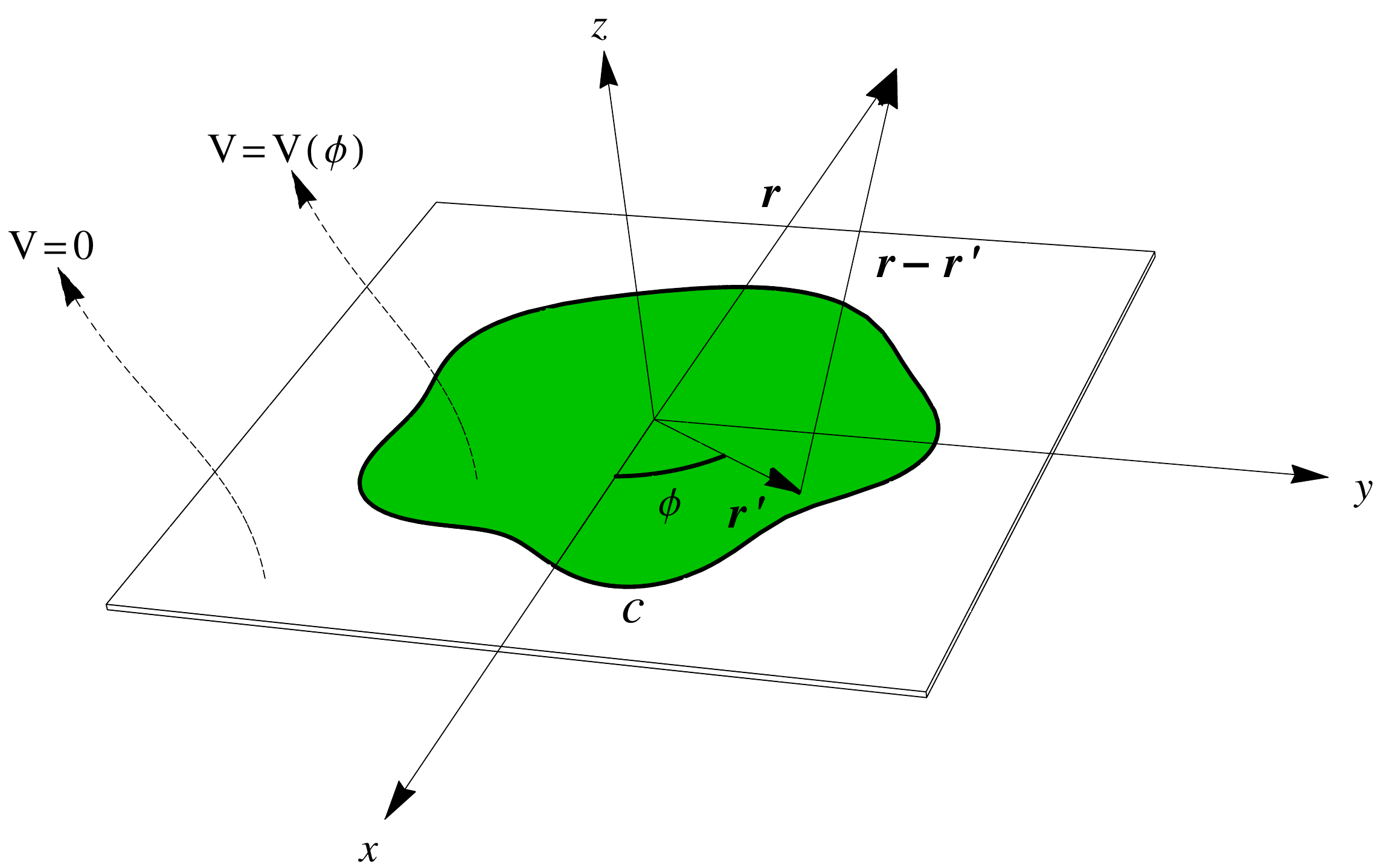}
   \caption*{(a)}
    
  \end{minipage} 
  \begin{minipage}[b]{0.5\linewidth}
    
    \includegraphics[width=0.65\textwidth]{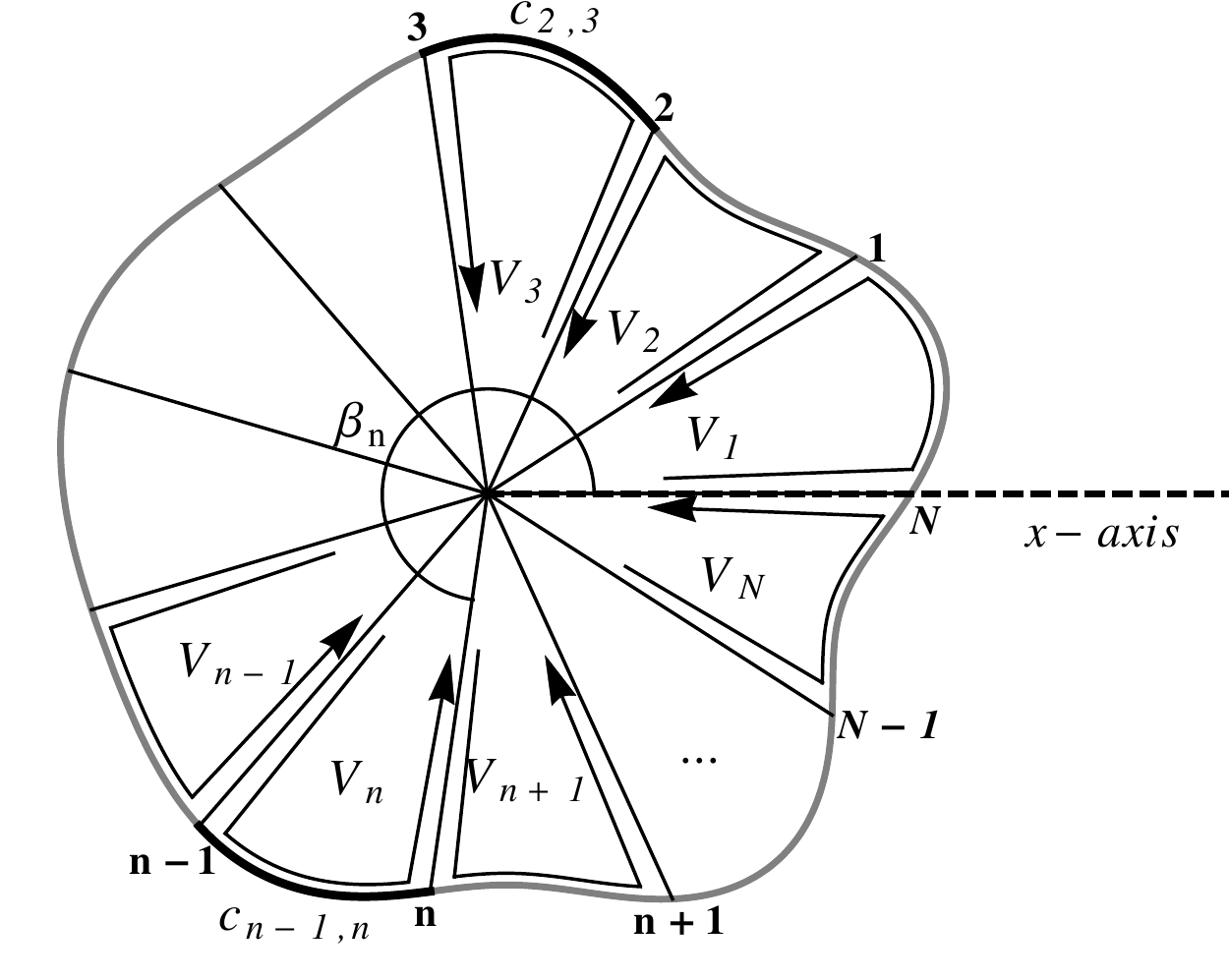}
    \caption*{(b) } 
  
  \end{minipage} 
  \caption[The system.]{(a) Planar region $\mathcal{A}$ of arbitrary contour $c$ with a $\phi$-dependent electric potential $V(\phi)$}
    (b) Discrete distribution of the potential on the sheet.
\label{theSystemFig} 
\end{figure}

In this article, we describe a technique to compute the electric field generated by a planar region with a non-uniform electric potential and its closure as grounded, as it is shown in Fig.~\ref{theSystemFig}. 
Using a cylindrical reference system, we define $\boldsymbol{r}=(r,\phi,z)$ to be any position inside the domain $\mathfrak{D}=\left\{\boldsymbol{r} \in \mathbb{R}^3 : z \geq 0 \right\}$. 
A non-uniform electric potential $V=V(\phi)$ is defined inside the planar region $\mathcal{A}$ which is located at $z=0$ and enclosed by the curve $c$.
The system in Fig.~\ref{theSystemFig} is commonly referred as the \textit{Gapless Plane Electrode} (GPE), and it plays an important role in the study of Surface-electrode (SE) Radio Frequency ion traps since those can be modelled as an infinite plane gaplessly covered by an array of electrodes of arbitrary shape. 
Indeed, SE ion traps are a promising approach to build ion-trap networks suitable for large-scale quantum processing \cite{chiaverini2005surface,leibfried2005surface,seidelin2006microfabricated,daniilidis2011fabrication,kim2011surface,hong2017experimental,mokhberi2017optimised,tao2018fabrication}, and there has been an increasing interest to determine the analytic electric field of these structures. 
An analytical exact solution of the electric potential of a rectangular strip electrode held at constant voltage is presented in \cite{house2008analytic}.
Ring-shaped SE traps have also been studied analytically for constant voltages in \cite{wesenberg2008electrostatics, schmied2010electrostatics}. 
In general, the computation of the electric field in those systems can be achieved by numerical approaches. 
However, if the electric potential in $\mathcal{A}$ is constant, say $V=V_o$, then the electric field can be computed from  
\begin{equation}
\boldsymbol{E}(\boldsymbol{r})  = \frac{V_o}{2\pi} \oint_{ c }  \frac{d\boldsymbol{r}' \times (\boldsymbol{r}-\boldsymbol{r}')  }{|\boldsymbol{r}-\boldsymbol{r}'|^3}, 
\label{oliveiraBiotEq}
\end{equation}
since the electric field at $z>0$ is proportional to the magnetic field that would be observed if a current runs along $c$ \cite{oliveira2001biot}. 
Hence, the Biot-Savart-like integral in Eq.~(\ref{oliveiraBiotEq}) can be employed to drastically simplify the problem as it is shown in \cite{wesenberg2008electrostatics} both for a strip electrode and a ring-shaped SE trap.         

If the potential in $\mathcal{A}$ is not constant, then Eq.~(\ref{oliveiraBiotEq}) cannot be applied. 
Finding analytic solutions for this system is still a challenging problem. 
Most approaches rely on applying numerical methods to approximate the solution of the Laplace's equation \cite{shortley1938numerical,rangogni1986numerical,gray1986program, li2011finite}. 

The main objective of this work is to provide a generalization of Eq.~(\ref{oliveiraBiotEq}) for angular dependent potentials $V=V(\phi)$ on $\mathcal{A}$ which enable to find analytic solutions of the electric field. We focus our study on two specific settings where it can be applied: a circular region and a general non-intersecting polygonal region.   

The remaining parts of this document are organized as follows: In Section 2 we derive an analytic expression to generate the electric field $\boldsymbol{E}(\boldsymbol{r})$ by using an analogy between this electrostatic problem and magnetostatics. 
In that section we shall demonstrate that the expression for the electric field can be written as  
\begin{equation}
\boldsymbol{E}(\boldsymbol{r})  = - \frac{1}{2\pi} \oint_{ c } V(\phi') \frac{(\boldsymbol{r}-\boldsymbol{r}') \times d\boldsymbol{r}'}{|\boldsymbol{r}-\boldsymbol{r}'|^3} + \frac{1}{2\pi} \int_{0}^{2\pi}  V(\beta) \partial_\phi \boldsymbol{f}(\beta,\boldsymbol{r}) d\beta \hspace{0.5cm}\mbox{for}\hspace{0.5cm}z>0,
\label{centralResultEq}    
\end{equation}
where the line integral at the right side of the previous expression is analogous to the problem of computing the magnetic field along a closed loop $c$ \footnote{Being only a an analogy between two different physical systems: the electrostatic where charges are static and the magnetostatic where current is time-independent.}. Here, the magnetostatic analogue corresponds to counter-intuitive situation where the loop carries a non-uniform electric current \footnote{Generally speaking, an electric current along a wire is considered as constant. However, there are devices that \textit{emulate} non-uniform distributed currents along wires (see for instance the setup in \cite{gonzalez2005novel} which can be used in the experimental evaluation of pipe-lines' cathodic protection, the DC version of this setup would also emulate a steady non-uniform current along a wire). Another case concerns the electric current of a point-like charge moving in a loop: the electric current associated to this charge is not uniform. }. 
The second term in the right of Eq.~$(\ref{centralResultEq})$ takes into account the electric field contributions due to the variations of $V(\phi)$ inside the region, where $\boldsymbol{f}$ is a vector field that depends on the shape of $c$. 
We stress out the fact that Eq.~$(\ref{centralResultEq})$ \textit{is valid for any closed loop $c$ independently of its shape}, but the exact solution of this expression for arbitrary contours can be difficult. 
This is especially the case of the Biot-Savart-Like term, which depending on the contour can be very hard to be solved.
Hence, in Section 3, we perform a exact analytic expansion in the case of circular contours with any arbitrary but periodic potential functions. 
The application of this technique on SE settings having polygonal boundaries is demonstrated next in Section \ref{BSLContributionOfPolygonal}. 
We devote considerable effort in Section \ref{NumericalComparisonSection} to perform systematic comparisons of the analytic results against the numerical ones. 
Finally, conclusions are stated at the end of the document.

\section{Electric field via Biot-Savart}
In this section we address the general expression to obtain the electric field inside the problem setting.
We begin this section by recalling the basic definitions of the electrostatic problem:
in that case, a steady electric field $\boldsymbol{E}$ has associated an electric potential $\Phi(\boldsymbol{r})$, such that $\boldsymbol{E} = - \nabla\Phi(\boldsymbol{r})$. 
Including the previous definition in the Gauss's law, leads to the Poisson's problem:
\[
\nabla^2 \Phi(\boldsymbol{r}) = -\frac{\rho(\boldsymbol{r})}{\epsilon_o}\hspace{0.5cm}\mbox{for}\hspace{0.5cm} \boldsymbol{r} \in \mathfrak{D},
\]
where $\rho(\boldsymbol{r})$ is the volume charge density \cite{jackson1999classical,sadiku2014elements,griffiths2005introduction}.  In this problem $\rho(\boldsymbol{r}) = 0$ for $z>0$, such that the Laplace's equation $\nabla^2 \Phi(\boldsymbol{r}) = 0$, $\boldsymbol{r} \in \mathfrak{D}$ subjected to the boundary conditions
\[
\Phi(\boldsymbol{r}) = V(\phi)  \hspace{0.5cm} \mbox{\textbf{if}} \hspace{0.5cm} \boldsymbol{r} \in \mathcal{A} \subset \left\{\boldsymbol{r} \in \mathbb{R}^2 : z = 0 \right\}, \hspace{0.25cm}\text{ and }\hspace{0.25cm}
\Phi(\boldsymbol{r}) = 0 \hspace{0.5cm} \mbox{\textbf{if}} \hspace{0.5cm} \boldsymbol{r} \in \left\{\boldsymbol{r} \in \mathbb{R}^2 : z = 0 \right\}\setminus\mathcal{A},
\]
has to be solved in the domain.
Naturally, the electric potential can be obtained from the solution of the Laplace's equation using Green's functions $G(\boldsymbol{r},\boldsymbol{r}')$ \cite{jackson1999classical}. The general solution for the Poisson's problem can written as
\[
\Phi(\boldsymbol{r}) = \frac{1}{4\pi\epsilon_o}\int_{\mathfrak{D}} \rho(\boldsymbol{r}) G(\boldsymbol{r},\boldsymbol{r}')d^3 \boldsymbol{r}' + \frac{1}{4\pi}\oint_S \left[G(\boldsymbol{r},\boldsymbol{r}')\frac{\partial\Phi}{\partial n'} - \Phi(\boldsymbol{r'}) \frac{\partial G(\boldsymbol{r},\boldsymbol{r}')}{\partial n'} \right]d^2 \boldsymbol{r}'.
\]
Here the first term at the right hand vanishes since $\rho(\boldsymbol{r})=0$ for $r \in \mathfrak{D}$. As usual, we may demand that $\left.G(\boldsymbol{r},\boldsymbol{r}')\right|_{z'=0} = 0$, by choosing the Green's function for the half-space $z>0$ and considering the potential as a punctual charge at $(x',y',z')$, with $z'>0$, plus the potential of an image charge placed in a symmetric position in the lower-half plane, at $(x',y',-z')$. The Green's function can be stated as
\[
G(\boldsymbol{r},\boldsymbol{r}') = \sum_{\sigma\in\{+1,-1\}}\frac{\sigma}{\sqrt{(x-x')^2+(y-y')^2+(z - \sigma z')^2}},
\]
Hence, the solution to the electrostatic problem reduces to evaluate the following integral
\begin{equation}
\Phi(\boldsymbol{r}) = - \frac{1}{4\pi}\int_{\mathcal{A}} V(\phi') \frac{\partial G(\boldsymbol{r},\boldsymbol{r}')}{\partial n'} d^2 \boldsymbol{r}',
\label{electricPotentialIntegralEq}
\end{equation}
where $\hat{n}'=-\hat{z}$ is the outward normal of $\partial\mathfrak{D}$ on the $(z=0)$-plane, and
\[
\left.\frac{\partial G(\boldsymbol{r},\boldsymbol{r}')}{\partial n'}\right|_{z=0} = - \left.\frac{2(z-z')}{|\boldsymbol{r}-\boldsymbol{r}'|^3}\right|_{z'=0}.
\]
Solving the previous expression given the electric potential is not easy and usually numerical integration is needed. 
However, if the function $V(\phi)$ is constant, say $V_o$, then, from Eq.~(\ref{electricPotentialIntegralEq}), the electric potential $\Phi_{unif.}(\boldsymbol{r})$ simplifies to
\[
\Phi_{unif.}(\boldsymbol{r}) = \frac{V_o}{2\pi}\int_{\mathcal{A}} \frac{(\boldsymbol{r}-\boldsymbol{r}')\cdot \hat{z}}{|\boldsymbol{r}-\boldsymbol{r}'|^3} d^2 \boldsymbol{r}' = \frac{V_o}{2\pi} \Omega(\boldsymbol{r}).
\]
Authors of Ref.\cite{oliveira2001biot} remarked that the physical quantity $\Omega(\boldsymbol{r})$ is proportional to the \textit{scalar potential} of a steady magnetic field $\boldsymbol{B}(\boldsymbol{r})$ \cite{eyges2012classical,vanderlinde2006classical} generated by the electric current $i_o$ that flows along a closed arbitrary loop $c$. 
In that problem, the magnetic field could be computed from
\[
\boldsymbol{B}(\boldsymbol{r}) = - \frac{\mu_o i_o}{4\pi} \nabla \Omega(\boldsymbol{r}) = \frac{\mu_o i_o}{4\pi} \oint_c  \frac{ d\boldsymbol{r}' \times (\boldsymbol{r}-\boldsymbol{r}') }{|\boldsymbol{r}-\boldsymbol{r}'|^3}.
\]
Similarly to the previous expression, one can compute the electric field from the definition of the scalar potential as
\begin{equation}
\boldsymbol{E}_{unif.}(\boldsymbol{r}) = -\nabla \Phi_{unif.}(\boldsymbol{r})= - \frac{V_o}{2\pi} \nabla \Omega(\boldsymbol{r}) = \frac{V_o}{2\pi} \oint_c  \frac{  d\boldsymbol{r}' \times (\boldsymbol{r}-\boldsymbol{r}')}{|\boldsymbol{r}-\boldsymbol{r}'|^3} \hspace{0.25cm}\mbox{for}\hspace{0.25cm}z>0,
\label{EUniformBiotEq}
\end{equation}
and therefore, the electric field can be computed by evaluating a Biot-Savart-Like (BSL) integral. 
Formally, Eq.~(\ref{EUniformBiotEq}) is only valid when $V$ is kept as a constant in the region $\mathcal{A}$. The challenging work is to generalize this result for any potential $V(\phi)$ with an arbitrary angular dependence. To this aim, let us first consider the problem where $V(\phi)$ is not uniform but varies discretely as follows $V(\phi) = V_n  \hspace{0.25cm}\mbox{\textbf{if}}\hspace{0.25cm}\phi \in (\beta_{n-1},\beta_{n})$ where $0<\beta_1 < \beta_2 < \ldots < \beta_N = 2\pi$ defines $N$ consecutive sectors $\{\mathcal{A}\}_{n=1,\dots,N}$ of uniform electric potential  $\left\{V_n\right\}_{n=1,\dots,N}$, where 
$\mathcal{A} = \mathcal{A}_1 \cup \mathcal{A}_2 \ldots \cup \mathcal{A}_N$. The boundary of each sector $A_n$ is the loop $c_n \cup c_{n-1,n} \cup \tilde{c}_{n-1}$. Let us introduce an angular function $\mathscr{R}(\phi)$ that returns the planar region contour's radius and helps to define the position of the external curve in polar coordinates $c_{n-1,n} := \left\{ (\mathscr{R}(\phi), \phi) : \beta_{n-1} < \phi \leq \beta_n \right\}$. Therefore, the complete closed curve $c$ can be defined as $c = \left\{ (\mathscr{R}(\phi), \phi) : 0 < \phi \leq 2\pi \right\} = c_{1,2}\cup c_{2,3} \ldots \cup c_{N-1,N} \cup c_{N,1}$. The path $c_n$ is a straight line from the point $(\mathscr{R}(\beta_n), \beta_n)$ to the origin and $\tilde{c}_n$ is its reversed trajectory. Since the potential of each sector is constant, we may use the superposition principle together with Eq.~(\ref{EUniformBiotEq}) to compute the electric field in the domain as
\[
\boldsymbol{E}(\boldsymbol{r}) = \sum_{n=1}^N \boldsymbol{E}_n(\boldsymbol{r})=  \frac{1}{2\pi} \sum_{n=1}^N \int_{c_n \cup c_{n-1,n} \cup \tilde{c}_{n-1} } V_{n} \frac{d\boldsymbol{r}' \times (\boldsymbol{r}-\boldsymbol{r}')}{|\boldsymbol{r}-\boldsymbol{r}'|^3}.
\]
We have denoted $\boldsymbol{E}_n(\boldsymbol{r})$ as the electric potential due to the $n$-th sector of $\mathcal{A}$. It is even possible to split the integral of the previous equation into three different contributing terms:
\[
\int_{c_n \cup c_{n-1,n} \cup \tilde{c}_{n-1} } V_{n} \frac{ d\boldsymbol{r}' \times  (\boldsymbol{r}-\boldsymbol{r}')}{|\boldsymbol{r}-\boldsymbol{r}'|^3} = \int_{c_{n-1,n}} V_{n} \frac{ d\boldsymbol{r}'\times (\boldsymbol{r}-\boldsymbol{r}')}{|\boldsymbol{r}-\boldsymbol{r}'|^3} +  \int_{c_n} V_{n} \frac{(\boldsymbol{r}-\boldsymbol{r}') \times d\boldsymbol{r}'}{|\boldsymbol{r}-\boldsymbol{r}'|^3} - \int_{c_{n-1} } V_{n} \frac{(\boldsymbol{r}-\boldsymbol{r}') \times d\boldsymbol{r}'}{|\boldsymbol{r}-\boldsymbol{r}'|^3} .
\]
Thus, the expression for the electric field expression can take the form
\begin{equation}
\boldsymbol{E}(\boldsymbol{r}) = \frac{1}{2\pi} \int_{ \bigcup_{n=1}^N c_{n-1,n} } V(\phi') \frac{d\boldsymbol{r}'\times (\boldsymbol{r}-\boldsymbol{r}')}{|\boldsymbol{r}-\boldsymbol{r}'|^3} + \frac{1}{2\pi} \sum_{n=1}^N V_{n}  \left[ \int_{c_n} \frac{(\boldsymbol{r}-\boldsymbol{r}') \times d\boldsymbol{r}'}{|\boldsymbol{r}-\boldsymbol{r}'|^3} - \int_{c_{n-1} }  \frac{(\boldsymbol{r}-\boldsymbol{r}') \times d\boldsymbol{r}'}{|\boldsymbol{r}-\boldsymbol{r}'|^3} \right].
\label{auxForEFieldEq}
\end{equation}
The first term on the right of Eq.~(\ref{auxForEFieldEq}) is the circulation over $c$, since $c=\bigcup_{n=1}^N c_{n-1,n}$. On the other hand, the sum of the straight lines' integrals can be written in a most convenient form by noting that
\[
\sum_{n=1}^N V_{n} \int_{c_{n-1} }  \frac{(\boldsymbol{r}-\boldsymbol{r}') \times d\boldsymbol{r}'}{|\boldsymbol{r}-\boldsymbol{r}'|^3} = - V_1 \int_{c_{0} } \frac{(\boldsymbol{r}-\boldsymbol{r}') \times d\boldsymbol{r}'}{|\boldsymbol{r}-\boldsymbol{r}'|^3} +  \sum_{n=1}^{N-1} V_{n+1} \int_{c_{n} } \frac{(\boldsymbol{r}-\boldsymbol{r}') \times d\boldsymbol{r}'}{|\boldsymbol{r}-\boldsymbol{r}'|^3} . 
\]
Therefore,
\[
\begin{split}
\boldsymbol{E}(\boldsymbol{r})  = & \frac{1}{2\pi} \oint_{ c } V(\phi') \frac{d\boldsymbol{r}' \times (\boldsymbol{r}-\boldsymbol{r}') }{|\boldsymbol{r}-\boldsymbol{r}'|^3} + \\ & \frac{1}{2\pi}  \left[ - V_1 \int_{c_{0} } \frac{(\boldsymbol{r}-\boldsymbol{r}') \times d\boldsymbol{r}'}{|\boldsymbol{r}-\boldsymbol{r}'|^3} + \sum_{n=1}^{N-1} (V_{n}-V_{n+1})\int_{c_n} \frac{(\boldsymbol{r}-\boldsymbol{r}') \times d\boldsymbol{r}'}{|\boldsymbol{r}-\boldsymbol{r}'|^3}  + V_N \int_{c_{N} } \frac{(\boldsymbol{r}-\boldsymbol{r}') \times d\boldsymbol{r}'}{|\boldsymbol{r}-\boldsymbol{r}'|^3}\right],
\end{split}
\]
and using the periodicity conditions
\begin{equation}
c_m = c_{N+m}, \hspace{0.5cm}\mbox{and}\hspace{0.5cm} V_{N+m} = V_{m},  \hspace{0.5cm}\forall\hspace{0.5cm}  m\in \mathbb{Z}^0,  
\label{periodicityConditionsEq}
\end{equation}
then the electric field can be written as
\begin{equation}
\boldsymbol{E}(\boldsymbol{r})  = - \frac{1}{2\pi} \oint_{ c } V(\phi') \frac{(\boldsymbol{r}-\boldsymbol{r}') \times d\boldsymbol{r}'}{|\boldsymbol{r}-\boldsymbol{r}'|^3} + \frac{1}{2\pi}  \sum_{n=1}^N (V_{n}-V_{n+1}) \boldsymbol{f}(\beta_n,\boldsymbol{r}).
\label{electricFieldBiotPlusStaircaseVContributionEq}
\end{equation}
\begin{figure}[h]
\centering
\includegraphics[width=0.325\textwidth]{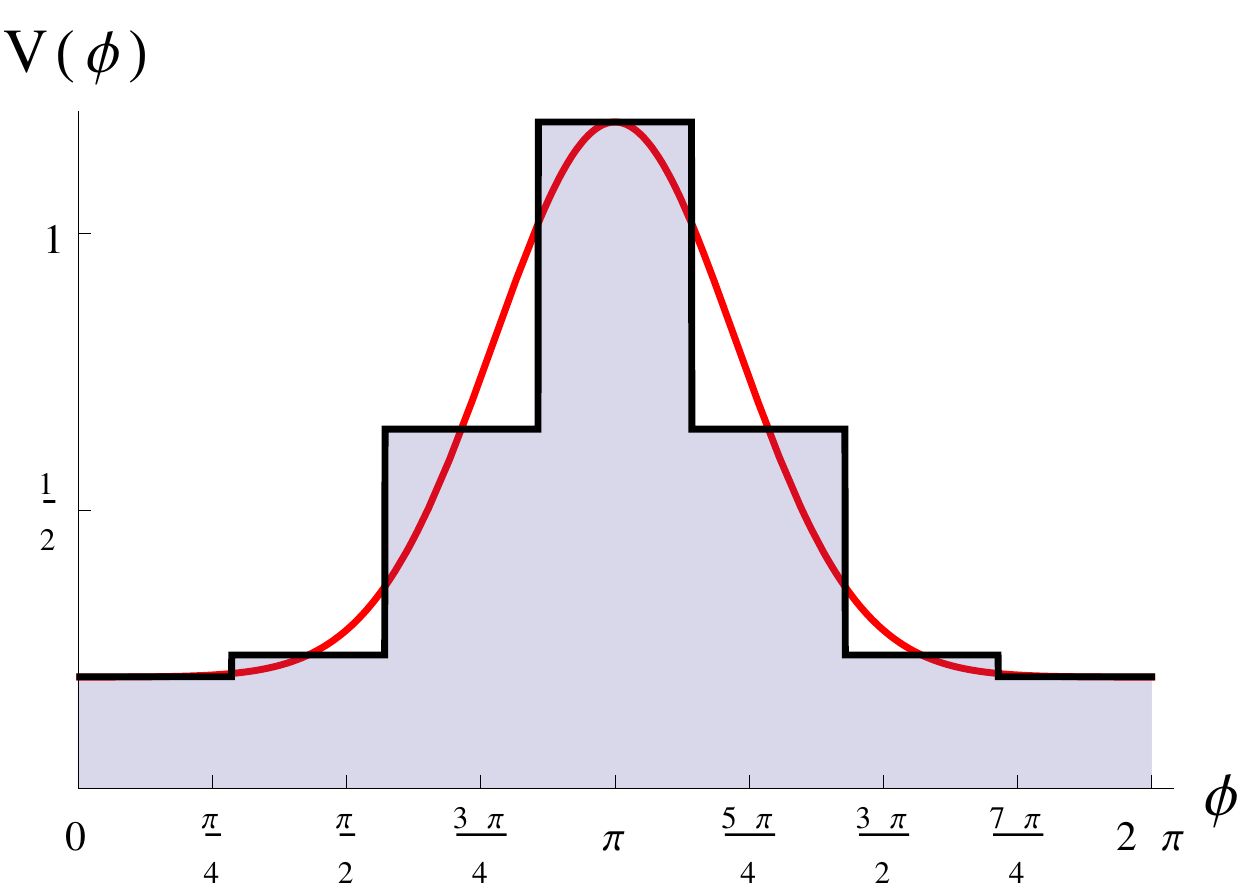}%PotentialN7
\includegraphics[width=0.325\textwidth]{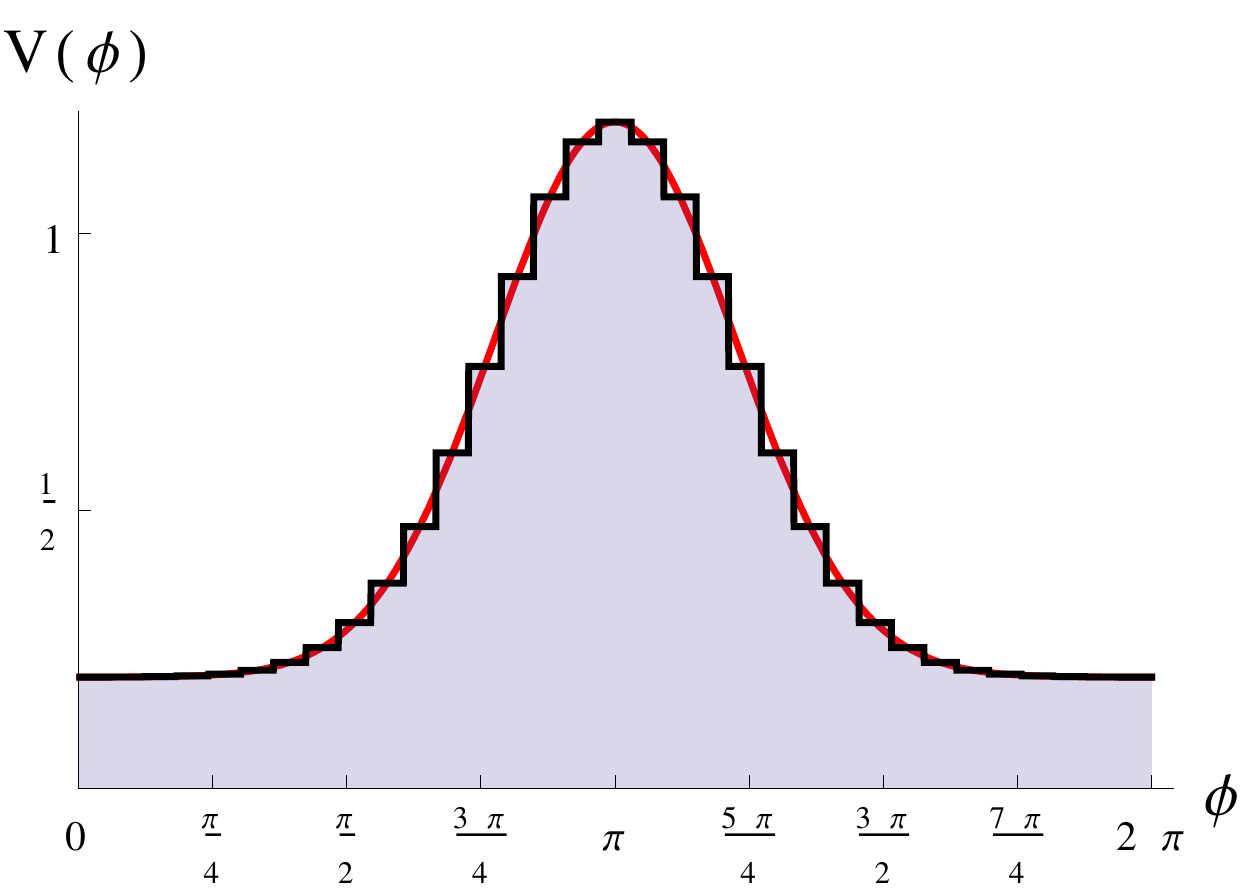}%potentialN33
\includegraphics[width=0.325\textwidth]{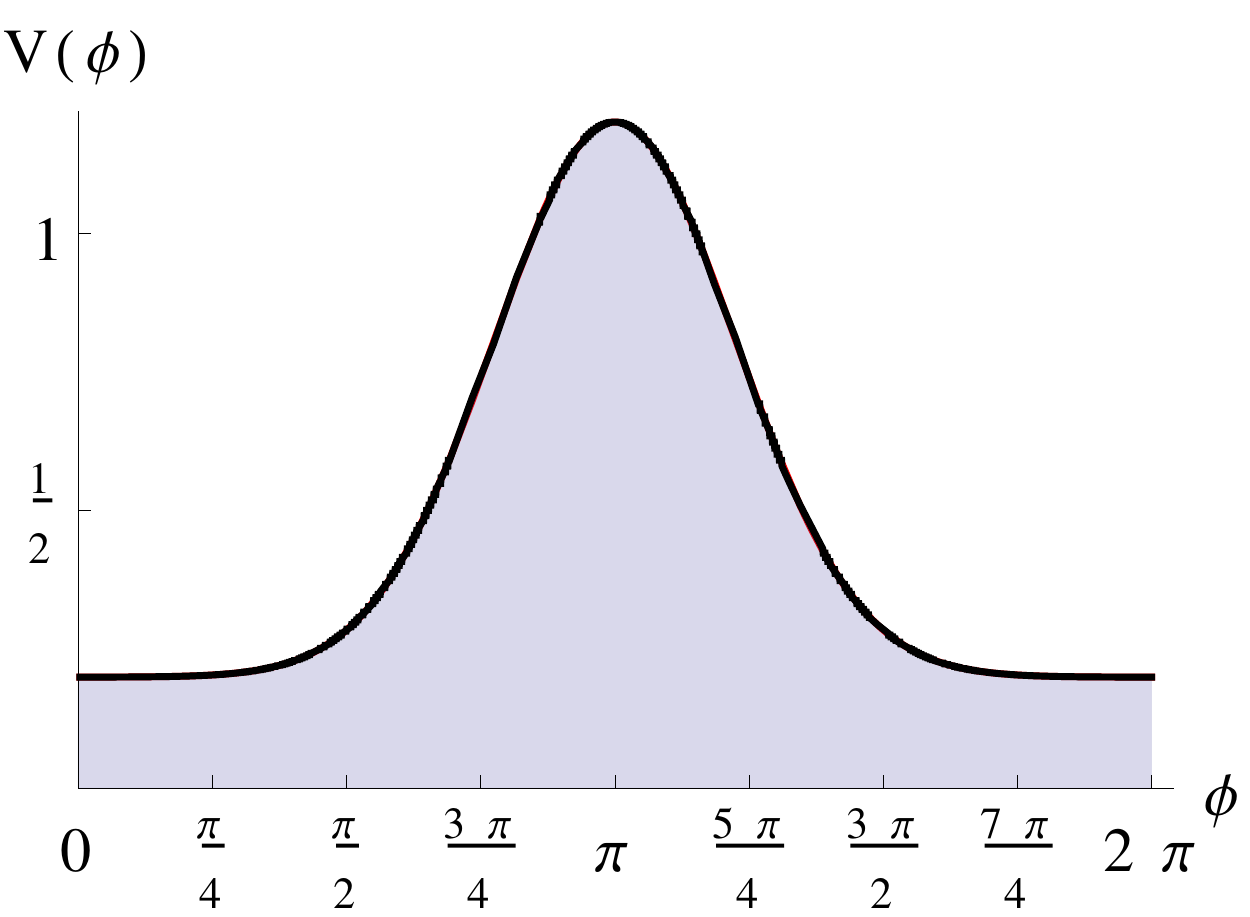}%potentialN444
    \caption{Smooth and continuous potential distribution $\mathcal{V}$ (red line) and staircase-like potential distribution $V$ (black line) for $N=6, 33,$ and $444$. $\mathcal{V}$ is chosen to be a smooth periodic function including the starting and and ending points $\mathcal{V}(0)=\mathcal{V}(2\pi)$. }
\label{discretePotentialLimitFig}
\end{figure}
Here we have defined the vector field 
\begin{equation}
\boldsymbol{f}(\phi',\boldsymbol{r}) := \int_{0}^{\mathscr{R}(\phi')} \frac{(\boldsymbol{r}-\boldsymbol{r}') \times d \rho' \hat{\rho}(\phi')}{|\boldsymbol{r}-\boldsymbol{r}'|^3}
\label{fVectorPrimaryDefEq}
\end{equation}
in Cartesian coordinates with $\hat{\rho}(\phi') = (\cos(\phi'),\sin(\phi'), 0) $. This term is a vector with dimensions of inverse length, which is analogous to the \textit{magnetic field} generated by a unitary current on the straight finite line defined from a point on $c$ located at $\left( \mathscr{R}(\beta) , \beta  \right)$ (in polar coordinates) to the origin. The integral in Eq.~(\ref{fVectorPrimaryDefEq}) can be evaluated with
\[
\boldsymbol{f}(\beta,\boldsymbol{r}) = - \hat{n}(\beta) \int_{0}^{\mathscr{R}(\beta)} \frac{\sin\gamma d \rho'}{|\boldsymbol{r}-\boldsymbol{r}'|^2}, \hspace{0.5cm}\mbox{ being }\hspace{0.5cm}\hat{n}(\beta) := \frac{\hat{\rho}(\beta) \times \boldsymbol{r}}{\sin\gamma_1},
\]
$\gamma$ the angle between $(\boldsymbol{r}-\boldsymbol{r}')$ and $\hat{\rho}(\beta)$, and  $\gamma_1$ the angle between $\boldsymbol{r}$ and $\hat{\rho}(\beta)$. Defining $\tilde{R} = |\boldsymbol{r}-\boldsymbol{r}'| \sin\gamma$ and $s'= - \tilde{R}\cot\gamma$, then $ds'=d\rho'=\tilde{R}\csc^2\gamma d\gamma$, and therefore,

\begin{equation}
\boldsymbol{f}(\beta,\boldsymbol{r}) = - \frac{\hat{n}(\beta)}{\tilde{R}} \int_{\gamma_1(\beta)}^{\gamma_2(\beta)} \sin\gamma d\gamma = \frac{1}{r} \frac{ \hat{\rho}(\beta) \times \hat{r} }{1-[\hat{\rho}(\beta) \cdot \hat{r}]^2} \left[ \frac{ \hat{\rho}(\beta) \cdot \boldsymbol{r} - \mathscr{R}(\beta) }{|\boldsymbol{r}-\mathscr{R}(\beta)\hat{\rho}(\beta)|} - \hat{\rho}(\beta) \cdot \hat{r}  \right],
    \label{fVectorWithVectorNotationEq}
\end{equation}

since $\cos\gamma_2(\beta) = (\hat{\rho}(\beta)\cdot \boldsymbol{r} - \mathscr{R}(\beta))/|\boldsymbol{r}-\boldsymbol{r}'|$. At this point we have found an analytic expression for the electric field generated by a non-uniform potential distribution inside a general planar region. Let us now suppose an specific potential field distribution in the planar region which fulfils, among other properties, the periodicity conditions. Indeed, some staircase-like (or piece-wise) distribution $V(\phi)$ of the type $V(\phi) = \mathcal{V}(\beta_{n-1} + \delta\beta_n /2 )  \hspace{0.25cm}\mbox{\textbf{if}}\hspace{0.25cm}\phi \in (\beta_{n-1},\beta_{n})$ would tend to a smooth and continuous potential $\mathcal{V}(\phi)$ as $N\rightarrow\infty$ (see Fig.~(\ref{discretePotentialLimitFig})).
If $\mathcal{V}(\phi)$ is defined as a fully periodic function in $\phi\in[0,2\pi]$, it can be expanded using the Taylor series as $V_{n+1} = \sum_{j=1}^\infty (\delta\beta^j/j!)\partial_\phi^j \mathcal{V}(\beta_{n})$. Note that not every distribution fulfill this condition \footnote{For instance, a smooth linear function $\mathcal{V}(\phi) = \phi$ can be approximated by selecting $V(\phi)$ as an staircase function satisfying periodicity conditions Eq.~(\ref{periodicityConditionsEq}). However, in that situation, there would exist a discontinuity in which $\partial_\phi \mathcal{V}$ is a Dirac delta function, and the Taylor series expansion would diverge, no matter how large becomes $N$.}. Assuming that $\mathcal{V}(\phi)$ is a suitable distribution, then the expression for a $N\rightarrow\infty$ distribution becomes
\[
\boldsymbol{E}(\boldsymbol{r})  = - \frac{1}{2\pi} \oint_{ c } V(\phi') \frac{(\boldsymbol{r}-\boldsymbol{r}') \times d\boldsymbol{r}'}{|\boldsymbol{r}-\boldsymbol{r}'|^3} - \frac{1}{2\pi} \lim_{N \to \infty} \sum_{n=1}^N \partial_\phi V(\beta_{n}) \boldsymbol{f}(\beta_n,\boldsymbol{r}) \delta \beta_n,
\]
which can be replaced by the following the continuous version:
\[
\boldsymbol{E}(\boldsymbol{r})  = - \frac{1}{2\pi} \oint_{ c } V(\phi') \frac{(\boldsymbol{r}-\boldsymbol{r}') \times d\boldsymbol{r}'}{|\boldsymbol{r}-\boldsymbol{r}'|^3} - \frac{1}{2\pi} \int_{0}^{2\pi} \partial_\phi V(\beta) \boldsymbol{f}(\beta,\boldsymbol{r}) d\beta.
\]
Since $V(0)\boldsymbol{f}(0,\boldsymbol{r})=V(2\pi)\boldsymbol{f}(2\pi,\boldsymbol{r})$, we can perform an integration by parts on the second term, resulting in the final expression for the electric field
\begin{equation}
\boxed{
\boldsymbol{E}(\boldsymbol{r})  = -\frac{1}{2\pi} \oint_{ c } V(\phi') \frac{(\boldsymbol{r}-\boldsymbol{r}') \times d\boldsymbol{r}'}{|\boldsymbol{r}-\boldsymbol{r}'|^3} + \langle V, \boldsymbol{f} \rangle,
}
\label{electricFieldBiotSavartWithCorrectionEq}
\end{equation}
including the term that accounts for the angular variations which is 
\[
\langle V, \boldsymbol{f} \rangle := \frac{1}{2\pi} \int_{0}^{2\pi}  V(\beta) \partial_\phi \boldsymbol{f}(\beta,\boldsymbol{r}) d\beta .
\]
As commented before, one can identify a \textit{magnetic analogy} or a correspondence between the line integral at the right hand of Eq.~(\ref{electricFieldBiotSavartWithCorrectionEq}) and the steady magnetic field generated by a non-uniform current circulating along $c$ in the Biot-Savart law.
In the case of the $\langle V, \boldsymbol{f} \rangle$ term, the \textit{analogous} term accounts for the magnetic contribution of the \textit{electric currents} coming from the region's center to the contour $c$ \footnote{At this point, it is necessary emphasize that we deal with the electrostatic problem which involves no currents. However, in the SE with variable potential $V(\phi)$ the lines of the electric field can be matched with the field lines of two superposed magnetic fields of its magnetostatic analogue.}. As it is expected, the $\langle V, \boldsymbol{f} \rangle$ term vanishes when $V$ is constant. More explicitly, the vector $\boldsymbol{f}$ can be written using spherical coordinates as
\[
\boldsymbol{f}(\phi',\boldsymbol{r}) = \mathscr{F}(r,\theta,\phi,\phi') \left[\sin\theta\sin(\phi-\phi')\hat{z} - \cos\theta\hat{\phi}(\phi') \right],
\]
with
\[
\mathscr{F}(r,\theta,\phi,\phi') = \frac{1}{r}\frac{1}{1-\sin^2\theta\cos^2(\phi-\phi')}\left[\frac{r\sin\theta\cos(\phi-\phi') - \mathscr{R}(\phi')}{\sqrt{r^2+\mathscr{R}^2(\phi')-2r\mathscr{R}(\phi')\sin\theta\cos(\phi-\phi')}}-\sin\theta\cos(\phi-\phi')\right],
\]
and defining $\hat{\phi}(\phi') = -\sin\phi' \hat{x} + cos\phi'\hat{y}$, as usual. Replacing the unitary vector of the Cartesian coordinates in terms of the spherical ones, and simplifying the relation, one obtains
\begin{equation}
 \boldsymbol{f}(\phi',\boldsymbol{r}) = \mathscr{F}(r,\theta,\phi,\phi') \left[\sin(\phi-\phi')\hat{\theta}(\boldsymbol{r}) + \cos\theta\cos(\phi-\phi')\hat{\phi}(\boldsymbol{r})\right].
 \label{fVectorSphericalCoordEq}
\end{equation}

Hence, $f_r = 0$, which implies that the \textit{radial component of the electric field $E_r(\boldsymbol{r})$ is only being contributed by the Biot-Savart-like term}, since $\langle V, \boldsymbol{f} \rangle$ has no effect on that component. 

\section{Circular region with an arbitrary staircase-like function $V(\phi)$}
In this section we demonstrate an exact expansion solution of the electric field generated by a circular region of radius $\mathscr{R}(\phi) = R$ with a staircase-like distribution of the electric potential $V(\phi)$.  
According to Eq.~(\ref{electricFieldBiotPlusStaircaseVContributionEq}), the general expression for the electric field is
\begin{equation}
\boldsymbol{E}(\boldsymbol{r})  = \pmb{\mathscr{E}}(\boldsymbol{r}) + \frac{\textbf{sgn}(z)}{2\pi}  \sum_{n=1}^N (V_{n}-V_{n+1}) \boldsymbol{f}(\beta_n,\boldsymbol{r}), \hspace{0.5cm}\mbox{where}\hspace{0.5cm} \pmb{\mathscr{E}}(\boldsymbol{r}) = - \frac{\textbf{sgn}(z)}{2\pi} \oint_{ c } V(\phi') \frac{(\boldsymbol{r}-\boldsymbol{r}') \times d\boldsymbol{r}'}{|\boldsymbol{r}-\boldsymbol{r}'|^3}
    \label{electricFieldBiotPlusStaircaseVContributionIIEq}
\end{equation}
is the Biot-Savart contribution whose spherical components for $z>0$ are given by 
\[
\mathscr{E}_r(\boldsymbol{r}) = \frac{R^2}{2\pi}\cos\theta \int_{ 0 }^{2\pi}   \frac{V(\phi')}{\mathscr{r}(r,\theta,\phi-\phi')^3} d\phi' ,
\]

\[
\mathscr{E}_\theta(\boldsymbol{r}) = \frac{R r}{2\pi} \int_{ 0 }^{2\pi}   \frac{V(\phi')\cos(\phi-\phi')}{\mathscr{r}(r,\theta,\phi-\phi')^3} d\phi' - \frac{R^2}{2\pi}\sin\theta \int_{ 0 }^{2\pi}   \frac{V(\phi')}{\mathscr{r}(r,\theta,\phi-\phi')^3} d\phi',
\]
and
\[
\mathscr{E}_\phi(\boldsymbol{r}) = -\frac{R r}{2\pi} \cos\theta \int_{ 0 }^{2\pi}   \frac{V(\phi')\sin(\phi-\phi')}{\mathscr{r}(r,\theta,\phi-\phi')^3} d\phi'. 
\]
Here we have denoted the magnitude of $\boldsymbol{r}-\boldsymbol{r}'$ by $\mathscr{r}(r,\theta,\phi-\phi') =\sqrt{R^2+r^2-2Rr\sin\theta\cos(\phi-\phi')}$. 
Now, the radial component can be simplified by including the definition of the staircase-like distribution of  $V(\phi)$, resulting in 
\[
\mathscr{E}_r(\boldsymbol{r}) = \frac{R^2}{2\pi}\cos\theta \sum_{n=1}^N V_n \int_{ \beta_{n-1} }^{\beta_n} \frac{1}{\mathscr{r}(r,\theta,\phi-\phi')^3} d\phi' .
\]
Similar expressions can be written for the other two components of the electric field in the sense of 
\[
\int_{ 0 }^{2\pi}   V(\phi') g(\boldsymbol{r},\phi') d\phi' \longrightarrow \sum_{n=1}^N V_n \int_{ \beta_{n-1} }^{\beta_n} g(\boldsymbol{r},\phi') d\phi'.
\]

Our analytic approach relies in the inclusion of an exact expansion to ease the evaluation of some integral terms.
Hence, we recall the expansion
\begin{equation}
\frac{1}{(1-\chi)^{\alpha/2}} = 1 + \frac{\alpha}{2}\chi + \frac{\alpha}{8}(\alpha+2)\chi^2 + \cdots = \sum_{n=0}^{\infty} (-1)^n \binom{-\alpha/2}{n} \chi^n,
\label{binomialTheoremSeriesExpansionEq}    
\end{equation}
for which application it is convenient to write the inverse $\mathscr{r}(r,\theta,\phi-\phi')^{-1} = 1/(\sqrt{R^2+r^2}\sqrt{1-\xi\cos(\phi-\phi')})$,
with $\xi$ defined as $\xi(r,\theta) := \frac{2 R r \sin\theta}{R^2+r^2}$.
Hence, the one term of the angular expression can be expanded with (\ref{binomialTheoremSeriesExpansionEq}), resulting in
\begin{equation}
\frac{\cos^m(\phi-\phi')}{\mathtt{r}(r,\theta,\phi-\phi')^\alpha} = \frac{1}{\left(\sqrt{R^2+r^2}\right)^\alpha} \sum_{n=0}^{\infty} (-1)^n \binom{-\alpha/2}{n} \xi^n \left\{ \zeta_{n+m} + \frac{2}{2^{n+m}} \sum_{k=0}^{\floor*{(n+m-1)/2}} \binom{n+m}{k} \cos[\nu_{nmk}(\phi-\phi')] \right\}
\label{cosBetaOverREq}
\end{equation}
for any positive integer $m$ (including zero), $\nu_{nmk}=n+m-2k$, $\floor*{z}$ the floor function, and
\[
\zeta_n := \frac{1}{2^n} \binom{n}{n/2} \hspace{0.5cm}\mbox{\textbf{if}}\hspace{0.5cm} n \in 2\mathbb{N}^0 \hspace{0.5cm}\mbox{\textbf{else}}\hspace{0.5cm} 0.
\]

\begin{figure}[h]
  \begin{minipage}[b]{0.33\linewidth}
    
   \includegraphics[width=1.0\textwidth]{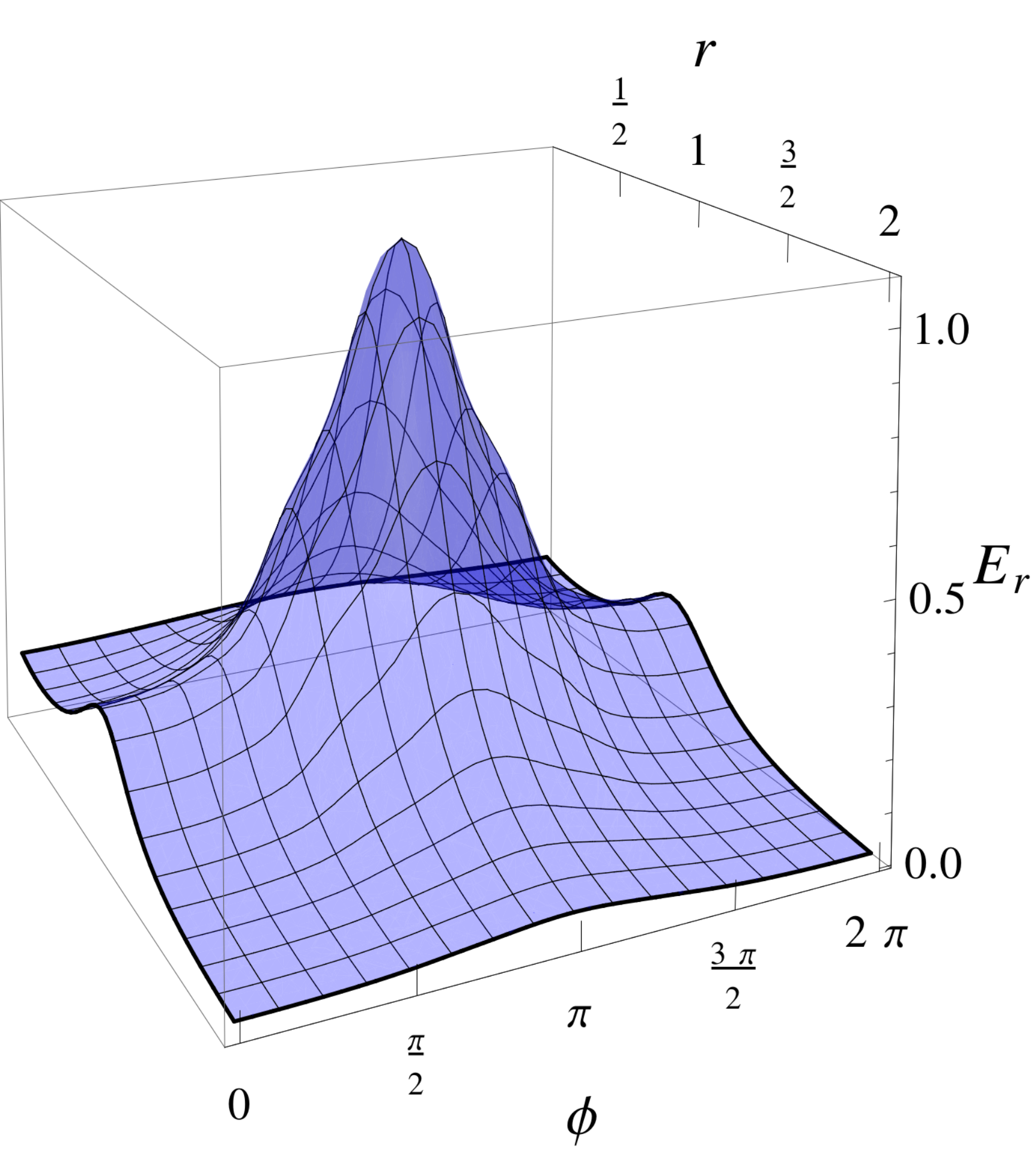}
   \caption*{(a)}%ErSurfaceN7.pdf
    
  \end{minipage} 
  \begin{minipage}[b]{0.33\linewidth}
    
    \includegraphics[width=1.0\textwidth]{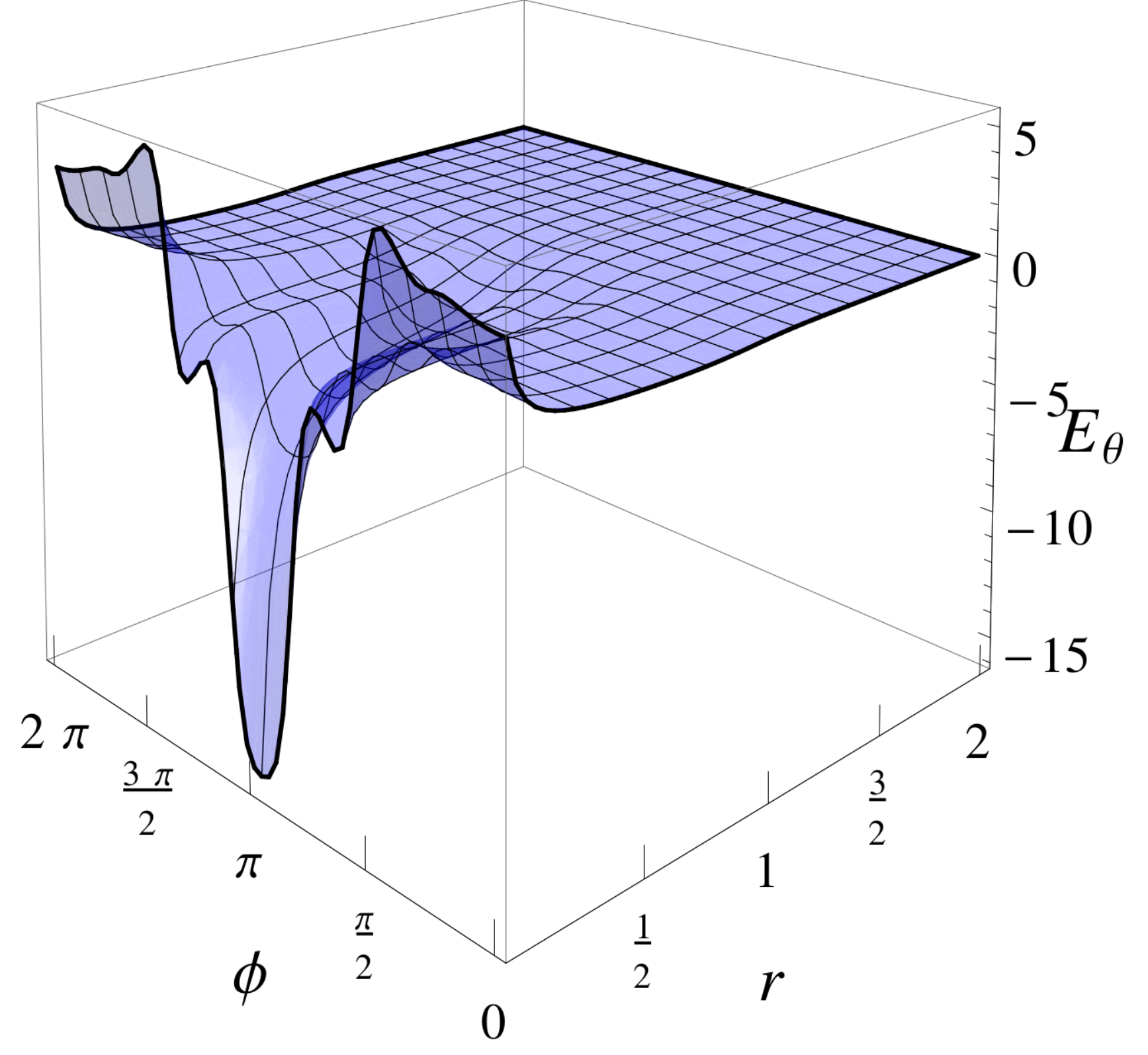}
    \caption*{(b) } %EThetaSurfaceN7.pdf
  
  \end{minipage} 
  \begin{minipage}[b]{0.33\linewidth}
    
    \includegraphics[width=1.0\textwidth]{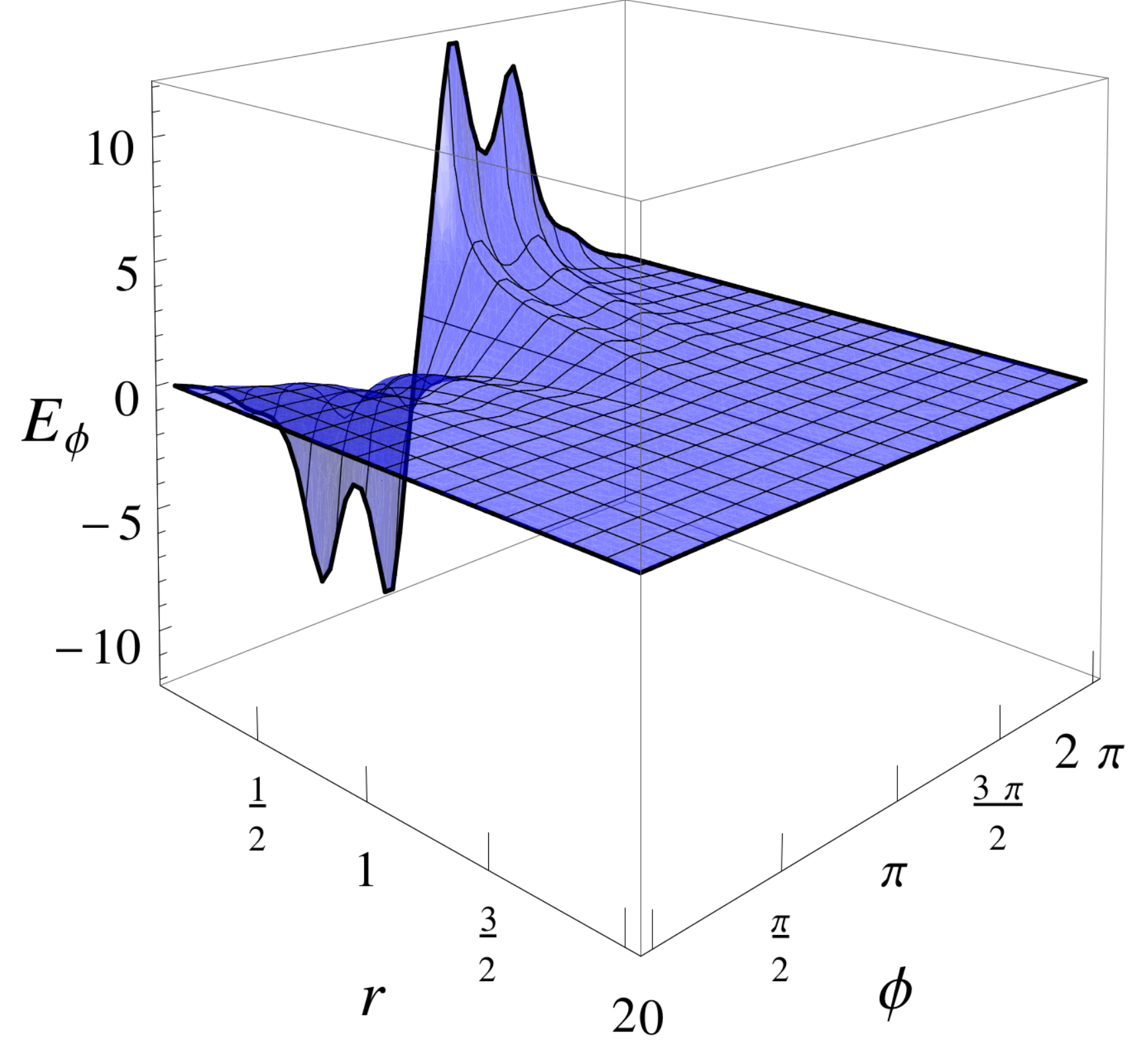}
    \caption*{(c) } %EphiSurfaceN7.pdf
  
  \end{minipage}
  \caption{Electric field components. (a) $E_r(\boldsymbol{r})$, (b) $E_\theta(\boldsymbol{r})$, and (c) $E_\phi(\boldsymbol{r})$ components results from setting $\theta = 2\pi/5$ and $V(\phi)$ as the one at the left side of  Fig.~\ref{discretePotentialLimitFig}.}
   \label{EComponentsSurfFig} 
\end{figure}

We use the previous expansion to evaluate the integrals of the Biot-Savart-like term contribution. For instance, to evaluate the integral  
\begin{equation}
    \sum_{n=1}^N V_n \int_{ \beta_{n-1} }^{\beta_n} \frac{\cos^m (\phi-\phi')}{\mathscr{r}(r,\theta,\phi-\phi')^3} d\phi'
    \label{cosIntegralDefEq}
\end{equation}
that is included in the computation of $\mathscr{E}_r(\boldsymbol{r})$ and $\mathscr{E}_\theta(\boldsymbol{r})$, where $m$ should be set as 0 or 1 depending on the case. This integral can be evaluated by applying (\ref{cosBetaOverREq}), which results in
\[
\begin{split}
\sum_{n=1}^N V_n \int_{ \beta_{n-1} }^{\beta_n} \frac{\cos^m (\phi-\phi')}{\mathscr{r}(r,\theta,\phi-\phi')^3} d\phi' & =  \frac{1}{\left(R^2+r^2\right)^{3/2}} \sum_{n=0}^{\infty} (-1)^s \binom{-3/2}{s} \xi^s \left\{  \zeta_{s+m} \sum_{n=1}^{N-1} V_n(\beta_{n} - \beta_{n-1}) + \right.  \\ & \left. \frac{2}{2^{s+m}} \sum_{k=0}^{\floor*{(s+m-1)/2}} \binom{s+m}{k}  \sum_{n=1}^N \frac{\sin[\nu_{smk}(\phi-\beta_{n-1})] -  \sin[\nu_{smk}(\phi-\beta_{n})]}{\nu_{smk}} \right\}.
\end{split}
\]

\begin{figure}[h] 
  \begin{minipage}[b]{0.33\linewidth}
    
   \includegraphics[width=1.0\textwidth]{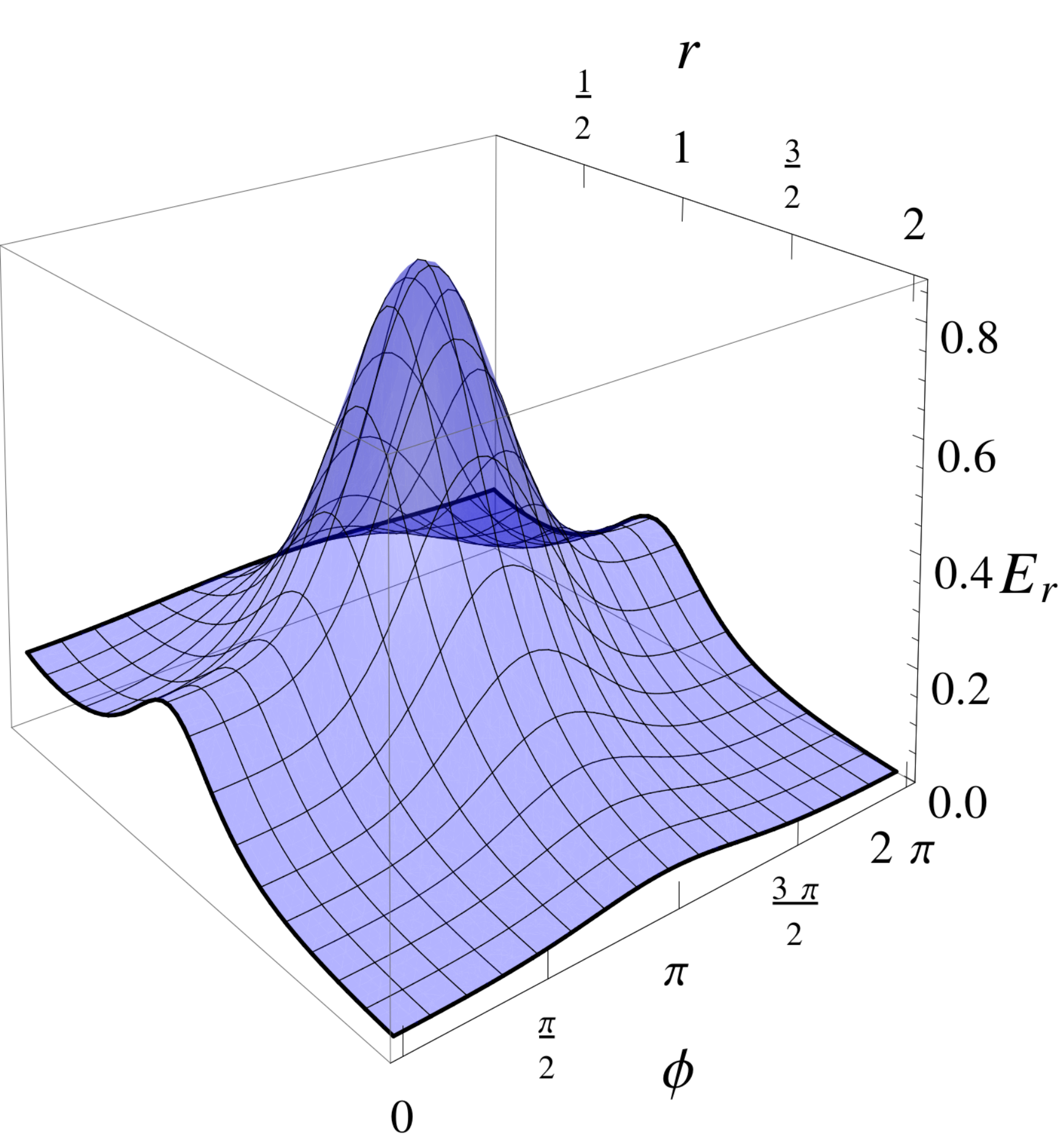}
   \caption*{(a)}%ErSurfaceN33.pdf
    
  \end{minipage} 
  \begin{minipage}[b]{0.33\linewidth}
    
    \includegraphics[width=1.0\textwidth]{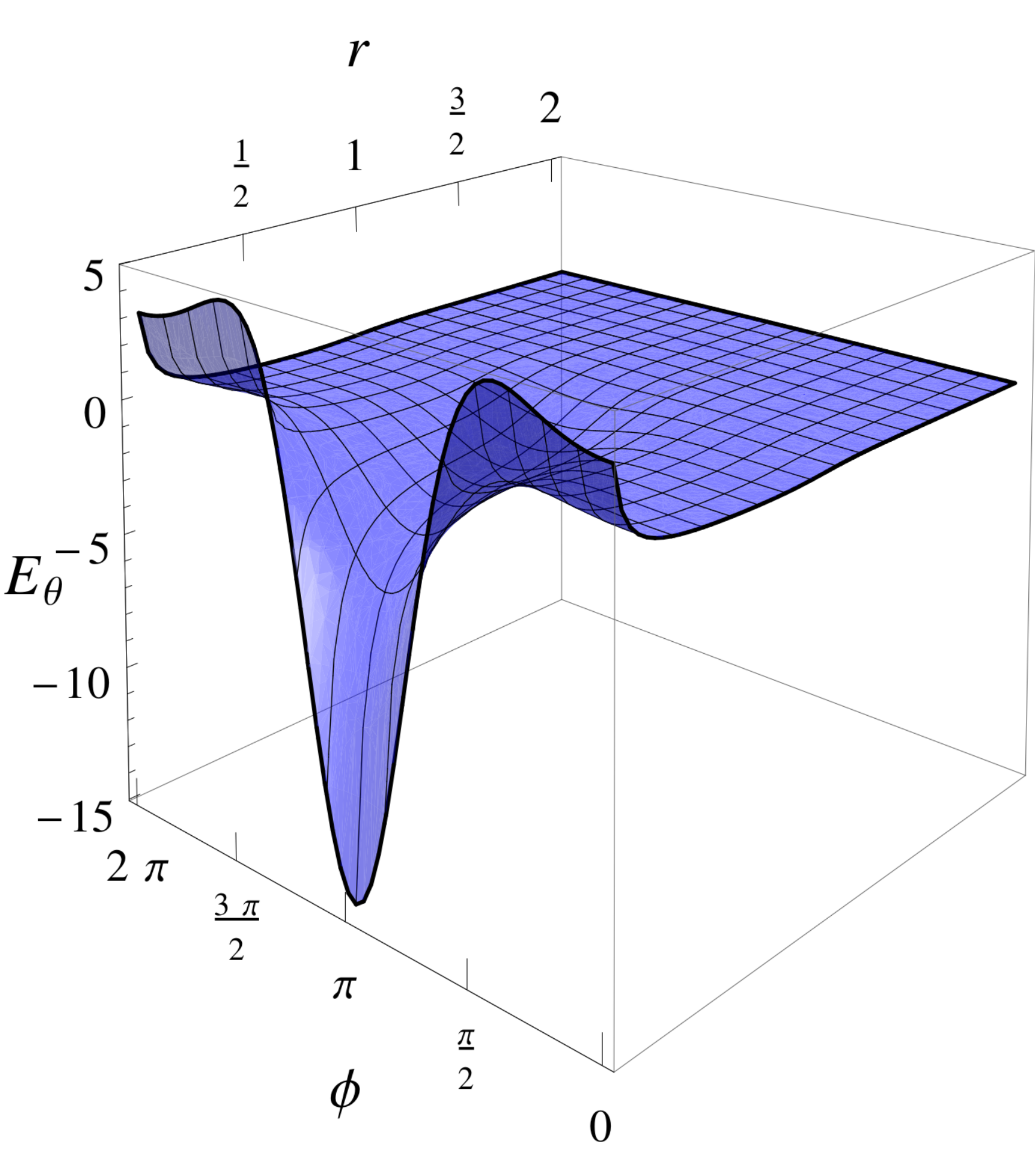}
    \caption*{(b) } %EThetaSurfaceN33.pdf
  
  \end{minipage} 
  \begin{minipage}[b]{0.33\linewidth}
    
    \includegraphics[width=1.0\textwidth]{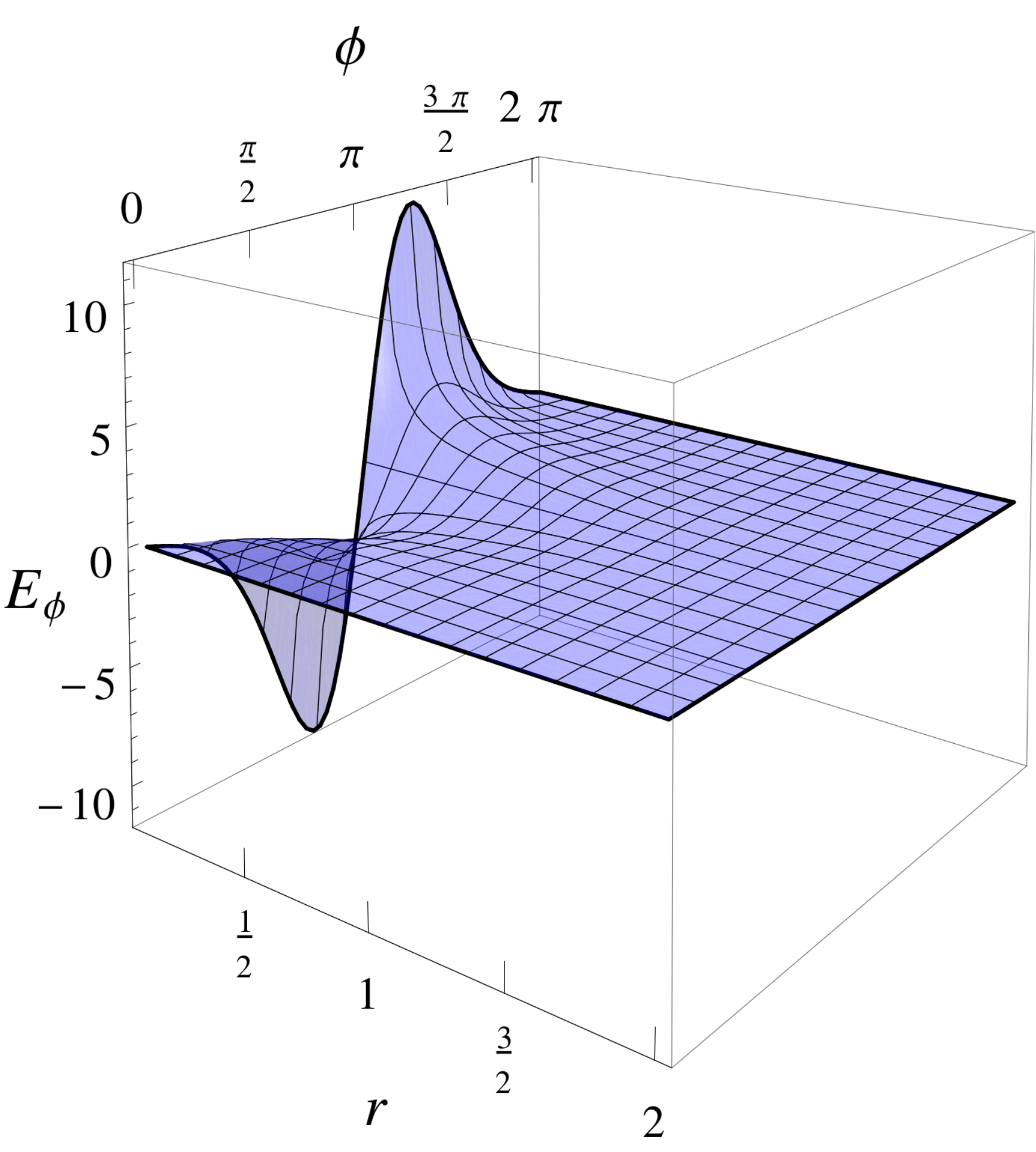}
    \caption*{(c) } %EphiSurfaceN33.pdf
  
  \end{minipage}
  \caption{Electric field components. (a) $E_r(\boldsymbol{r})$, (b) $E_\theta(\boldsymbol{r})$, and (c) $E_\phi(\boldsymbol{r})$ components results from setting $\theta = 2\pi/5$ and $V(\phi)$ as the one at the right side of  Fig.~\ref{discretePotentialLimitFig}.}
  
  \label{EComponentsSurfIIFig} 
\end{figure}

Since $\beta_N = 2\pi$ and $\beta_0 = 0$, we operate the first term at the right hand side of the previous expression such that
\[
\sum_{n=1}^N V_n(\beta_n - \beta_{n-1}) = \sum_{n=1}^N V_n \beta_n - \sum_{n=0}^{N-1} V_{n+1}\beta_n = 2\pi V_N + \sum_{n=1}^{N-1} (V_n - V_{n+1})\beta_n.
\]
Similarly, using the periodicity conditions of Eq.~(\ref{periodicityConditionsEq}), one obtains the following expression for the second term,
\[
\sum_{n=1}^N \left\{\sin[\nu(\phi-\beta_{n})] -  \sin[\nu(\phi-\beta_{n-1})]\right\} = \sum_{n=1}^N (V_n - V_{n+1}) \sin[\nu(\phi-\beta_{n})]  \hspace{0.5cm}\forall\hspace{0.5cm}  \nu\in \mathbb{Z}^0.
\]

Including the previous results into the integral expression (\ref{cosIntegralDefEq}), one obtains
\[
\begin{split}
\sum_{n=1}^N V_n \int_{ \beta_{n-1} }^{\beta_n} \frac{\cos^m (\phi-\phi')}{\mathscr{r}(r,\theta,\phi-\phi')^3} d\phi' & =  \frac{1}{\left(R^2+r^2\right)^{3/2}} \sum_{n=0}^{\infty} (-1)^s \binom{-3/2}{s} \xi^s \left\{  \zeta_{s+m} \left[\sum_{n=1}^{N-1} (V_n - V_{n+1})\beta_n + 2\pi V_N \right] - \right.  \\ & \left. \frac{2}{2^{s+m}} \sum_{k=0}^{\floor*{(s+m-1)/2}} \binom{s+m}{k}  \frac{1}{\nu_{smk}} \sum_{n=1}^N (V_n - V_{n+1}) \sin[\nu_{smk}(\phi-\beta_{n})]   \right\}   .
\end{split}  
\]
Note that in the case of uniform potential distributions $V_n = V_1 \hspace{0.25cm} \forall \hspace{0.25cm} n \in \{1,2,\ldots,N\}$, the previous integral (divided by $V_1$) takes the form
\begin{equation}
\int_{0}^{2\pi} \frac{\cos^m (\phi-\phi')}{\mathscr{r}(r,\theta,\phi-\phi')^3} d\phi' = \frac{1}{\left(R^2+r^2\right)^{3/2}} \sum_{n=0}^{\infty} (-1)^s \binom{-3/2}{s} \xi^s \zeta_{s+m} (2\pi) ,
\label{integralExpansionForUniformRingEq}
\end{equation}
and therefore,
\begin{equation}
\begin{split}
\sum_{n=1}^N V_n \int_{ \beta_{n-1} }^{\beta_n} \frac{\cos^m (\phi-\phi')}{\mathscr{r}(r,\theta,\phi-\phi')^3} d\phi' & =  V_1 \int_{0}^{2\pi} \frac{\cos^m (\phi-\phi')}{\mathscr{r}(r,\theta,\phi-\phi')^3} d\phi' + \\ & \left. \frac{1}{\left(R^2+r^2\right)^{3/2}} \sum_{n=0}^{\infty} (-1)^s \binom{-3/2}{s} \xi^s \sum_{n=1}^N (V_n - V_{n+1}) \tau_{m,n}^s(\phi)   \right\},
\end{split}  
\label{cosIntegralExpansionEq}
\end{equation}
with the introduction of the term
\[
\tau_{m,n}^s (\phi) :=  \zeta_{s+m}\beta_n - \frac{2}{2^{s+m}} \sum_{k=0}^{\floor*{(s+m-1)/2}} \binom{s+m}{k}  \frac{1}{\nu_{smk}} \sin[\nu_{smk}(\phi - \beta_{n})].
\]
Another integration that is required in the computation of $\mathscr{E}_\phi(\boldsymbol{r})$ is
\begin{equation}
\sum_{n=1}^N V_n \int_{ \beta_{n-1} }^{\beta_n} \frac{\sin (\phi-\phi')}{\mathscr{r}(r,\theta,\phi-\phi')^3} d\phi'  .
\label{sinIntegralDefEq}
\end{equation}
We may employ the same strategy presented before for Eq.~(\ref{cosIntegralDefEq}) in order to evaluate Eq.~(\ref{sinIntegralDefEq}).
By setting $m=0$, and using
\[
\int_{\beta_{n-1}}^{\beta_n} \cos[\nu(\phi-\phi')] \sin(\phi-\phi')d\phi' = \frac{1}{2}\sum_{\sigma\in\left\{-1,1\right\}} \frac{\sigma}{\nu+\sigma}\left\{ \cos[(\nu+\sigma)(\beta_n-\phi)] - \cos[(\nu+\sigma)(\beta_{n-1}-\phi)] \right\},
\]
if $|\nu| \neq 1$, otherwise
\[
\int_{\beta_{n-1}}^{\beta_n} \cos[(\phi-\phi')] \sin(\phi-\phi')d\phi' = \frac{1}{4}\left\{\cos[2(\phi-\beta_{n})]-\cos[2(\phi-\beta_{n-1})]\right\} ,
\]
that term can be now written as 
\begin{equation}
\sum_{n=1}^N V_n \int_{ \beta_{n-1} }^{\beta_n} \frac{\sin (\phi-\phi')}{\mathscr{r}(r,\theta,\phi-\phi')^3} d\phi' = \frac{1}{\left(R^2+r^2\right)^{3/2}} \sum_{n=0}^{\infty} (-1)^s \binom{-3/2}{s} \xi^s \sum_{n=1}^N (V_n - V_{n+1}) \tau^{(s)}(\phi - \beta_{n}).
\label{sinIntegralExpansionEq}
\end{equation}
Here we have used the periodicity condition given by Eq.~(\ref{periodicityConditionsEq}) and the introduction of $\tau^{(s)}(\phi)$ which is defined as follows:
\[
\tau^{(s)}(\phi) =  \zeta_{s}\cos(\phi) + \frac{1}{2^{s}} \sum_{k=0}^{\floor*{(s-1)/2}} \binom{s}{k} \left\{ \sum_{\sigma\in\left\{-1,1\right\}}  \left(\frac{\sigma}{\nu_{sk}+\sigma}\right) \cos[(\nu_{sk}+\sigma)\phi] \hspace{0.25cm}\mbox{\textbf{if}}\hspace{0.25cm} |s-2k| \neq 1 \hspace{0.25cm}\mbox{\textbf{else}}\hspace{0.25cm} \frac{\cos(2\phi)}{2} \right\}.
\]

Hence, the radial component of the Biot-Savart-like contribution takes the form
 \[
 \mathscr{E}_r(\boldsymbol{r}) = \frac{R^2 \cos\theta}{2\pi} \left\{ V_1\int_{0}^{2\pi} \frac{d\phi'}{\mathscr{r}(r,\theta,\phi-\phi')^3} + \frac{1}{\left(R^2+r^2\right)^{3/2}} \sum_{n=0}^{\infty} (-1)^s \binom{-3/2}{s} \xi^s \sum_{n=1}^N (V_n - V_{n+1}) \tau_{0,n}^s(\phi) \right\} .
 \]

Note that the integral term
\[
\left( \mathscr{E}_r(\boldsymbol{r}) \right)_{unif.} = \frac{R^2 \cos\theta}{2\pi} V_1\int_{0}^{2\pi} \frac{d\phi'}{\mathscr{r}(r,\theta,\phi-\phi')^3}
\]
is the radial component of a electric field generated by a circular region with a fixed potential $V_1$ and zero at the remaining part of the $(z=0)$-plane. Similarly, 
\[
 \mathscr{E}_{\theta}(\boldsymbol{r}) =  \left( \mathscr{E}_{\theta}(\boldsymbol{r} ) \right)_{unif.} +\frac{1}{2\pi} \frac{R}{\left(R^2+r^2\right)^{3/2}} \sum_{n=0}^{\infty} (-1)^s \binom{-3/2}{s} \xi^s \sum_{n=1}^N (V_n - V_{n+1})\left[ r \tau_{1,n}^s(\phi) - R\sin\theta \tau_{0,n}^s(\phi) \right],
 \]
where
\[
\left( \mathscr{E}_{\theta}(\boldsymbol{r}) \right)_{unif.} = \frac{R r V_1}{2\pi} \int_{ 0 }^{2\pi}   \frac{\cos(\phi-\phi')}{\mathscr{r}(r,\theta,\phi-\phi')^3} d\phi' - \frac{R^2 V_1}{2\pi}\sin\theta \int_{ 0 }^{2\pi}   \frac{1}{\mathscr{r}(r,\theta,\phi-\phi')^3} d\phi'.
\]
Also,
\[
 \mathscr{E}_{\phi}(\boldsymbol{r}) = \left( \mathscr{E}_{\phi}(\boldsymbol{r}) \right)_{unif.} -\frac{Rr\cos\theta}{2\pi} \frac{1}{\left(R^2+r^2\right)^{3/2}} \sum_{n=0}^{\infty} (-1)^s \binom{-3/2}{s} \xi^s \sum_{n=1}^N (V_n - V_{n+1}) \tau^{(s)}(\phi-\beta_n),
\]
with $\left( \mathscr{E}_{\phi}(\boldsymbol{r}) \right)_{unif.} = 0$ since the circular region at constant potential is axially symmetric. We may use the fact that $\pmb{\mathscr{E}}(\boldsymbol{r})_{unif.}$  is analogous to the magnetic field $\boldsymbol{B}(\boldsymbol{r})_{ring}$ of a circular loop carrying a uniform current $i_o$ by a proportional constant equal to $\mu_o i_o / 2 V_1$ with $\mu_o$ as the magnetic permeability. In general, the magnetic field can be obtained via the vector potential. In our case, we may also associate a vector potential $\boldsymbol{\Theta}(\boldsymbol{r})$ such that $\pmb{\mathscr{E}}(\boldsymbol{r})_{unif.} = \textbf{sgn}(z) \mbox{curl} \boldsymbol{\Theta}(\boldsymbol{r})$ where     
\[
 \boldsymbol{\Theta}(\boldsymbol{r}) = \frac{V_1}{2\pi} \frac{4  R}{\sqrt{R^2+r^2+2Rr\sin\theta}} \left[ \frac{(2-\gamma^2)K(\gamma^2)-2\b{E}(\gamma^2)}{\gamma^2} \right] \hat{\phi}(\boldsymbol{r}),\hspace{0.2cm}\mbox{with}\hspace{0.2cm}\gamma^2 = \frac{4 R r \sin\theta}{R^2+r^2+2Rr\sin\theta},
\]
and $K(\gamma^2)$, $\b{E}(\gamma^2)$ the complete elliptic integrals of the first and second kind, respectively \cite{jackson1999classical,abramowitz1965handbook}. The bar under letter E is used to avoid confusions between the electric field and the elliptic integral. The application of the rotational on the vector potential can be demanding algebraically but it is a standard problem that has been already solved in \cite{garrett1963calculation,simpson2001simple}. An alternative may be to use the series expansions of Eq.~(\ref{integralExpansionForUniformRingEq}) to evaluate $\left( \mathscr{E}_{r}(\boldsymbol{r}) \right)_{unif.}$ and $\left( \mathscr{E}_{\theta}(\boldsymbol{r}) \right)_{unif.}$ by setting $m=1$ or $m=0$, depending on the case.

\begin{figure}[H]  
  \begin{minipage}[b]{0.45\linewidth}
    
   \includegraphics[width=1\textwidth]{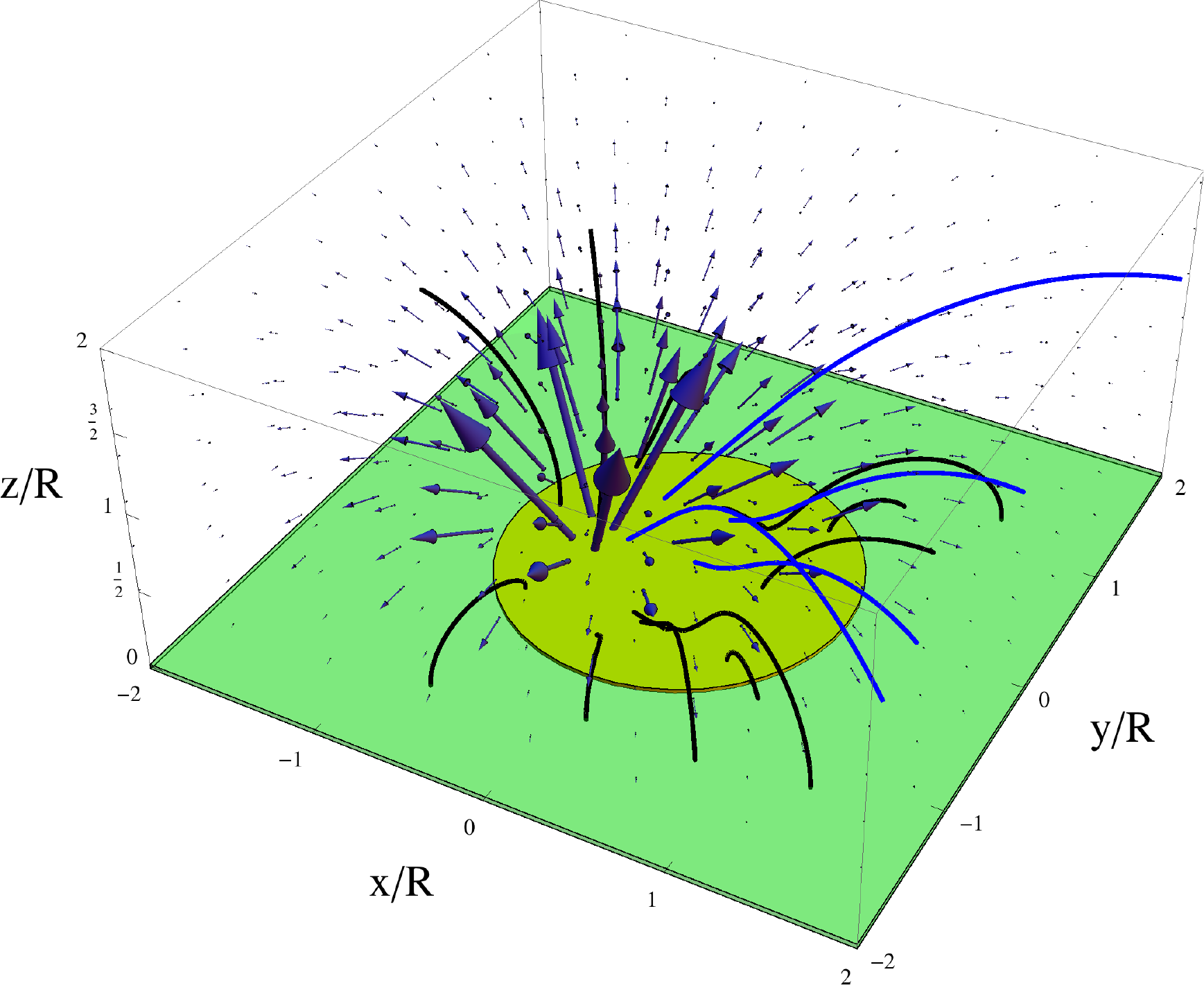}
   \caption*{(a)}%EVectorFieldPlotN33.pdf
    
  \end{minipage} 
  \begin{minipage}[b]{0.5\linewidth}
    
    \includegraphics[width=1\textwidth]{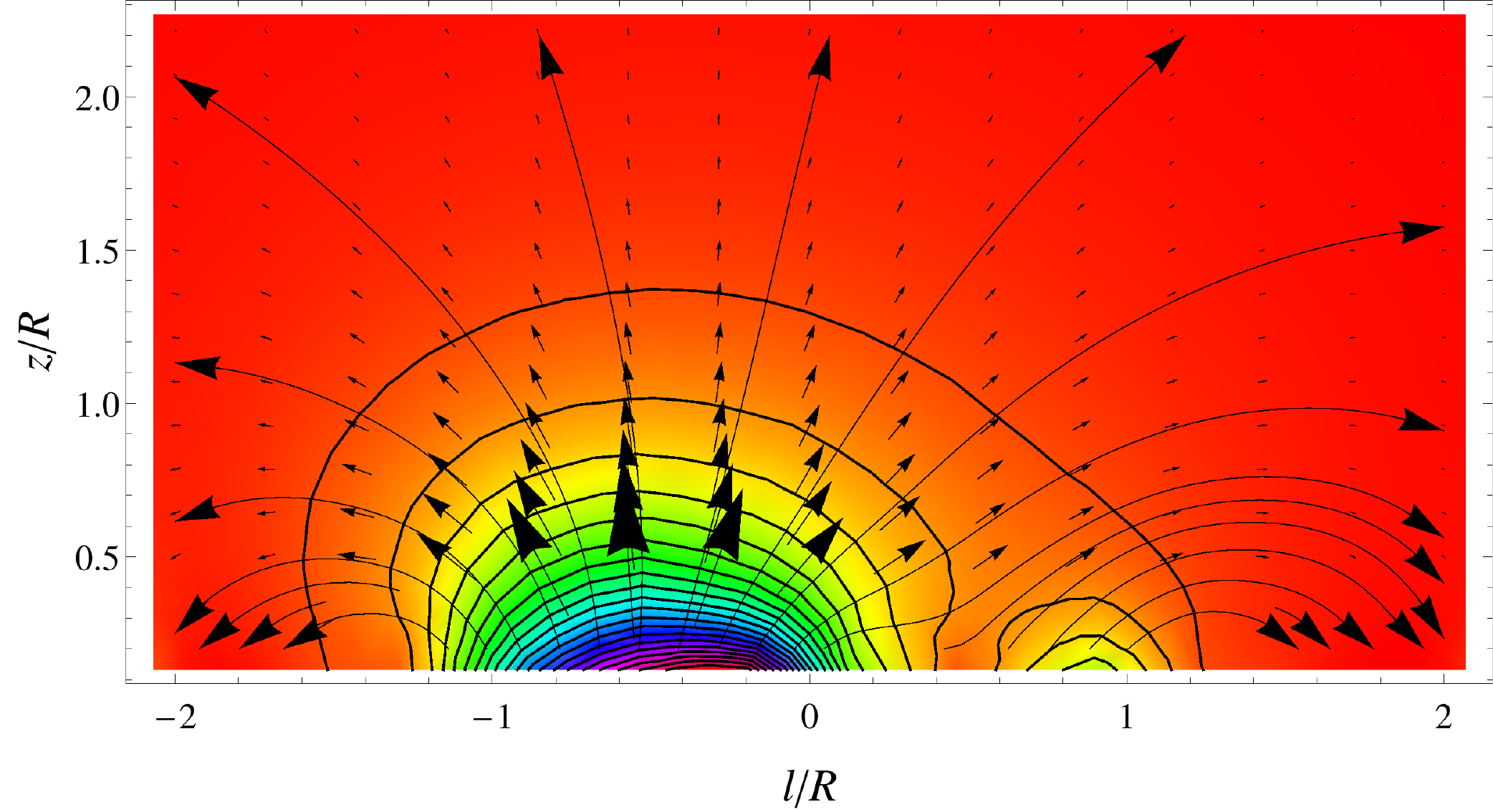}
    \caption*{(b) $\phi=0$} %vectorFieldPhiZeroN33.pdf
  
  \end{minipage} 
  \begin{minipage}[b]{0.5\linewidth}
    
    \includegraphics[width=1\textwidth]{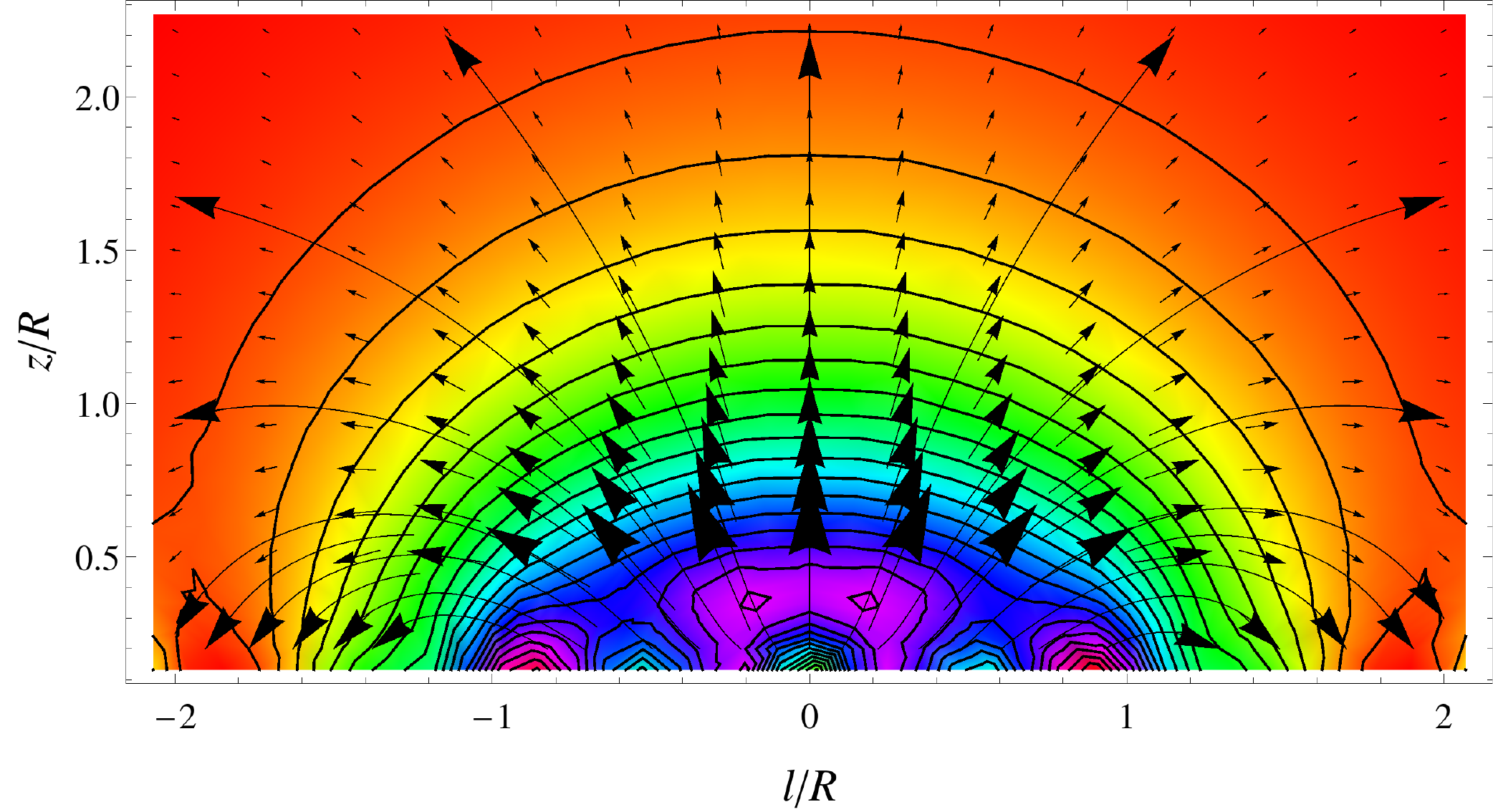}
    \caption*{(c) $\phi=\pi/2$} %vectorFieldPhi05PiN33.pdf
  
  \end{minipage}
  \hfill
  \begin{minipage}[b]{0.5\linewidth}
    \includegraphics[width=1\textwidth]{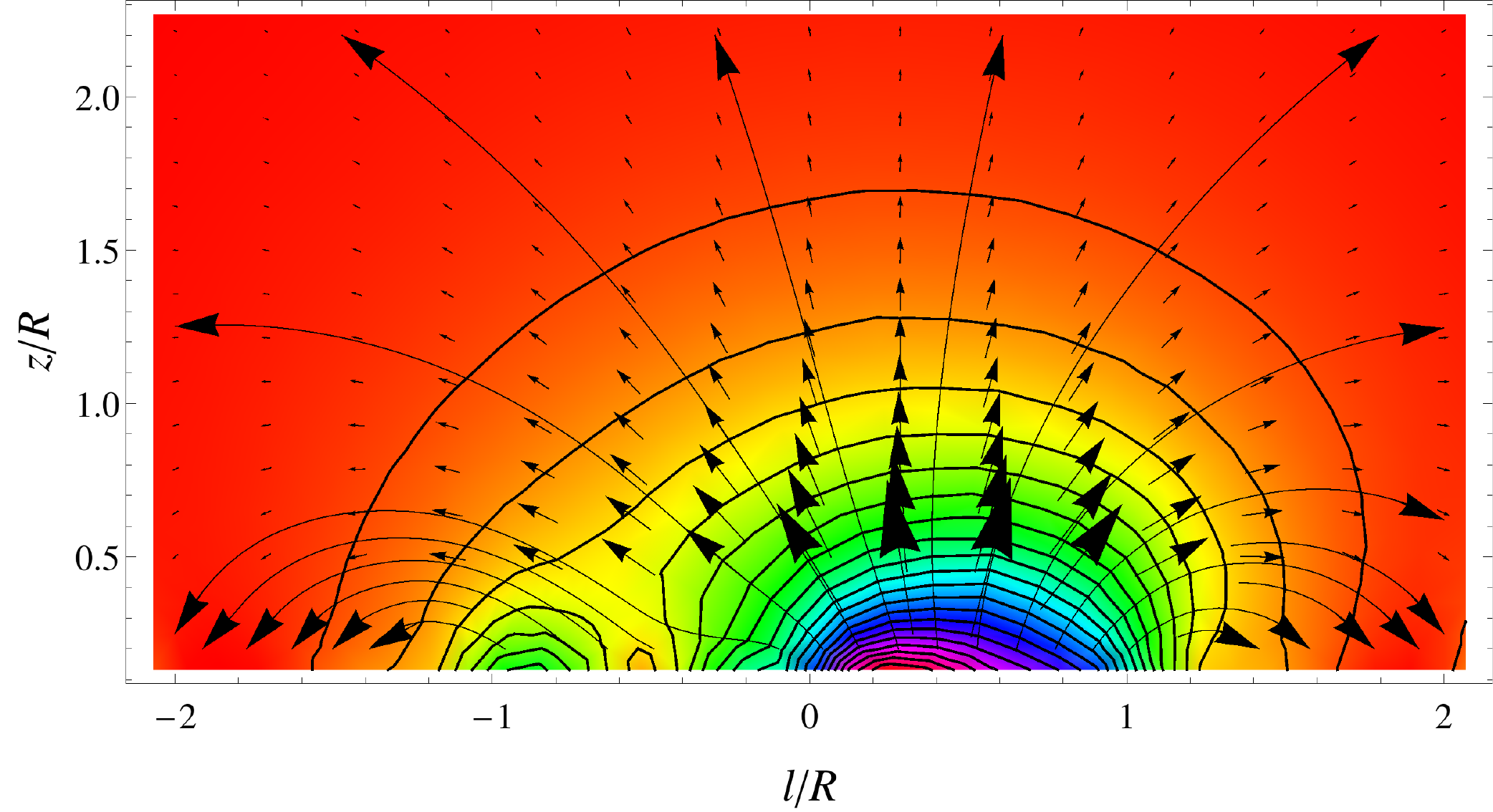}
    \caption*{(d) $\phi=3\pi/4$} %vectorFieldPhi3Pi4N33.pdf
    
  \end{minipage}
  
  \caption{ Electric field. (a) Vector field and few stream lines computed with Eq.~(\ref{EVectorFieldStairCaseVExpansionEq}). Computations are performed using the Gaussian distribution for $\mathcal{V}$ in Eq.~(\ref{gaussianContinousVEq}) and setting $N=33$. Plots (b), (c) and (d) are projections of $\boldsymbol{E}(\boldsymbol{r})$ onto the planes $\phi=0, \pi/2$, and $3\pi/4$, respectively.  }
  
  \label{vectorFieldEPhiConstFig}
  
\end{figure}

We obtain that the components of $\left(\pmb{\mathscr{E}}(\boldsymbol{r})\right)_{unif.}$ for $z > 0$ are  
\[
\left( \mathscr{E}_{r}(\boldsymbol{r}) \right)_{unif.} = \frac{R^2 V_1}{2\pi} \frac{4\cos\theta}{\mathscr{r}_{-}(r,\theta)^2 \mathscr{r}_{+}(r,\theta)} \b{E}\left[ \frac{4rR\sin\theta}{ \mathscr{r}_{+}(r,\theta)^2 }\right],
\]
and  
\[
\left( \mathscr{E}_{\theta}(\boldsymbol{r}) \right)_{unif.} = \frac{V_1}{\pi} \frac{\csc\theta}{\mathscr{r}_{-}^2 \mathscr{r}_{+}} \left[  (r^2+R^2\cos(2\theta))\b{E}\left( \frac{4rR\sin\theta}{ \mathscr{r}_{+}^2 }\right) - \mathscr{r}_{-}^2 K\left( \frac{4rR\sin\theta}{ \mathscr{r}_{+}^2 }\right)\right]
\]
where we have defined $\mathscr{r}_{\pm}(r,\theta) = \sqrt{r^2 + R^2 \pm 2rR\sin\theta}$. With all the previous results in hand, the Biot-Savart contribution can be written as
\[
\boldsymbol{\mathscr{E}}(\boldsymbol{r}) = \left( \boldsymbol{\mathscr{E}}(\boldsymbol{r}) \right)_{unif.} + \frac{1}{2\pi} \frac{R}{\left(R^2+r^2\right)^{3/2}} \sum_{n=1}^N (V_n - V_{n+1}) \sum_{n=0}^{\infty} (-1)^s \binom{-3/2}{s} \xi^s \boldsymbol{L}_n^{(s)}(\boldsymbol{r}),
\]
where $\boldsymbol{L}_n^{(s)}(\boldsymbol{r})$ is a vector with length units given by 
\[
\boldsymbol{L}_n^{(s)}(\beta_n,\boldsymbol{r}) = R\cos\theta\tau_{0,n}^s(\phi) \hat{r}(\boldsymbol{r}) + \left[ r \tau_{1,n}^s(\phi) - R\sin\theta \tau_{0,n}^s(\phi) \right]\hat{\theta}(\boldsymbol{r}) - r\cos\theta \tau^{(s)}(\phi-\beta_n)\hat{\phi}(\boldsymbol{r}) 
\]
and defined using spherical coordinates. We replace this last result in Eq.~(\ref{electricFieldBiotPlusStaircaseVContributionIIEq}) in order to obtain the total electric field written in the more convenient way
\begin{equation}
\boxed{
\boldsymbol{E}(\boldsymbol{r}) = \left( \boldsymbol{\mathscr{E}}(\boldsymbol{r}) \right)_{unif.} + \frac{1}{2\pi}  \sum_{n=1}^N (V_n - V_{n+1}) \left[\boldsymbol{f}(\beta_n,\boldsymbol{r}) + \frac{R}{\left(R^2+r^2\right)^{3/2}}\sum_{s=0}^{\infty} (-1)^s \binom{-3/2}{s} \xi^s \boldsymbol{L}_n^{(s)}(\beta_n,\boldsymbol{r}) \right],
}
\label{EVectorFieldStairCaseVExpansionEq}
\end{equation}
where $\left( \boldsymbol{\mathscr{E}}(\boldsymbol{r}) \right)_{unif.} = (\mathscr{E}_{r}(\boldsymbol{r}))_{unif.} \hat{r} + (\mathscr{E}_{\theta}(\boldsymbol{r}))_{unif.} \hat{\theta}$ and $z>0$. Finally, if $V(\phi)$ is a continuous and fully periodic function, then we can write the electric field expression as

\[
\boldsymbol{E}(\boldsymbol{r}) = \left( \boldsymbol{\mathscr{E}}(\boldsymbol{r}) \right)_{unif.} - \frac{1}{2\pi}  \int_{0}^{2\pi} \partial_\phi V(\beta) \left[\boldsymbol{f}(\beta,\boldsymbol{r}) + \frac{R}{\left(R^2+r^2\right)^{3/2}}\sum_{n=0}^{\infty} (-1)^s \binom{-3/2}{s} \xi^s \boldsymbol{L}_n^{(s)}(\beta,\boldsymbol{r}) \right]d\beta .
\]

Since $\boldsymbol{L}_n^{(s)}(\beta,\boldsymbol{r})$ is periodic with respect $\beta$, then
\[
\boldsymbol{E}(\boldsymbol{r}) = \left( \boldsymbol{\mathscr{E}}(\boldsymbol{r}) \right)_{unif.} + < V \boldsymbol{h} > \hspace{0.2cm}\mbox{with}\hspace{0.2cm}< V \boldsymbol{\kappa} > = \frac{1}{2\pi} \int_{0}^{2\pi} V(\beta) \partial_\beta \boldsymbol{\kappa}(\beta,\boldsymbol{r}) d\beta
\]
and
\[
\boldsymbol{\kappa}(\beta,\boldsymbol{r}) =  \boldsymbol{f}(\beta,\boldsymbol{r}) + \frac{R}{\left(R^2+r^2\right)^{3/2}}\sum_{n=0}^{\infty} (-1)^s \binom{-3/2}{s} \xi^s \partial_\phi\boldsymbol{L}_n^{(s)}(\beta,\boldsymbol{r}). 
\]

In practice, to use Eq.~(\ref{EVectorFieldStairCaseVExpansionEq}) requires to compute the complete elliptic integral implicit in $\left( \boldsymbol{\mathscr{E}}(\boldsymbol{r}) \right)_{unif.}$. To this aim, we use the series representations in \cite{radon1950sviluppi},
\begin{align*}
\b{E}(\chi) = \frac{\pi}{2} + \frac{\pi}{2}\sum_{m=1}^\infty \left[\frac{(2m-1)!!}{(2m)!!}\right]^2\frac{1}{1-2m} \chi^{2m}, &&\mbox{and} &&K(\chi) = \frac{\pi}{2} + \frac{\pi}{2}\sum_{m=1}^\infty  \frac{e_m}{1-2m} \chi^{2m},
\end{align*}
with $n!!$ the double factorial.
Another option could be to use a programming package where those functions are implemented. In general, the evaluation of Eq.~(\ref{EVectorFieldStairCaseVExpansionEq}) is not challenging since this analytic expression can be coded in a high-level language program. In particular, we have written a short notebook in Wolfram Mathematica 9.0 \cite{wolfram2012version} in order to explore the solution of Eq.~(\ref{EVectorFieldStairCaseVExpansionEq}) for an arbitrary potential $V(\phi)$. In this sense, we have chosen $\mathcal{V}(\phi)$ as follows 
\begin{equation}
\mathcal{V}(\phi) = U_o\left\{\frac{1}{5} + \exp\left[-(\phi-\pi)^2\right]\right\},
\label{gaussianContinousVEq}    
\end{equation}

with $U_o=1$. However, other definitions of $\mathcal{V}(\phi)$ are also valid if those fulfill the periodic condition $\mathcal{V}(0)=\mathcal{V}(2\pi)$. In Fig.~\ref{EComponentsSurfFig} we present some results corresponding to the electric field for $\theta = 2\pi/5$, which are presented relatively near to the $(z=0)$-plane, where $\theta \rightarrow \pi/2$. Plots of Fig.~\ref{EComponentsSurfFig} have been obtained using the staircase like potential $V(\phi)$ with $N=7$ sectors that is shown at the left part of Fig.~\ref{discretePotentialLimitFig}. Such potential distribution introduces $N-1$ discontinuities that can drastically affect the electric field near the $(z=0)$-plane, as we can observe in Fig.~\ref{EComponentsSurfFig}. These sudden changes on the electric field components with respect the $\phi$-coordinate and near the $(z=0)$-plane must disappear in the limit $N \rightarrow \infty$ where $\mathcal{V}(\phi)$ is a continuous smooth function. We start to observe this bounded behaviour by choosing a larger value of $N$. For instance, we present in Fig.~\ref{EComponentsSurfIIFig} the results for $N=33$. Finally, some plots of the vector field are shown in Fig.~\ref{vectorFieldEPhiConstFig}.    

\section{BSL contribution of polygonal interconnected contours held at arbitrary potential $V=V(\phi)$}
\label{BSLContributionOfPolygonal}
Let us consider now a contour $c=\bigcup_{n=1}^N\overline{P_{n}P_{n+1}}$ composed by rectilinear segments connecting a set of points $P_1,\ldots,P_N$ on the plane $z=0$, with $P_{N+1}=P_{1}$.  The length of arc of the corresponding polygon is
\[
s(\phi) = s^{(n+1)}(\phi) + \mathcal{L}_n = \int_{0}^\phi \sqrt{\mathscr{R}^2(\phi') + \dot{\mathscr{R}}^2(\phi')}d\phi' ,
\]
with $\mathcal{L}_n = \sum_{j=1}^{n} l_j$, $l_n$ the length of $\overline{P_{n}P_{n+1}}$ and $s^{(n)}(\phi) \in [0, l_{n}]$. The BSL contribution of the $n$-th segment is 
\[
\pmb{\mathscr{E}}_{n}(\boldsymbol{r}) = -\textbf{sgn}(z)\frac{1}{2\pi} \int_{ c_{n,n+1} } V(s^{(n+1)}) \frac{(\boldsymbol{r}-s^{(n+1)}\hat{t}_{n}) \times d\boldsymbol{s}^{(n+1)}}{|\boldsymbol{r}-s^{(n+1)}\hat{t}_{n}|^3},
\]
with $\hat{t}_{n}:=P_{n+1}-P_{n}$ the tangent vector. We may use the following generating function 
\begin{equation}
\sum_{n=0}^{\infty} C_n^{(\lambda)}\left(\chi\right)\xi^n = \frac{1}{\left(1-2\xi\chi+\chi^2\right)^\lambda} ,
\label{genertingFunctionEq}
\end{equation}
where $\chi\in\mathbb{C}$ with absolute value $|\chi|<1$, $\lambda\in\left(-1/2,\infty\right) \setminus \left\{0\right\}$, and   $C_n^{(\lambda)} :  \mathbb{C} \rightarrow \mathbb{C}$ the Gegenbauer Polynomials.
Indeed, we can write the BSL contribution to the electric field for each segment as
\begin{equation}
\pmb{\mathscr{E}}_{n}(\boldsymbol{r}) = \frac{r\sin\theta \hat{\phi}_n(\boldsymbol{r})}{2\pi} \sum_{m=0}^{\infty} C_m^{(\frac{3}{2})}\left(\cos\theta\right)\left[\frac{1}{r^{m+3}}\mathcal{A}_{m}^{(n)}(\boldsymbol{r}) + r^{m}\mathcal{B}_{m}^{(n)}(\boldsymbol{r}) \right]    ,
\label{expansionStrightLineEEq}
\end{equation}
with $\theta$ the angle between $\hat{t}_n$ and $\boldsymbol{r}$ with origin at $P_n$, and $\hat{\phi}_n(\boldsymbol{r})$ the corresponding $\phi$-vector of a cylindrical coordinate with origin at $P_n$ and longitudinal axis along $\hat{t}_n$. The integral coefficients are 
\[
\mathcal{A}^{(n)}_{m}(r) = \int_{0}^{\min(r,l_n)} (s^{n+1})^m V(s^{(n+1)}-\mathcal{L}_n) ds^{(n+1)}, \hspace{0.15cm}\mbox{and}\hspace{0.15cm} \mathcal{B}^{(n)}_{m}(r) =\left[1-\mathscr{U}(r-l_n)\right] \int_{r}^{l_n} \frac{V(s^{(n+1)}-\mathcal{L}_n)}{(s^{(n+1)})^{m+3}}  ds^{(n+1)},
\]
where 
\[
\mathscr{U}(z) = 0 \hspace{0.25cm} \mbox{\textbf{if}}\hspace{0.25cm} z<0 \hspace{0.25cm} \mbox{\textbf{otherwise}} \hspace{0.25cm} 1 
\]
is the unit step function, and $\min(z_1,z_2)$ is the minimum function which yields the numerically smallest of the $\{z_1,z_1\}$. Hence, the total electric field is given by 
\[
\pmb{\mathscr{E}}(\boldsymbol{r}) = \sum_{n=1}^N R_z(\gamma_n)\pmb{\mathscr{E}}_{n}(\hat{e}_1 \cdot R_z(\gamma_n)^{T}(\boldsymbol{r}-P_n), \hat{e}_2 \cdot R_z(\gamma_n)^{T}(\boldsymbol{r}-P_n), \hat{e}_3 \cdot R_z(\gamma_n)^{T}(\boldsymbol{r}-P_n)) = \sum_{m=0}^\infty \pmb{\mathscr{E}}^{(m)}(\boldsymbol{r}),
\]
where $R_z(\gamma)$ is a counter-clock wise rotation matrix around the z-axis, $R_z(\gamma)^{T}$ the transpose matrix, $\hat{e}_i$ the Cartesian unit vectors, and $\gamma_n$ the angle between the $x$-axis and the $n$-th tangen vector $\hat{t}_n$. Defining $P_n=(x_n, y_n, 0)$ in cartesian coordinates, then the field $\pmb{\mathscr{E}}(\boldsymbol{r})$ for $z>0$ can be written as follows:

\begin{equation}
\boxed{
\pmb{\mathscr{E}}(\boldsymbol{r}) = \frac{1}{2\pi} \sum_{n=1}^{N} \boldsymbol{D}_n \sum_{m=0}^{\infty} \left.C_m^{(\frac{3}{2})}\left(\cos\theta\left(\boldsymbol{r}\right)\right)  \right|_{\change} \left[\frac{\mathcal{A}_{m}^{(n)}(\lambda_n(\boldsymbol{r}))}{\lambda_n(\boldsymbol{r})^{m+3}} + \lambda_n(\boldsymbol{r})^{m}\mathcal{B}_{m}^{(n)}(\lambda_n(\boldsymbol{r})) \right],
}
\label{polySolEq}
\end{equation}

with $\lambda_n(\boldsymbol{r}):=\sqrt{r^2+|P_n|^2-2(x x_n + y y_n)}$, $\Lambda_n(\boldsymbol{r}):=(y-y_n)\cos\gamma_n-(x-x_n)\sin\gamma_n$, $r^2 = x^2 + y^2 + z^2$, and $\boldsymbol{D}_n(\boldsymbol{r}) = (z\cos\gamma_n,z\sin\gamma_n,-(x-x_n)\cos\gamma_n+(y-y_n)\sin\gamma_n)$ a vector with dimensions of length \footnote{The angle in the Gegenbauer Polynomial is computed as usual, $\theta(\boldsymbol{r}) = \arctan(\rho(x,z)/y)$ if  $y>0$,  $\pi - \arctan(-\rho(x,z)/y)$  if  $y<0$,  otherwise $\pi/2$, with $\rho(x,z)^2 = x^2 + z^2$ accounting for the coordinate transformation. In other words,  $\theta(\boldsymbol{r}) =(1-\mathscr{U}(\Upsilon_n(\boldsymbol{r}))\pi + \Theta_n(\boldsymbol{r})$ if $\Lambda_n(\boldsymbol{r}) \neq 0$, otherwise $\pi/2$, where $\Theta_n(\boldsymbol{r})=\arctan(\Upsilon(\boldsymbol{r})/\Lambda_n(\boldsymbol{r}))$ and $\Upsilon_n(\boldsymbol{r})=\sqrt{\lambda_n(\boldsymbol{r})^2-\Lambda_n(\boldsymbol{r})^2}$.}. If $N$ is not large and the point of evaluation is near the contour, then it is necessary to consider many terms of the infinite sum in Eq.~(\ref{polySolEq}). However, if $N$ large, we may obtain a good approximation by only taking the $m=0$ term:
\begin{equation}
\pmb{\mathscr{E}^{(0)}}(\boldsymbol{r}) = \frac{1}{2\pi} \sum_{n=1}^{N} \boldsymbol{D}_n(\boldsymbol{r}) \left[\frac{\mathcal{A}_{0}^{(n)}}{\lambda_n(\boldsymbol{r})^{3}} + \mathcal{B}_{0}^{(n)}(\lambda_n(\boldsymbol{r}))\right] .
\end{equation}
Of course, if $N$ is not large enough, the variation of $V(s)$ along $\overline{P_{n}P_{n+1}}$ can be significant and the electric field must be necessarily computed from Eq.~(\ref{polySolEq}). 

\subsection{Exponentially decaying distribution of the electric potential along the contour}
\begin{figure}[H]
\centering
\includegraphics[width=0.15\textwidth]{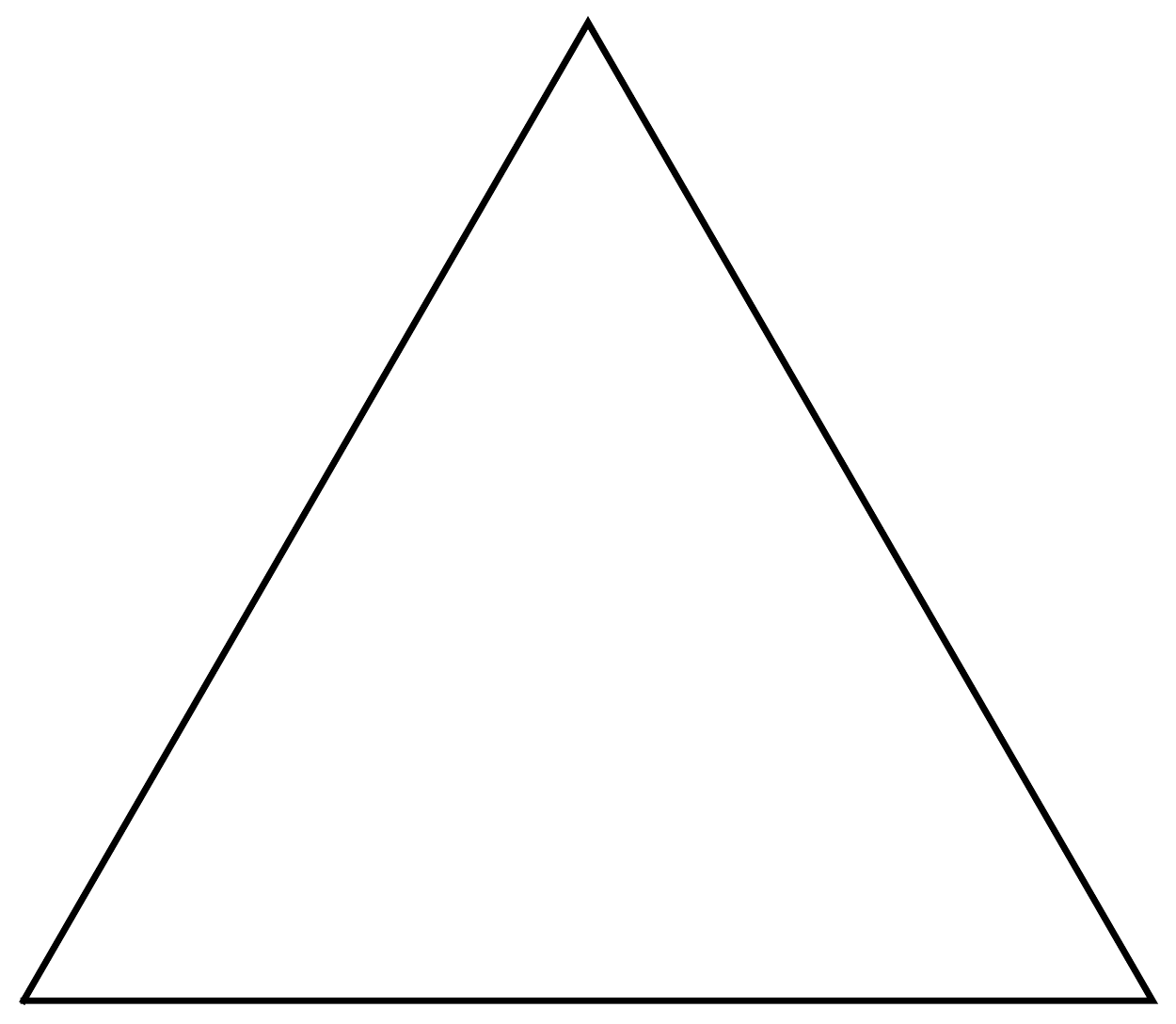} \hspace{0.2cm}%file.pdf
\includegraphics[width=0.15\textwidth]{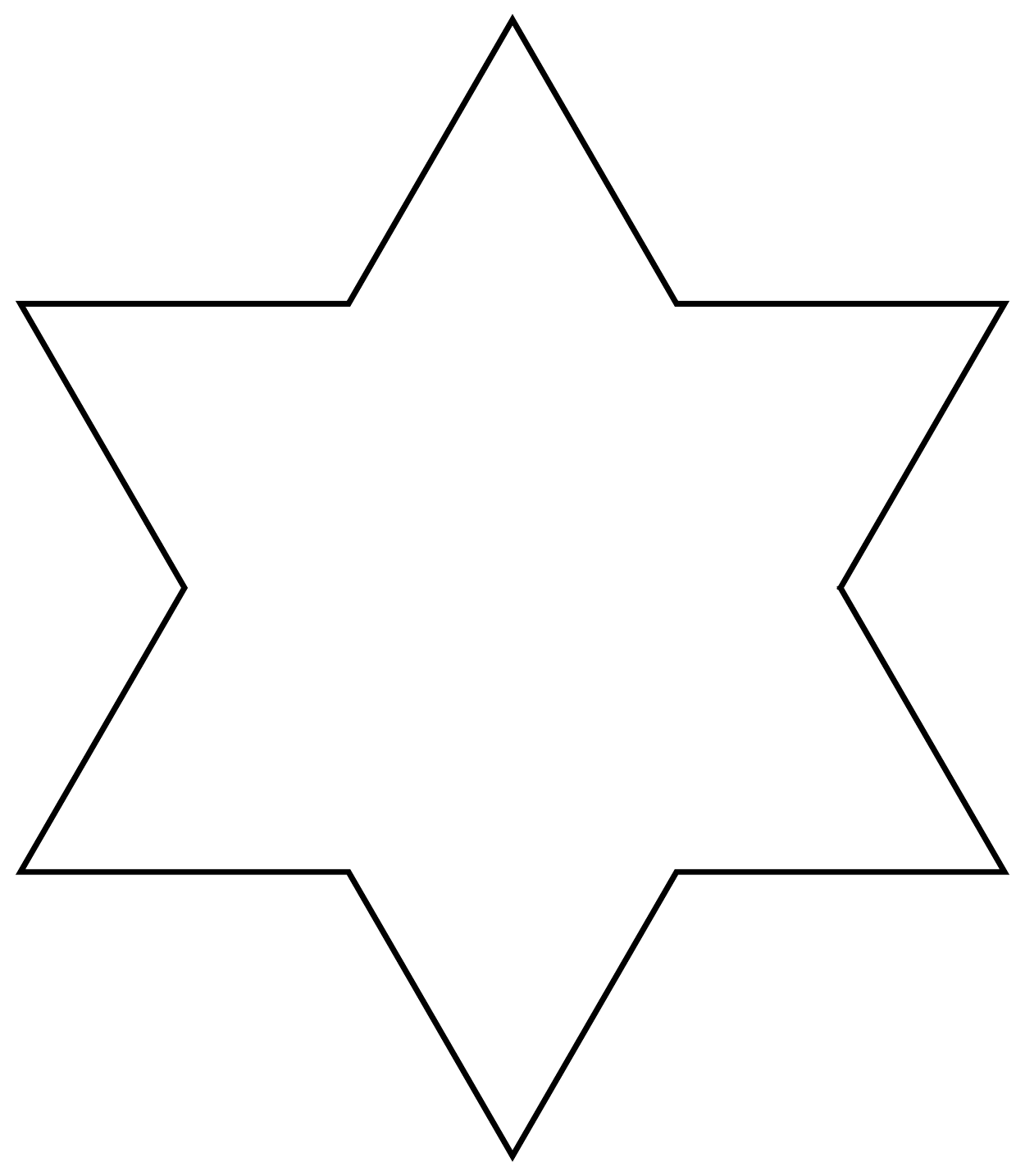} \hspace{0.2cm}%file.pdf
\includegraphics[width=0.15\textwidth]{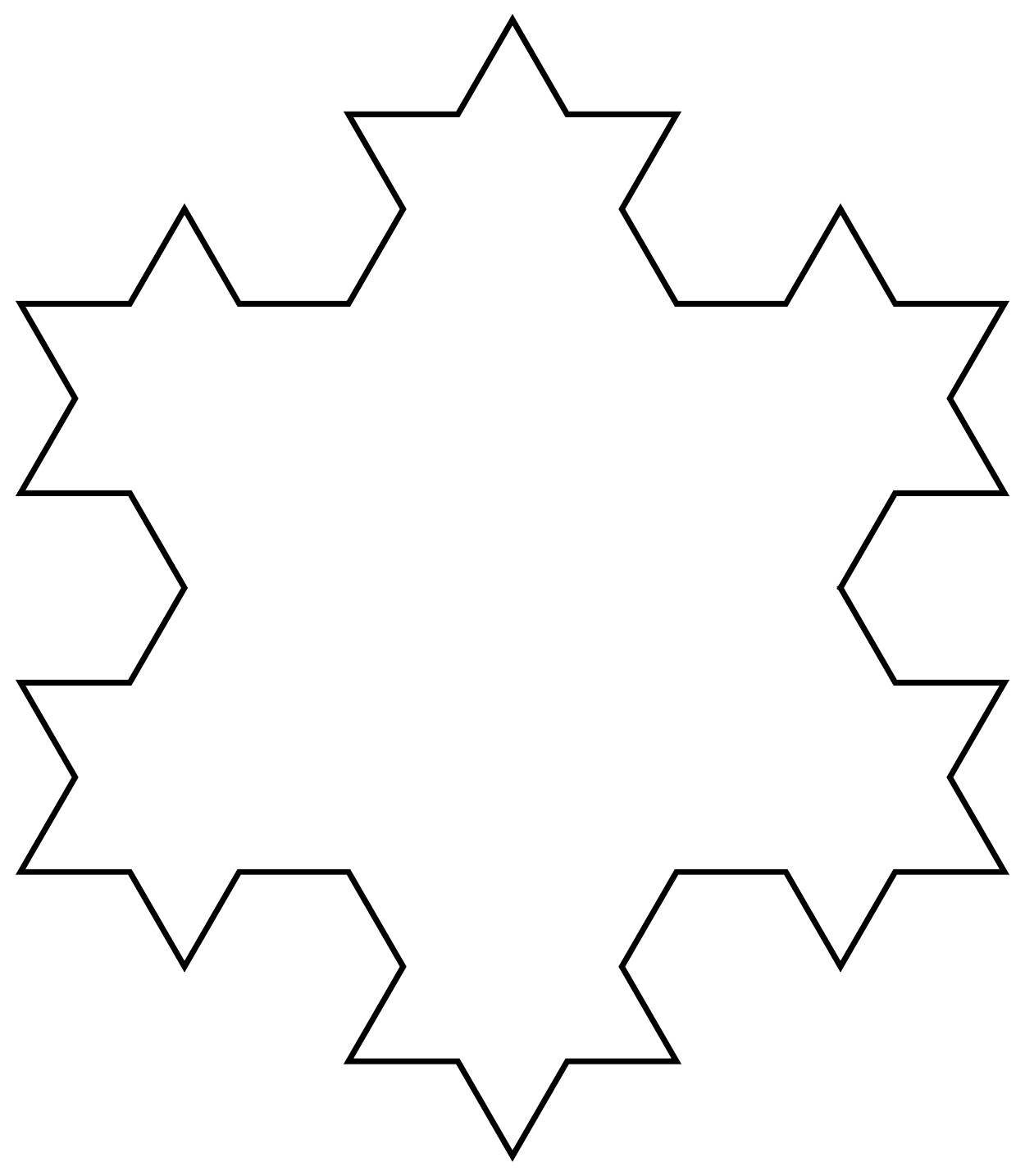} \hspace{0.2cm}%file.pdf
\includegraphics[width=0.15\textwidth]{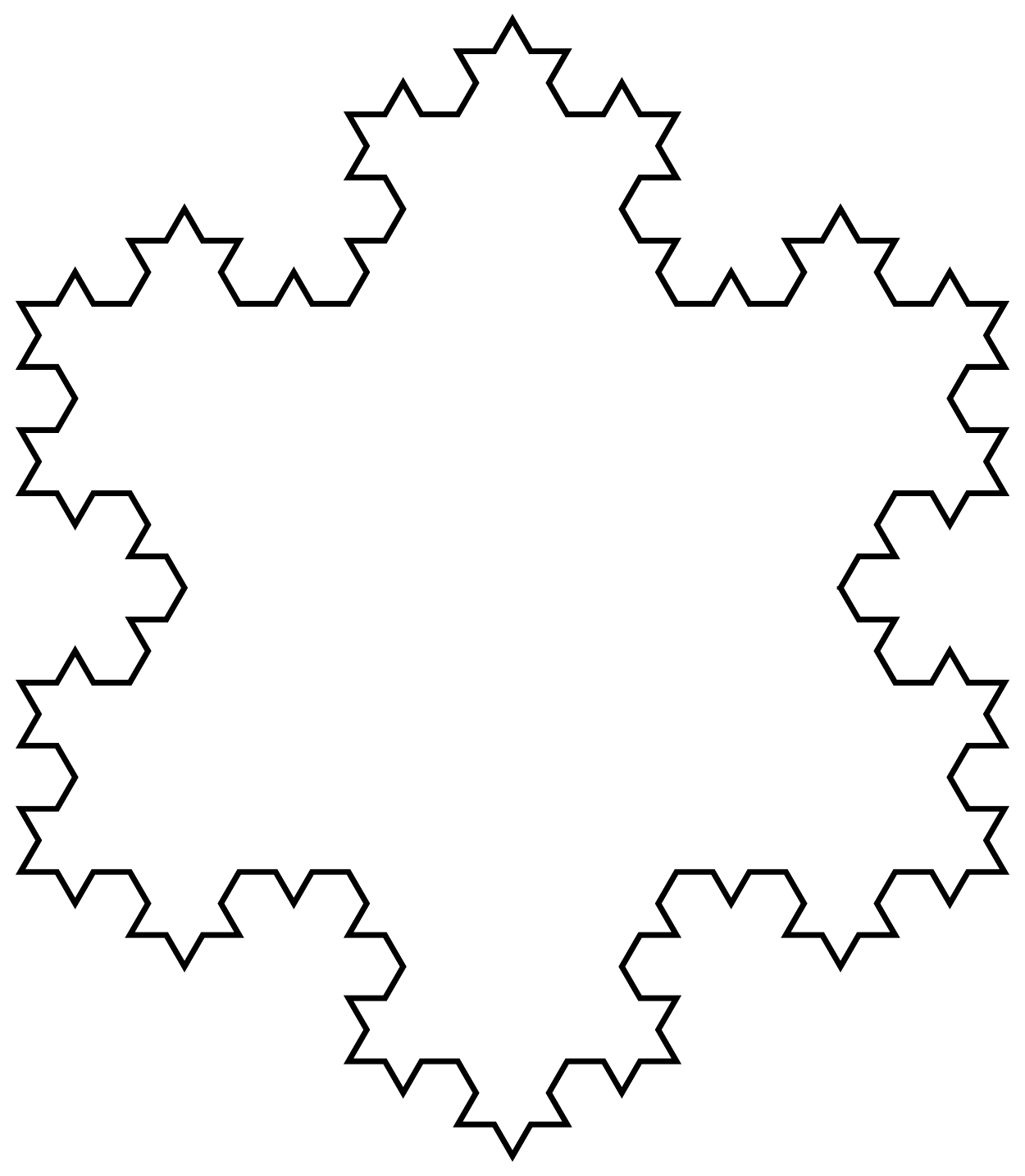} \hspace{0.2cm}%file.pdf
\includegraphics[width=0.15\textwidth]{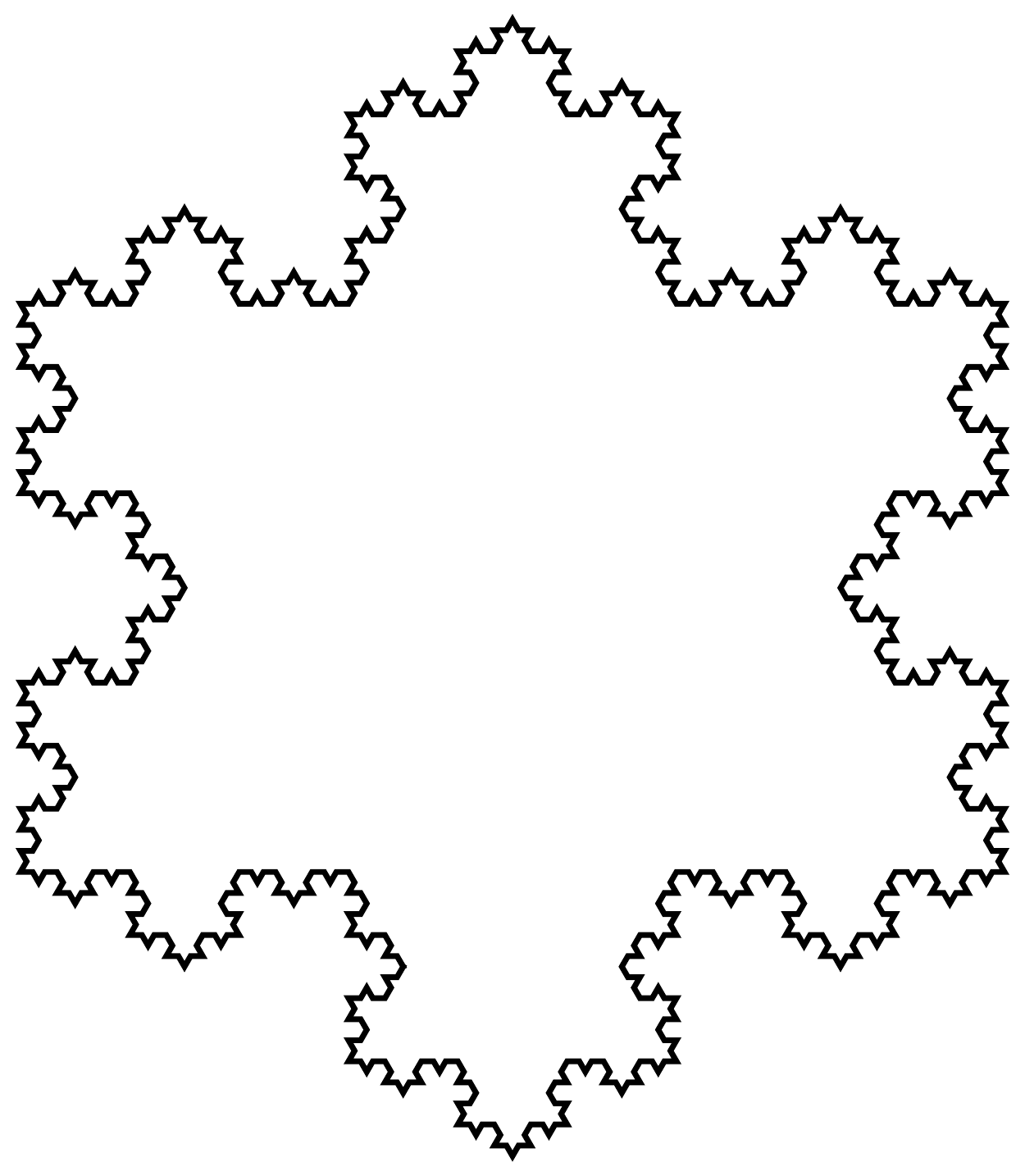} %file.pdf
\caption{Left to right. The $\nu$-level Koch snowflake for $\nu=0,1,2,3$ and 4.}
\label{snowFlakeContourFig}
\end{figure}

As an illustrative example, let us consider a $\nu$-level Koch snowflake curve. It is quite popular curve built by starting with an equilateral triangle, removing the inner third of each side, and building another equilateral triangle at the removed side (see Fig.~\ref{snowFlakeContourFig}). The procedure is repeated $\nu$-times.

If $a$ is the length of the side of the starting equilateral triangle, then 
\[
l_n(\nu) = \left(\frac{1}{3}\right)^{\nu} a,  \hspace{0.5cm} \mbox{and}  \hspace{0.5cm} \mathcal{L}_n(\nu) = n \left(\frac{1}{3}\right)^{\nu} a .
\]
The total number of points is $N(\nu) = 3 (4)^{\nu}$, and the perimeter is $s(2\pi) = 3(4/3)^{\nu} a$. The potential along the curve is set to be 
\[
V(s) = V_o \exp(-\kappa s),
\]
where $V_o$ and $\kappa$ are constants with units of electric potential and inverse length, respectively \footnote{We chose this potential because the coefficients $\mathcal{A}^{(n)}_{m}(r)$ and $\mathcal{B}^{(n)}_{m}(r)$ are found easily. However, other type of potentials portray as good options e.g. gaussian-like potentials, power-law potentials,  etc.}. The coefficients $\mathcal{A}^{(n)}_{m}(r)$ and $\mathcal{B}^{(n)}_{m}(r)$ can be evaluated straightforwardly for this potential. The resulting contribution to the electric field is the following 
\begin{equation}
\begin{split}
\pmb{\mathscr{E}}(\boldsymbol{r}) = \frac{\kappa V_o}{2\pi} \sum_{n=1}^{3 (4)^{\nu}} \boldsymbol{\eta}_n \sum_{m=0}^{\infty} & \left.  C_m^{(\frac{3}{2})}\left(\cos\theta\left(\boldsymbol{r}\right)\right) \Bigg|_{\change}  \left[\frac{m!-\Gamma(m+1,-\kappa \lambda_n(\boldsymbol{r}))}{(\kappa \lambda_n(\boldsymbol{r}))^{m+3}} + \right. \right. \\ & \left. \left. (\kappa \lambda_n(\boldsymbol{r}))^{m}\left[\Gamma(-(m+2),\kappa \lambda_n(\boldsymbol{r}))-\Gamma(-(m+2), n \left(1/3\right)^{\nu} \kappa a)\right] \right]\right. 
\end{split},
\label{muLevelSknowFlakeEBiotEq}
\end{equation}
with $\Gamma(n,z)$ the upper incomplete gamma function, and $\boldsymbol{\eta}_n(\boldsymbol{r}) = \kappa \boldsymbol{D}_n(\boldsymbol{r})$ a dimensionless vector in Cartesian coordinates. The $m=0$ contribution is
\[
\pmb{\mathscr{E}^{(0)}}(\boldsymbol{r}) = \frac{\kappa V_o}{2\pi} \sum_{n=1}^{3 (4)^{\nu}} \boldsymbol{\eta}_n (\boldsymbol{r}) \left[\frac{1-\Gamma(1,-\kappa \lambda_n(\boldsymbol{r}))}{(\kappa \lambda_n(\boldsymbol{r}))^{3}} + \Gamma(-2,\kappa \lambda_n(\boldsymbol{r})) - \Gamma(-2,n \left(1/3\right)^{\nu} \kappa a))\right].      
\]

\begin{figure}[H]
\centering
\includegraphics[width=0.24\textwidth]{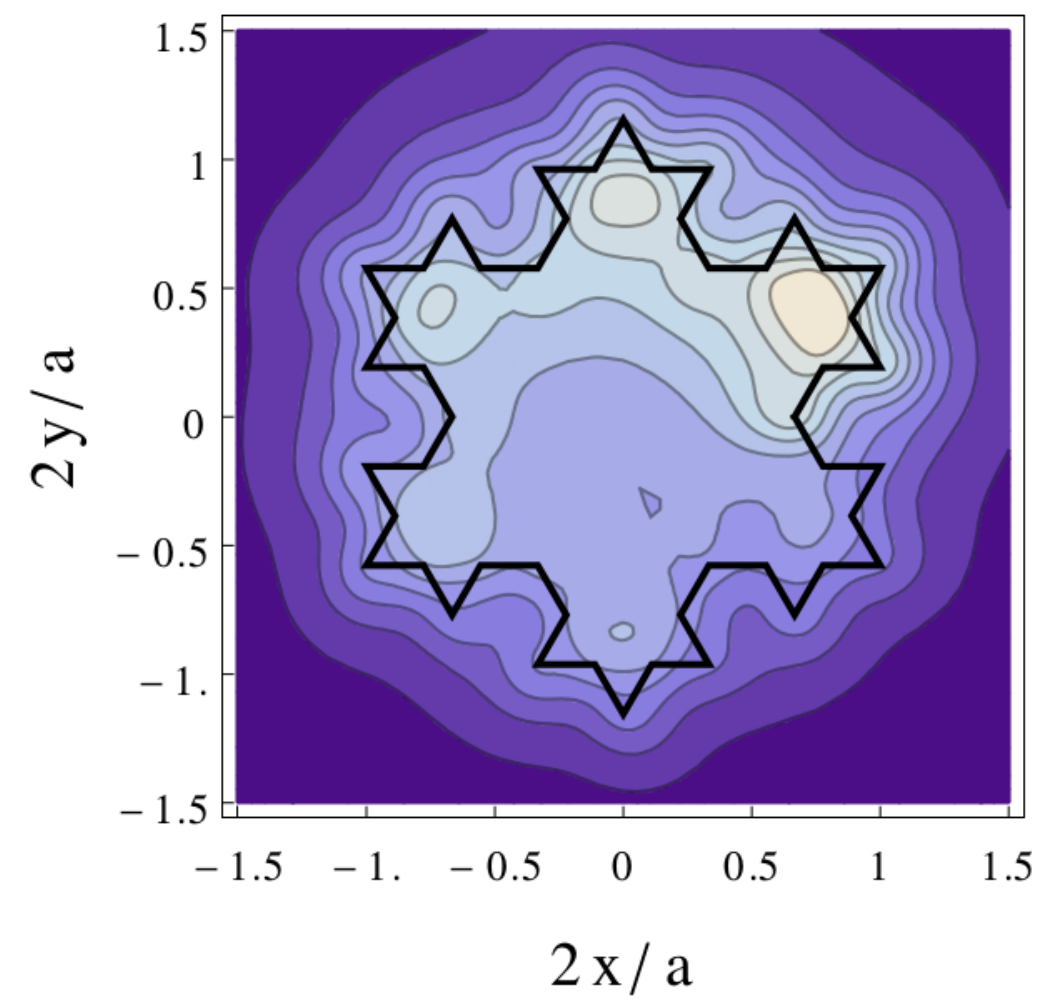} \includegraphics[width=0.24\textwidth]{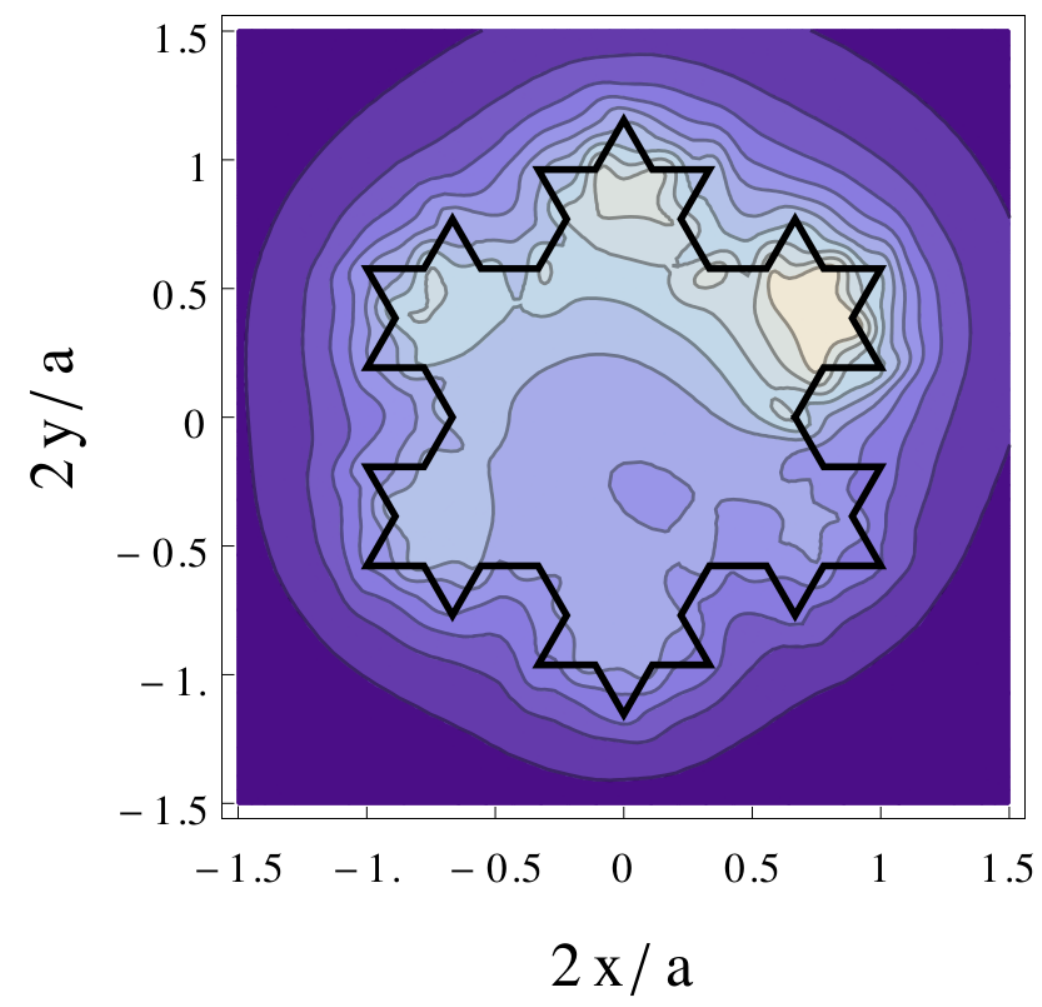} \includegraphics[width=0.24\textwidth]{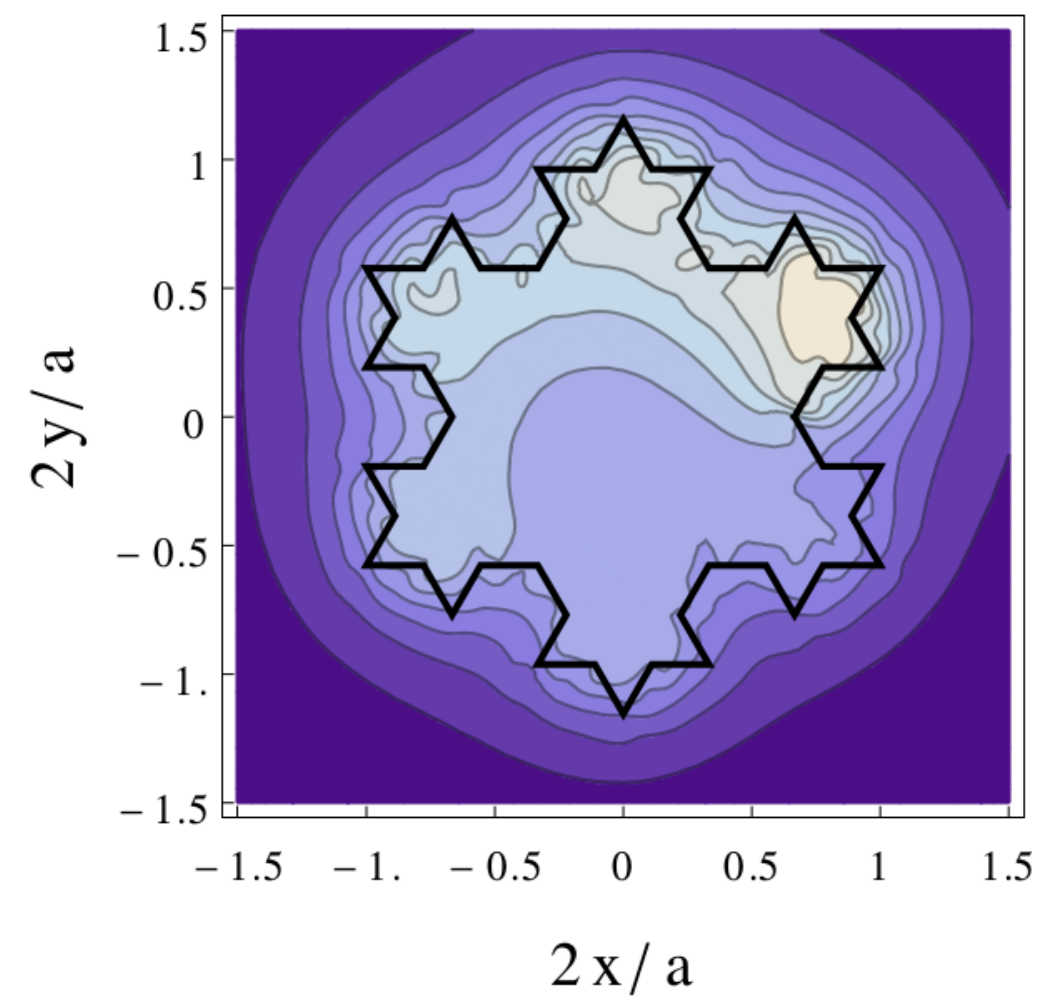} \includegraphics[width=0.24\textwidth]{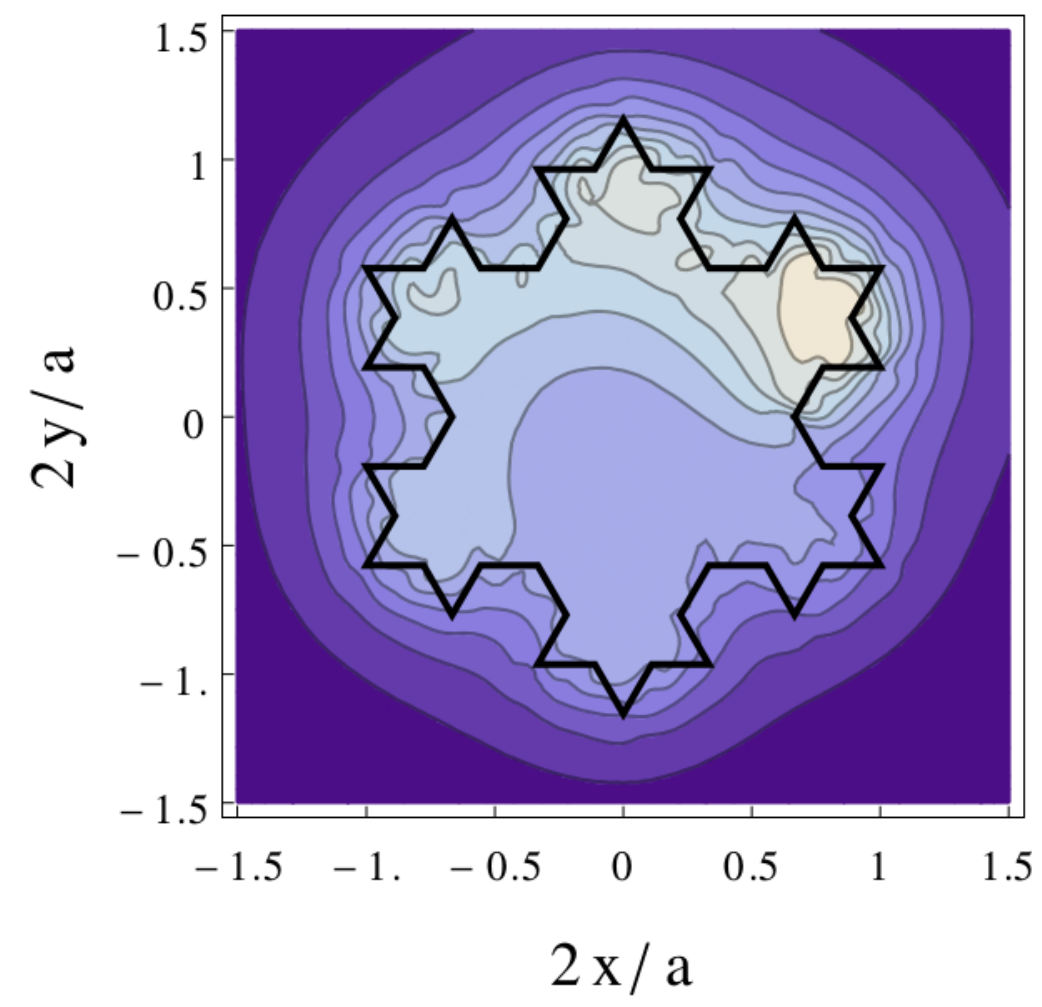} %file.pdf
\caption{Left to right. Magnitude of $\pmb{\mathscr{E}}(\boldsymbol{r})$ generated by a 2-level Koch snowflake keeping $m=0,4,6$ and 20 terms. $|\pmb{\mathscr{E}}(\boldsymbol{r})|$ is evaluated on the plane $z=l_1(2)=a/9$ and using Eq.~(\ref{muLevelSknowFlakeEBiotEq}).}
\label{2LevelsnowFlakeContourFig}
\end{figure}

A plot of $|\pmb{\mathscr{E}^{(0)}}(\boldsymbol{r})|$ for a 2-level Koch snowflake is shown in Fig.~\ref{2LevelsnowFlakeContourFig}. The BSL field is computed with Eq.~(\ref{muLevelSknowFlakeEBiotEq}) at $z=a/9$ and truncating the infinite sum. We observed that  series solution truncated up to 6 terms gives a good a approximation of the field for $z \geq a/9$ at level $\nu=2$. 

Highest levels, eventually requires less series terms. For instance, the BSL field of the 4-Snowflake is well represented by the $m=0$ term of the series solution. 

\begin{figure}[H]
\centering
\includegraphics[width=0.24\textwidth]{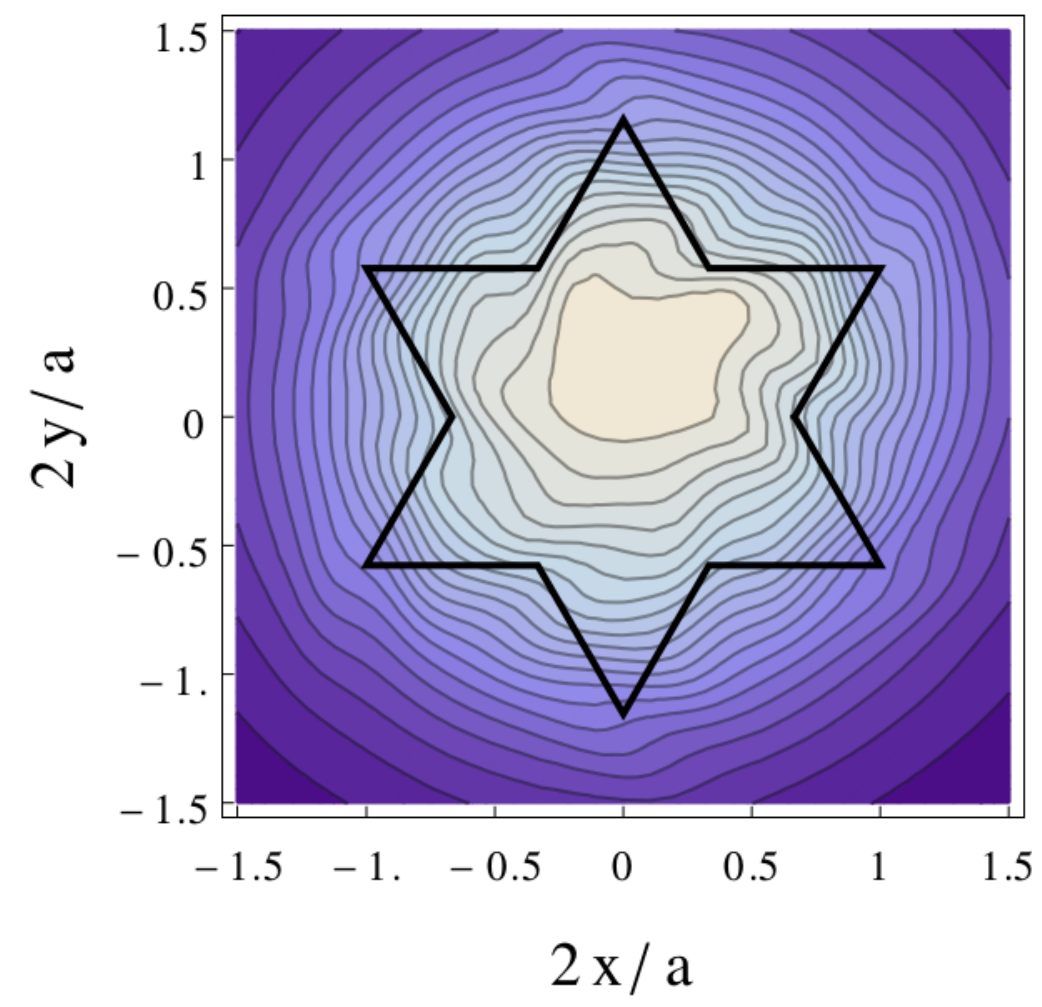}
\includegraphics[width=0.24\textwidth]{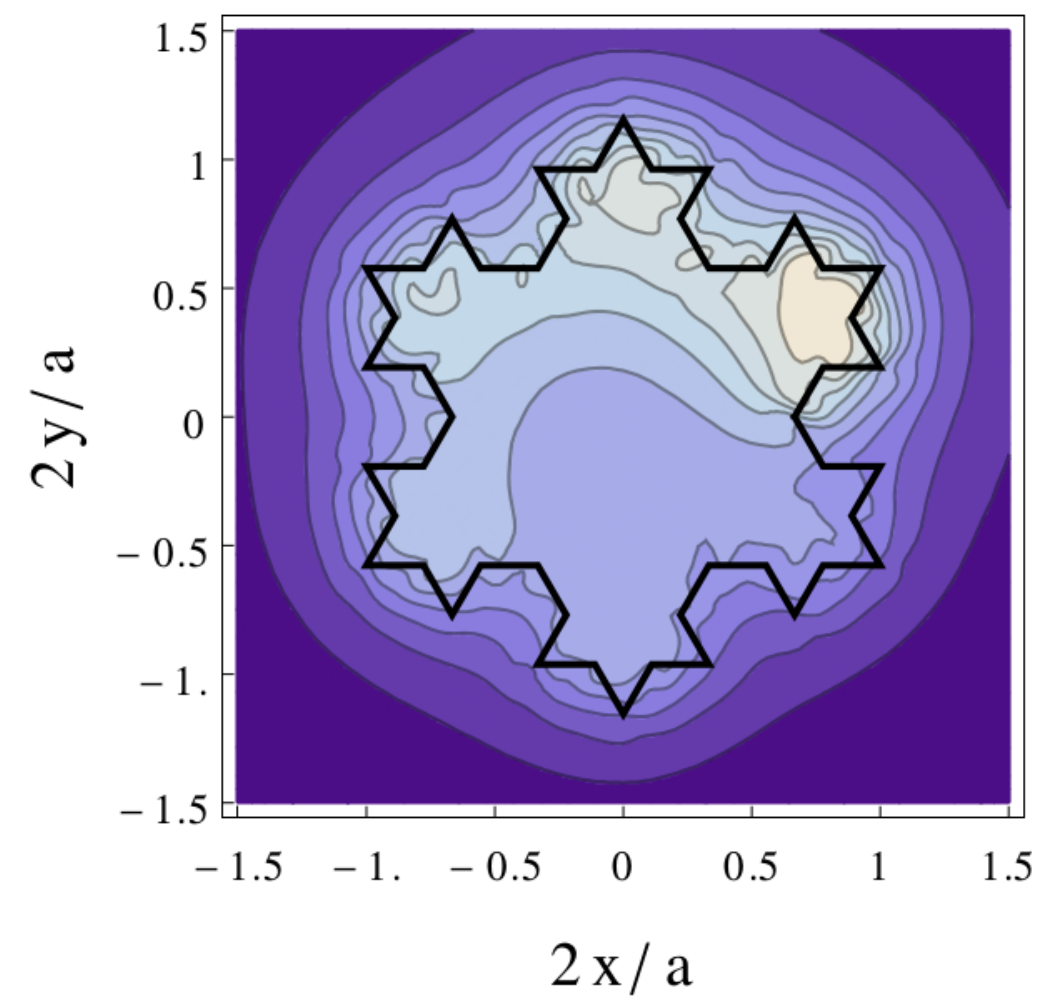}
\includegraphics[width=0.24\textwidth]{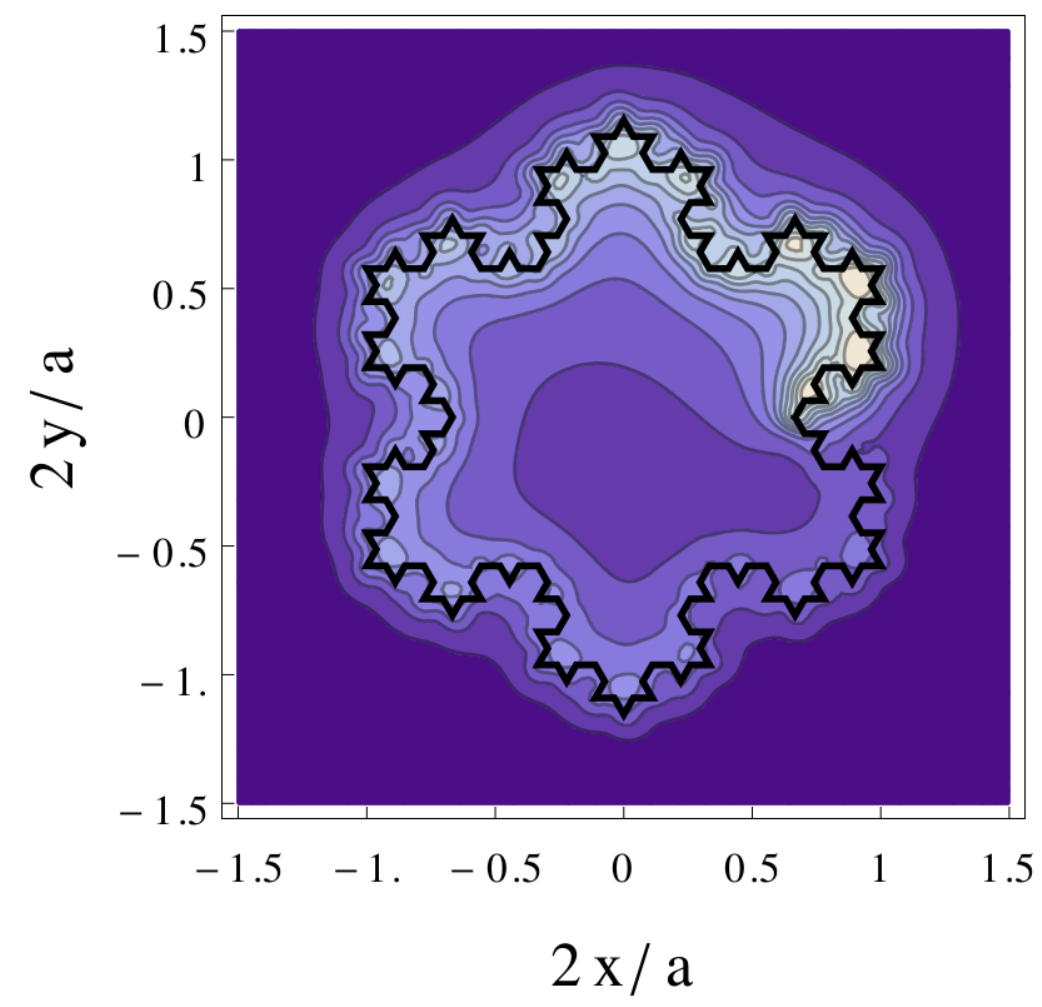}
\includegraphics[width=0.24\textwidth]{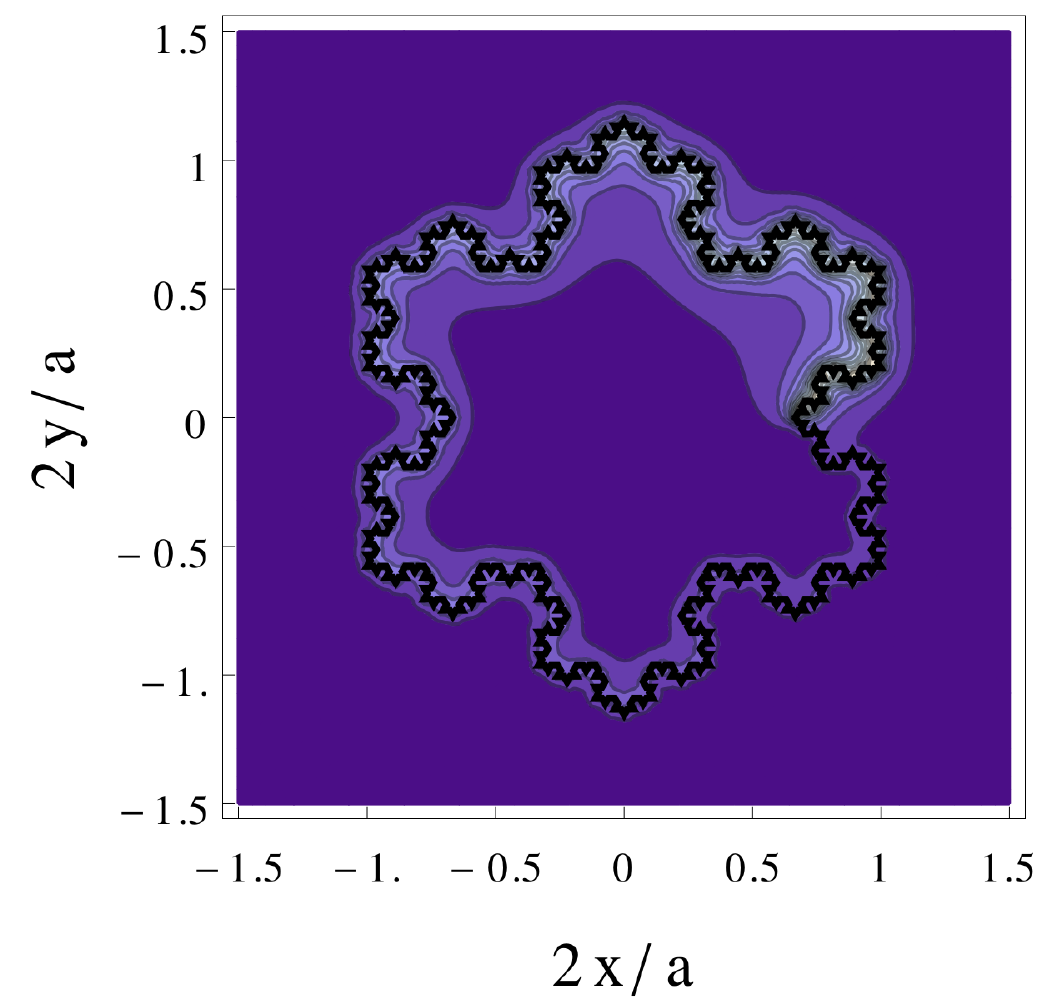}
%file.pdf
\caption{Left to right. Magnitude of $\pmb{\mathscr{E}}(\boldsymbol{r})$ generated by a $\nu$-level Koch snowflake for $\nu=1,2,3$ and $4$. We kept up to $m=10$ and 6 contribution terms in the 1 and 2 - level snowflakes. The 3-level snowflake corresponds to the $m=0$ term contribution. $|\pmb{\mathscr{E}}(\boldsymbol{r})|$ is evaluated on the plane $z=(1/3)^{\nu} a$ and using Eq.~(\ref{muLevelSknowFlakeEBiotEq}).}
\label{severalLevelsnowFlakeContourFig}
\end{figure}

\begin{figure}[H]
\centering
\includegraphics[width=0.28\textwidth]{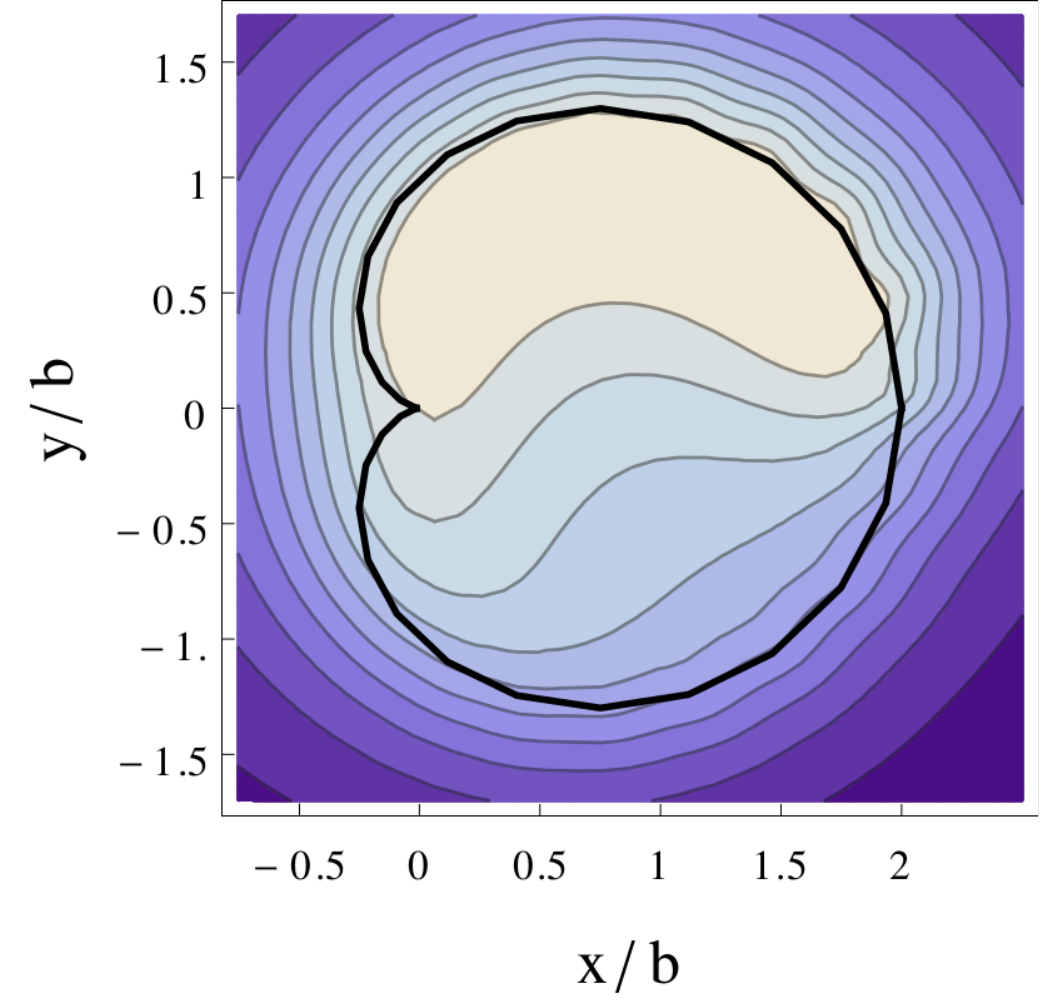}
\hspace{0.1cm}%file.pdf
\includegraphics[width=0.4\textwidth]{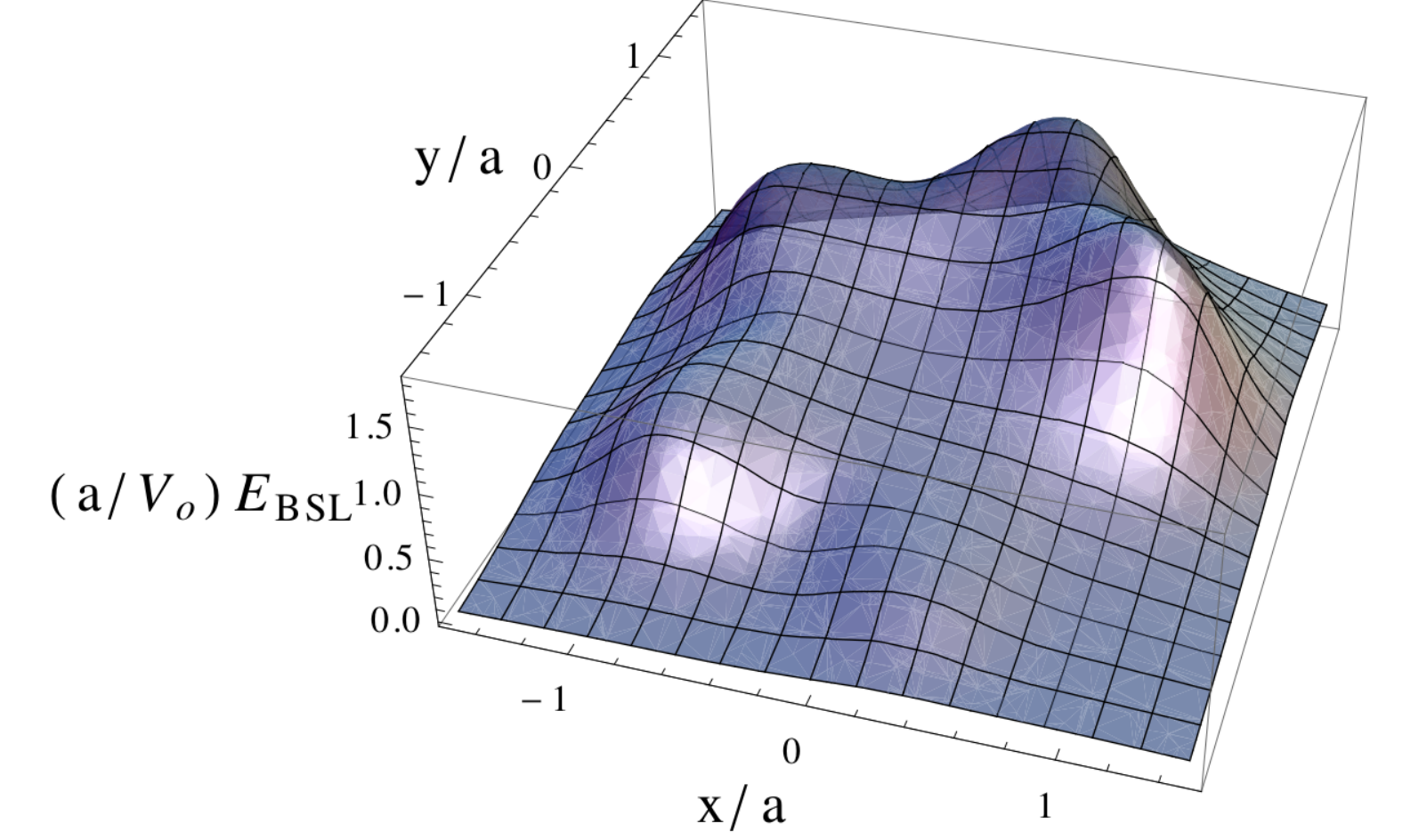} \hspace{0.1cm}%file.pdf
\includegraphics[width=0.28\textwidth]{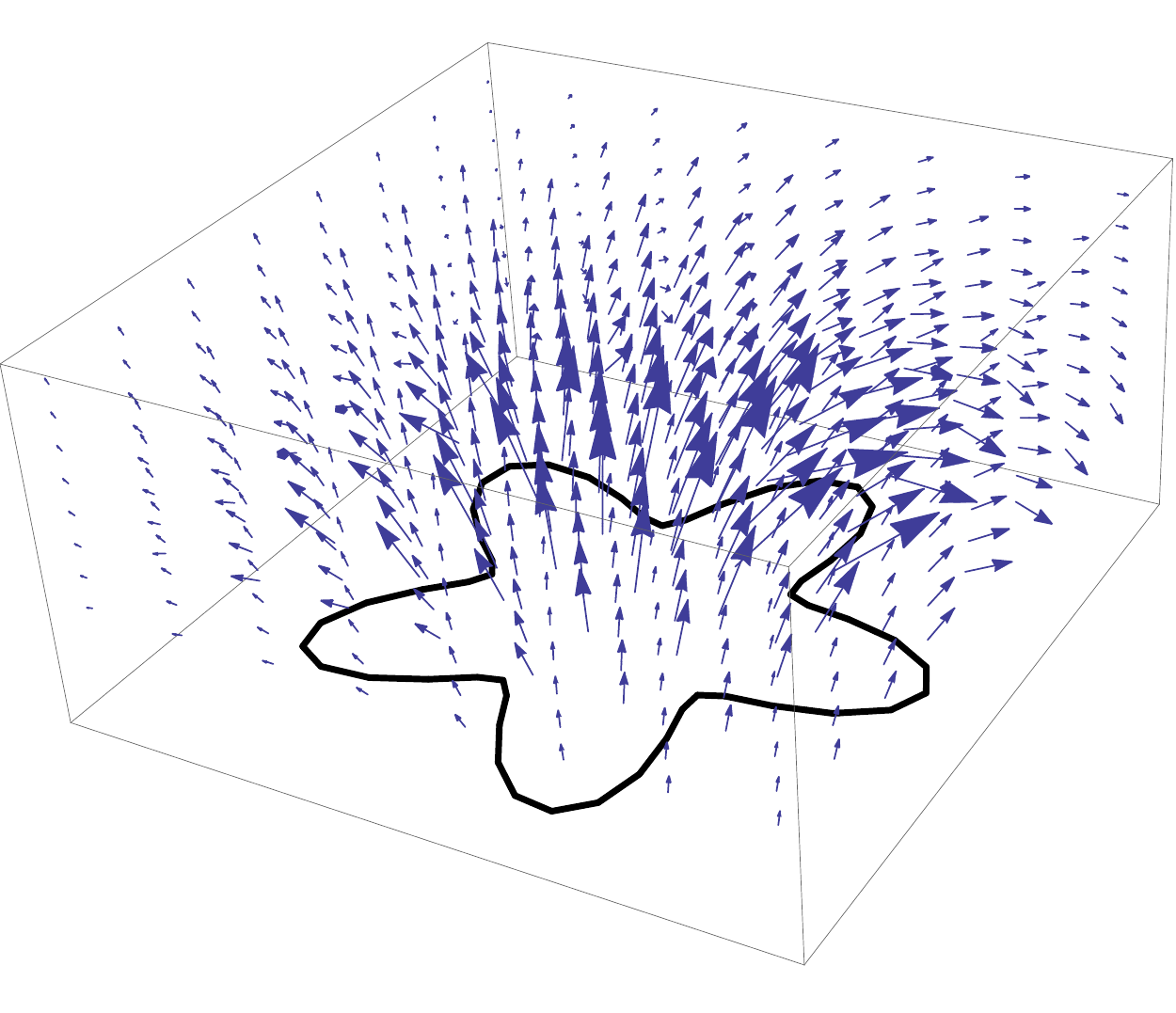} \hspace{0.2cm}%file.pdf
\caption{(left) Cardioid (center and right) 5-fold Harmonically deformed curve. }
\label{severalPlotsFig}
\end{figure}

Let us suppose that set $\{P_n\}_{1 \leq n \leq N}$ lie on a curve whose polar equation is known $\mathscr{R}=\mathscr{R}(\phi)$, then
\[
\lambda_n(\boldsymbol{r}) = \sqrt{r^2 + \mathscr{R}_n[\mathscr{R}_n - 2(x\cos\gamma_n+y\sin\gamma_n)] },\hspace{0.5cm}\Lambda_n(\boldsymbol{r}) = y\cos\gamma_n - x\sin\gamma_n + \mathscr{R}_n \sin(\gamma_n-\beta_n),
\]
and
\[
\boldsymbol{D}_n(\boldsymbol{r})=\componentsII,
\]

with $\beta_n=2n\pi/N$ and $\mathscr{R}_n=\mathscr{R}(\gamma_n)$. Again, Eqs.~(\ref{polySolEq}) and (\ref{muLevelSknowFlakeEBiotEq}) can be employed to compute $\pmb{\mathscr{E}}(\boldsymbol{r})$ for other geometries including non-intersecting polygons. For instance, a polygon with points lying on a cardioid curve $\mathscr{R}(\phi)=b(1-\cos\phi)$ with $b$ any constant and the potential distributed decaying exponentially along the contour. This example is shown in Fig.\ref{severalPlotsFig}-left. 

\subsection{Variable potential distribution but constant in each polygon side}
In the case of the polygonal curve, the simplest distribution concerns the case where the potential only changes between the segments $V(\phi) = V_n  \hspace{0.25cm}\mbox{\textbf{if}}\hspace{0.25cm}\phi \in (\beta_{n},\beta_{n+1})$. In that case, the expansion coefficients become
\[
\mathcal{A}^{(n)}_{m}(r) = \frac{V_n}{m+1}(\min(r,h))^{m+1}  \hspace{0.5cm}\mbox{and}\hspace{0.5cm} \mathcal{B}^{(n)}_{m}(r) = \frac{V_n}{m+2}(1-\mathscr{U}(r-h))\left(\frac{1}{r^{m+2}}-\frac{1}{h^{m+2}}\right),
\]
with $h$ the length of the segment. Then, according to Eq.~(\ref{expansionStrightLineEEq}), the electric field contribution of this segment in the limit $h\rightarrow\infty$ is
\[
\pmb{\mathscr{E}}_{n}^{+}(\boldsymbol{r}) = \frac{V_n \sin\theta \hat{\phi}_n(\boldsymbol{r})}{2\pi r} \sum_{m=0}^{\infty} C_m^{(\frac{3}{2})}\left(\cos\theta\right)\left(\frac{1}{m+1} + \frac{1}{m+1}  \right).   
\]
Or, written in another way,
\[
\pmb{\mathscr{E}}_{n}^{+}(\boldsymbol{r}) = \frac{V_n \hat{\phi}_n(\boldsymbol{r})}{2\pi r\sin\theta} \sin^2{\theta}\left[\sum_{m \in 2\mathbb{N}^{0}}\frac{2m+3}{(m+1)(m+2)} C_m^{(\frac{3}{2})}\left(\cos\theta\right) + \sum_{m \in 2\mathbb{N}^{0}+1}\frac{2m+3}{(m+1)(m+2)} C_m^{(\frac{3}{2})}\left(\cos\theta\right)\right],    
\]
with $2\mathbb{N}^{0}$ and $2\mathbb{N}^{0}+1$ the set of even and odd positive integers including 0, respectively. Now, the Gegenbauer polynomials $C^{\lambda}_{n}(\chi)$ are orthogonal polynomials with weighting function $w^{\lambda}(\chi)=(1-\chi^2)^{\lambda-1/2}$ in the interval $\chi \in [-1, 1]$, and they can be used as basis functions as follows
\[
f(\chi) = \sum_{n=0}^{\infty} q_{n}^{(\lambda)}\left\{f\right\} C_{n}^{(\lambda)} (\chi),
\]
with $q_{n}^{(\lambda)}\left\{f\right\}$ the expansion coefficients of $f$, which can be obtained from the normalization condition of the Gegenbauer polynomials.
For $\lambda=3/2$, they give
\[
q_{m}^{(3/2)}\left\{f\right\} = \frac{m+3/2}{(m+1)(m+2)} \int_{-1}^{1} C_{m}^{\left(\frac{3}{2}\right)}(\cos\theta) \sin^2\theta f(\theta) d(\cos\theta).
\]

Now, 
\[
q_{m}^{(3/2)}\left\{ \csc^2\theta \right\} = \frac{m+3/2}{(m+1)(m+2)} \int_{-1}^{1} C_{m}^{\left(\frac{3}{2}\right)}(\chi) d\chi = \frac{m+3/2}{(m+1)(m+2)}\left( 2 \hspace{0.25cm}\mbox{\textbf{if}}\hspace{0.25cm}m\in 2\mathbb{N}^0\hspace{0.25cm}\mbox{\textbf{else}}\hspace{0.25cm}0  \right) ,
\]
and similarly, 
\[
q_{m}^{(3/2)}\left\{ \csc^2\theta \cos\theta \right\} = \frac{m+3/2}{(m+1)(m+2)} \int_{-1}^{1} C_{m}^{\left(\frac{3}{2}\right)}(\chi) \chi d\chi = \frac{m+3/2}{(m+1)(m+2)}\left( 2 \hspace{0.25cm}\mbox{\textbf{if}}\hspace{0.25cm}m\in 2\mathbb{N}^0+1\hspace{0.25cm}\mbox{\textbf{else}}\hspace{0.25cm}0  \right) .
\]
Therefore, we can obtain an expression for the electric field contribution of each polygon segment that is given by
\[
\pmb{\mathscr{E}}_{n}^{+}(\boldsymbol{r}) = \frac{V_n \hat{\phi}_n(\boldsymbol{r})}{2\pi r\sin\theta} \sin^2{\theta}\left[\sum_{m \in \mathbb{N}^{0}} q_{m}^{(3/2)}\left\{ \csc^2\theta \right\} C_m^{(\frac{3}{2})}\left(\cos\theta\right) + \sum_{m \in \mathbb{N}^{0}} q_{m}^{(3/2)}\left\{ \cos\theta \csc^2\theta \right\} C_m^{(\frac{3}{2})}\left(\cos\theta\right)\right].
\]
Simplifying the previous result, we find that $\pmb{\mathscr{E}}_{n}^{+}(\boldsymbol{r}) = (V_n \hat{\phi}_n)(\boldsymbol{r})(/2\pi r\sin\theta) \left[1 +  z/r\right]$. Hence, the electric field given by a single segment can be obtained from computing $\pmb{\mathscr{E}}_{n}^{+}(x,y,z) - \pmb{\mathscr{E}}_{n}^{+}(x,y,z-l_n)$. This condition results in the following expression:
\[
\pmb{\mathscr{E}}_{n}(\boldsymbol{r}) = \frac{V_n \hat{\phi}_n(\boldsymbol{r})}{2\pi r\sin\theta} \left[\frac{l_n-z}{\sqrt{x^2+y^2+(z-l_n)^2}}+ \frac{z}{r}\right].
\]
Finally, the electric field given by the contribution of all segments is
\begin{equation}
\pmb{\mathscr{E}}(\boldsymbol{r}) =\frac{1}{2\pi}\sum_{n=1}^N \frac{V_n \boldsymbol{D}_n(\boldsymbol{r})}{\lambda_n(\boldsymbol{r})^2-\Lambda_n(\boldsymbol{r})^2} \left[\frac{l_n-\Lambda_n(\boldsymbol{r})}{\sqrt{\lambda_n(\boldsymbol{r})^2+l_n^2 - 2 l_n \Lambda_n(\boldsymbol{r})}}+ \frac{\Lambda_n(\boldsymbol{r})}{\lambda_n(\boldsymbol{r})}\right],
\label{EBiotConstantVPerSectorsEq}
\end{equation}
with $\boldsymbol{D}_n = (z\cos\gamma_n,z\sin\gamma_n,-(x-x_n)\cos\gamma_n+(y-y_n)\sin\gamma_n)$ vector with dimensions of length. In Fig.~\ref{severalPlotsFig}, we present the resulting electric field $\pmb{\mathscr{E}}(\boldsymbol{r})$ for the polygonal-interconnected harmonically-deformed curve $\mathscr{R}(\phi)=a(1+b\cos(n\phi))$, with $a=1$, $b=a/3$, and $n=5$.

\section{Numerical comparison}
\label{NumericalComparisonSection}
In this document we describe an analytic technique to compute the electric field due to a planar region hold at some non-uniform potential $V(\phi)$. However, there are also numerical alternatives to solve this problem. One of them is to evaluate numerically Eq.~(\ref{electricPotentialIntegralEq}) and to apply a numerical central differences method to compute an approximated gradient \cite{brezillon1981numerical,burden2001numerical} together with a numerical integration method. This is, to numerically approximate the electric contribution components as $E_r(\boldsymbol{r}) \approx  \frac{-1}{h} \left[\Phi\left(r+\frac{h}{2},\theta,\phi\right)-\Phi\left(r-\frac{h}{2},\theta,\phi\right)\right]$, $E_\theta(\boldsymbol{r}) \approx  \frac{-1}{r h} \left[\Phi\left(r,\theta+\frac{h}{2},\phi\right)-\Phi\left(r,\theta-\frac{h}{2},\phi\right)\right]$,
and $E_\phi(\boldsymbol{r}) \approx - \frac{1}{r \sin\theta h} \left[\Phi\left(r,\theta,\phi+\frac{h}{2}\right)-\Phi\left(r,\theta,\phi-\frac{h}{2}\right)\right]$. To this aim, we use the function NIntegrate of Wolfram Mathematica 9.0 \cite{wolfram2012version} in order to numerically perform the integration  $\Phi\left(r,\theta,\phi\right)$ and their shifted values over a small increment size $h$.

\begin{figure}[H]
\centering
\includegraphics[width=0.325\textwidth]{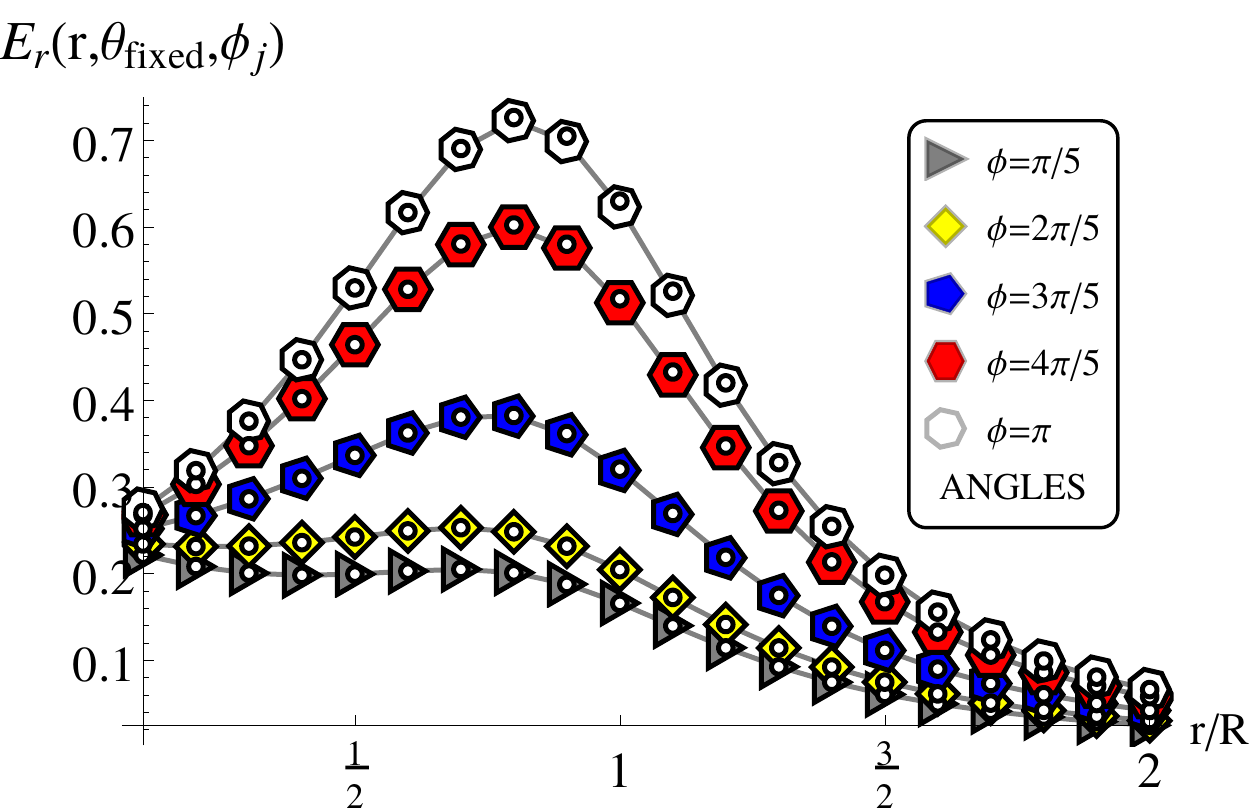}%ErComparison.pdf
\includegraphics[width=0.325\textwidth]{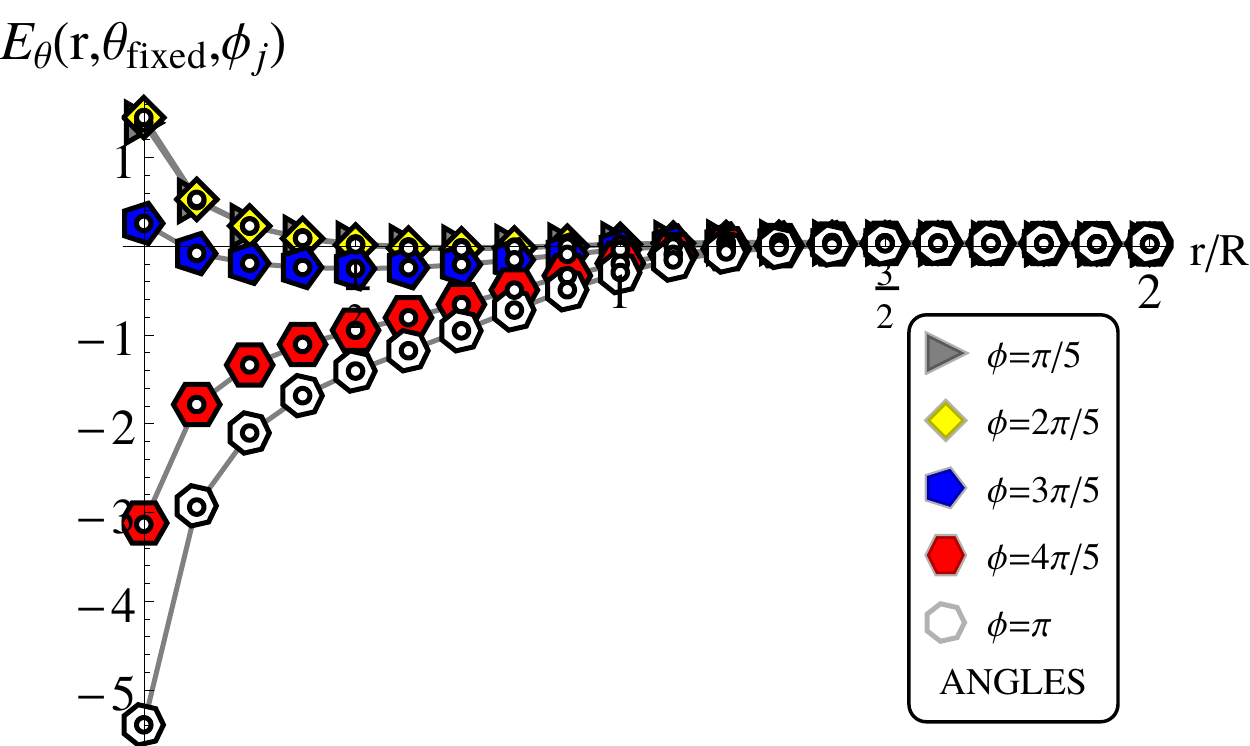}%EThetaComparison.pdf
\includegraphics[width=0.325\textwidth]{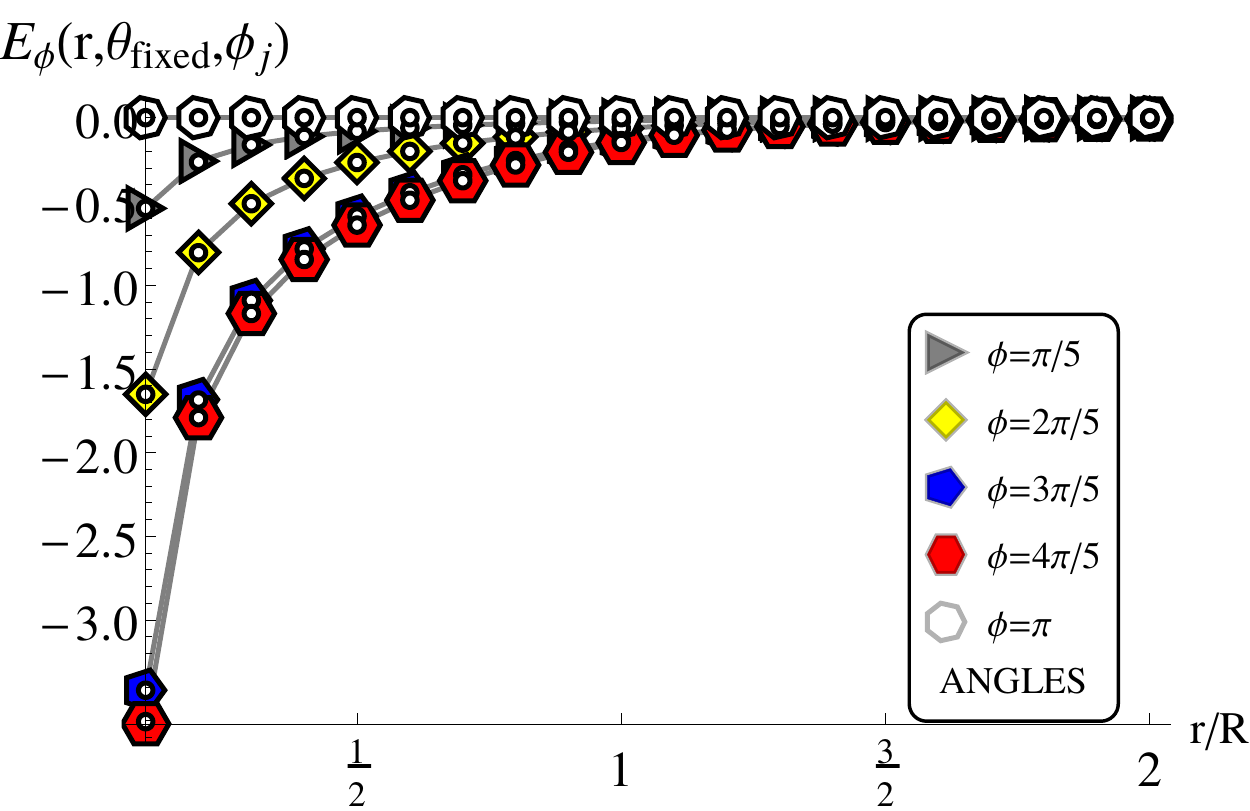}%EPhiComparison.pdf
    \caption{Numerical approximations and analytic series expansions of $\boldsymbol{E}(r,\theta,\phi)$ at $\theta = \pi/3$. Left to right. Components of the electric field. Polygonal symbols refer to the electric field computed with Eq.~(\ref{EVectorFieldStairCaseVExpansionEq}) considering $M=20$ terms. Open dots correspond to the numerical approximation of Eq.~(\ref{electricPotentialIntegralEq}). Gray solid lines are only included as guidance.}
\label{EComparisonPhipi3Fig}
\end{figure}

Even when this strategy is easy to implement, there are some sources of error that must be taken into account. The first one emerges from the numerical integration, since the integrand in Eq.~(\ref{electricPotentialIntegralEq}) may become highly oscillatory depending on the potential function $V(\phi)$ and the point of evaluation. Another source of error comes from the numerical differentiation. Typically, the error of central derivatives is proportional to $h^2$, however in numerical computations $h$ cannot be set arbitrarily small because the numerator of the finite derivative can be cancelled \cite{squire1998using}. This error depends on machine precision.

\begin{figure}[h]
\centering%analyticErFunctionOfNPhi2pi5.pdf
\includegraphics[width=0.4\textwidth]{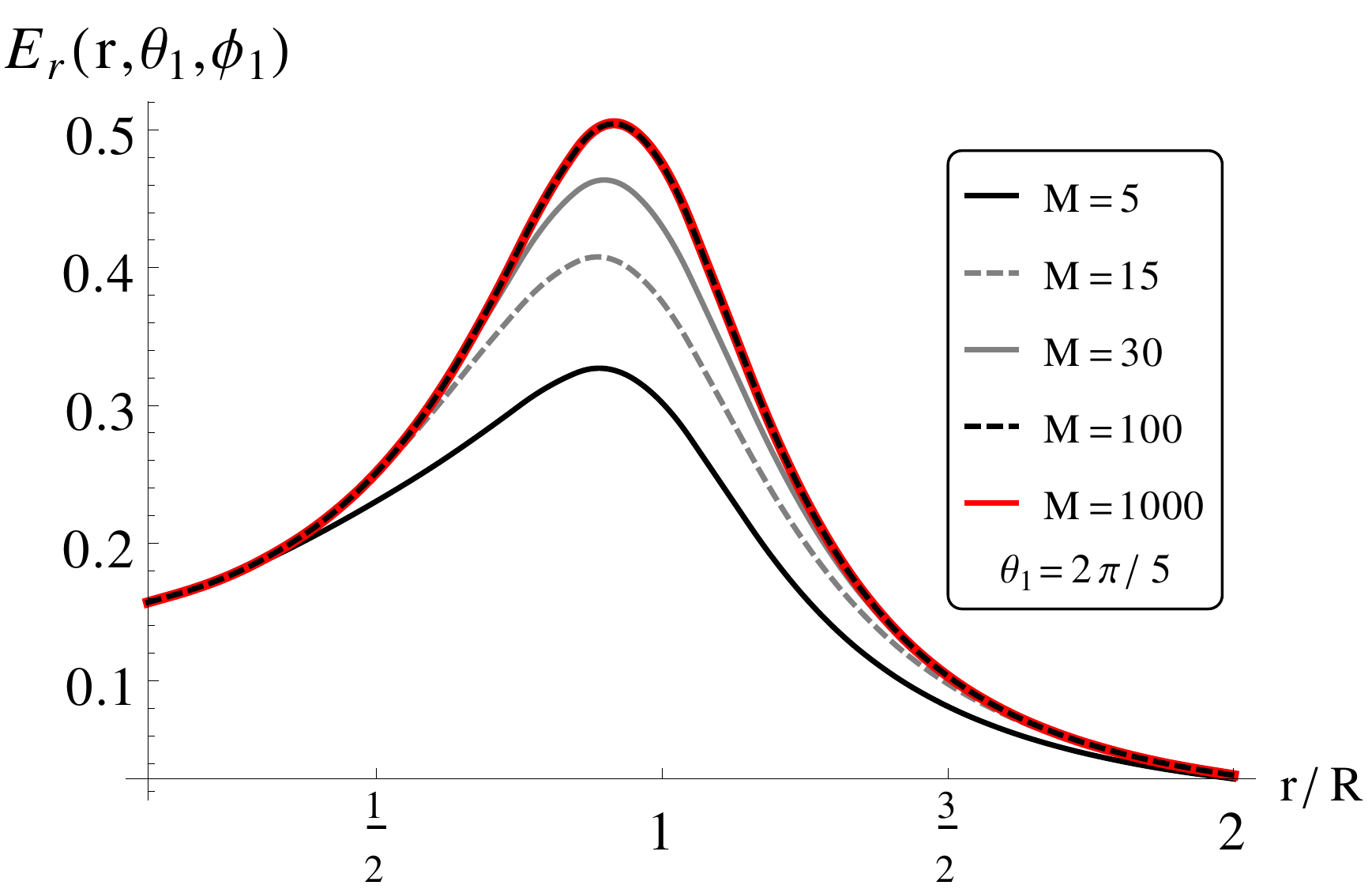}\hspace{0.5cm}%analyticErFunctionOfNPhipi3.pdf
\includegraphics[width=0.4\textwidth]{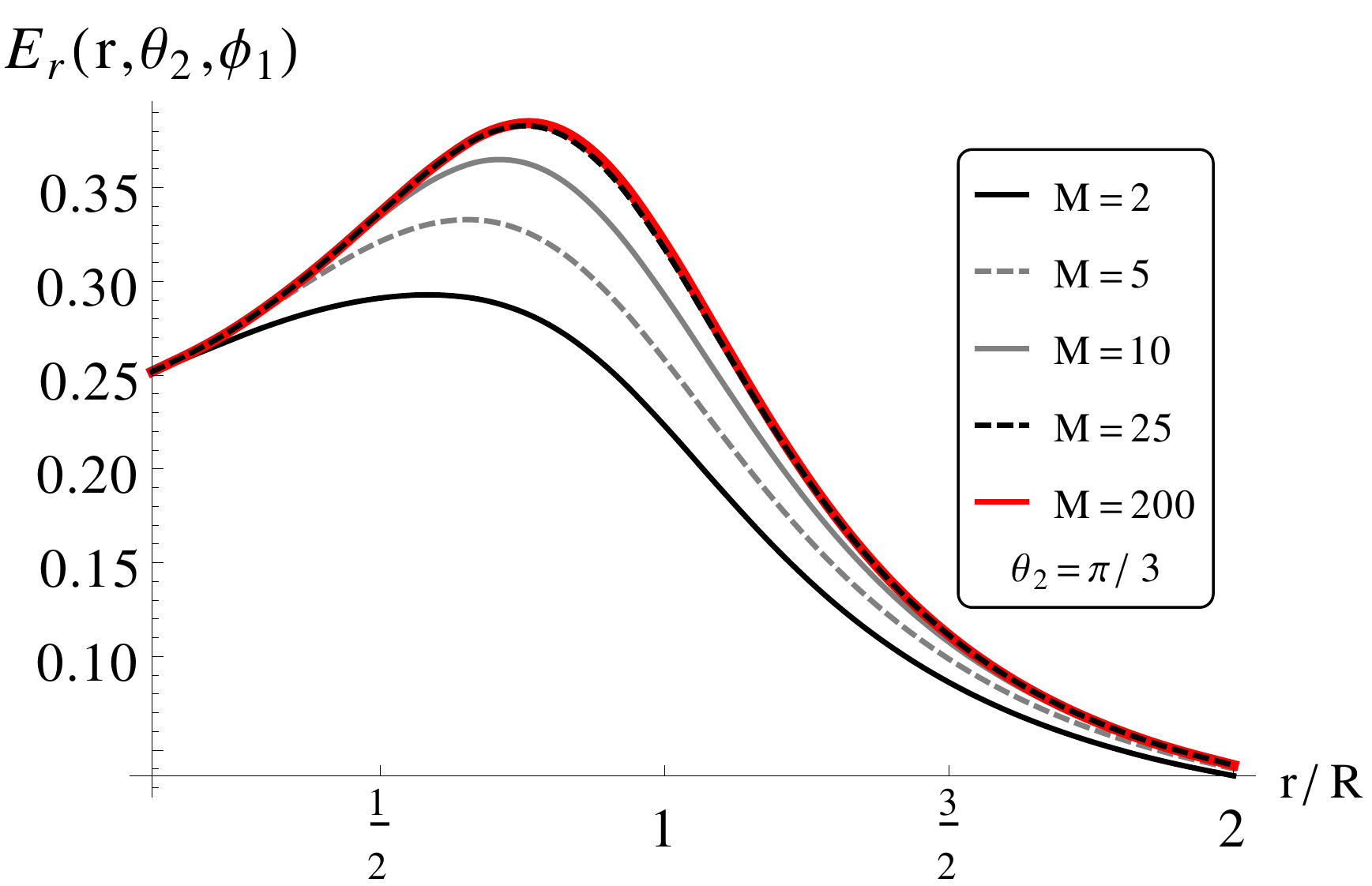}
    \caption{Series solution. Radial electric field at (left) $\theta_1=2\pi/5$ and (right) $\theta_2=\pi/3$. The value of $\phi$ was fixed at $\phi_1=\pi/5$ in both plots. }
\label{EComparisonNBehaviourFig}
\end{figure}

\begin{figure}[H]
\centering %ERROR.pdf
\includegraphics[width=0.45\textwidth]{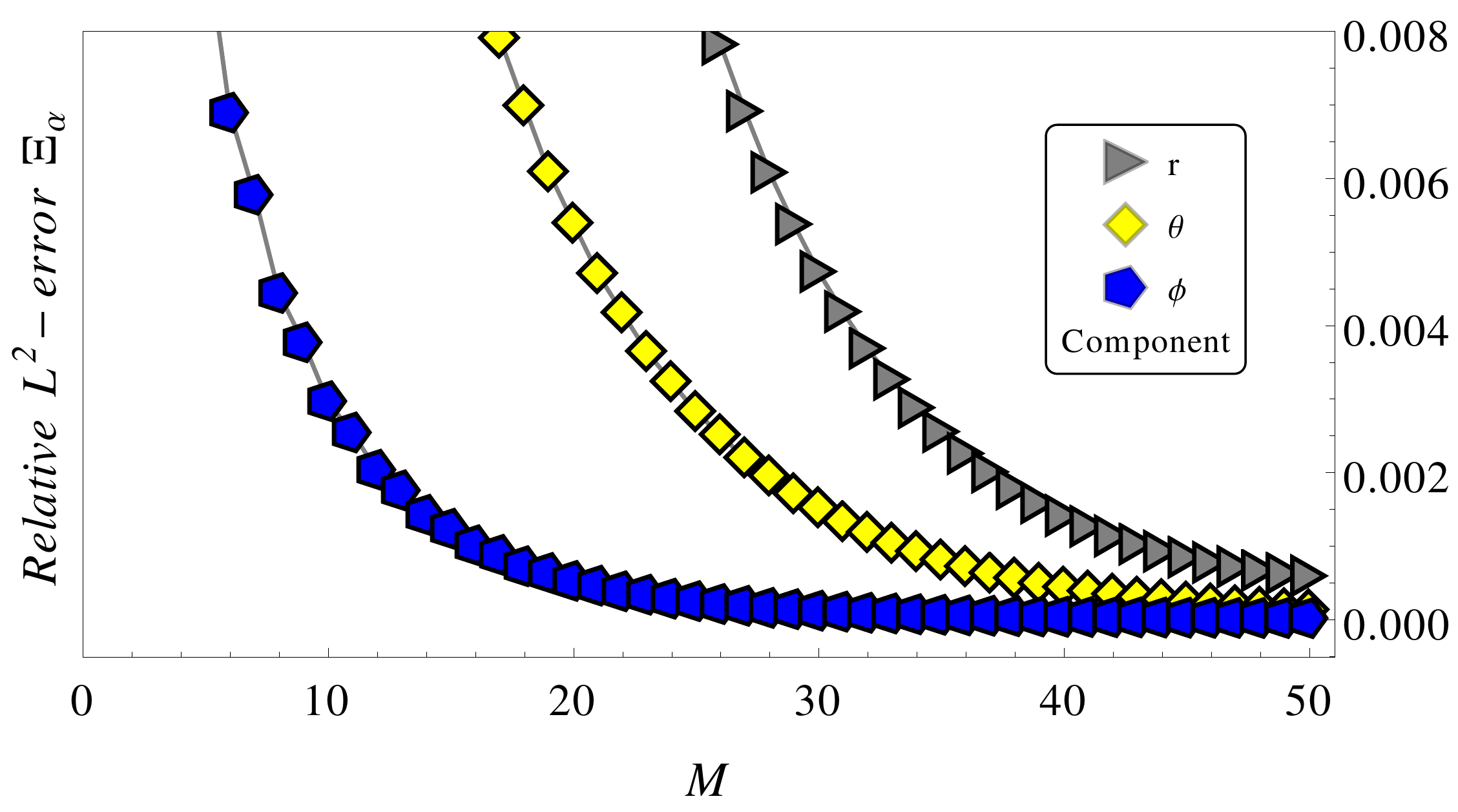}
    \caption{$L^2$ Relative error norm between the exact and the numerical computed electric field.}
\label{EErrorFig}
\end{figure}

On Fig.~\ref{EComparisonPhipi3Fig}, we show the computation of the electric field when using numerical integration and differentiation. The potential function $V(\phi)$ is chosen, in this case, as a staircase-like function with $N=33$ and $\mathcal{V}$ Gaussian, as shown at the center of Fig.~\ref{discretePotentialLimitFig}. We also show on Fig.~\ref{EComparisonPhipi3Fig} the analytic solutions of the electric field according to Eq.~(\ref{EVectorFieldStairCaseVExpansionEq}) at $\theta=\pi/3$ for some values of $\phi$ in $[0,\pi]$. Numerical and analytic computations of the electric field components are in good agreement. One remark to be done is that the evaluation of the analytic expression requires a truncation of the infinite sum of Eq.~(\ref{EVectorFieldStairCaseVExpansionEq}). Such infinite sum comes from the Taylor series of Eq.~(\ref{binomialTheoremSeriesExpansionEq}), which converges for $|\chi|<1$ for any complex number $\alpha$. However, if the series is truncated at the first $M$ terms, then $|\chi|$ approaches to one and $M$ has to be increased in order to obtain accurate solutions. This error dependence of the truncated series with $M$ is inherited on the computation of the electric field from Eq.~(\ref{EVectorFieldStairCaseVExpansionEq}).  Since $\chi(\boldsymbol{r},\boldsymbol{r}') = \frac{2 R r \sin\theta}{R^2+r^2} \cos(\phi-\phi')$, then it is mandatory to include several terms in the series in order to be able to evaluate the electric field on the $(z=0)$-plane (this is, at $\theta=\pi/2$) and near $r=R$, since $\chi$ approaches to one if $\phi'\rightarrow\phi$. This dependence can be observed on Fig.~(\ref{EComparisonNBehaviourFig}), where we present the analytic solution depending on $M$ at $\theta = 2\pi/5$ and $\theta = \pi/3$. There we require at least $M=100$ terms for $\theta = 2\pi/5$. On the other hand, $M=25$ terms are enough to compute $E_r$ at $\theta = \pi/3$. For this reason, we have been able to use $M=20$ terms to obtain the results in Fig.~\ref{EComparisonPhipi3Fig} without having large differences between the truncated series and the numeric results, even that $M$ is relatively small. We complete our comparisons between the analytic and the numerical approach by calculating the $L^2$ relative error norm
\[
\Xi_{\alpha} = \sqrt{\frac{1}{\sum_{(r_i,\theta_j,\phi_k)\in\mathcal{D}}  E_{\alpha}^2( \boldsymbol{r}_{ijk} )}\sum_{(r_i,\theta_j,\phi_k)\in\mathcal{D}} \Delta E_{\alpha}^2( \boldsymbol{r}_{ijk} ) } 
\]
for each $\alpha$-th component of the electric field in spherical coordinates. Here $\Delta E_{\alpha} = E_{\alpha} - E_{\alpha}^{(num.)}$ where $E_{\alpha}$ and $E_{\alpha}^{(num.)}$ are the components of $\boldsymbol{E}$ computed with Eq.~(\ref{EVectorFieldStairCaseVExpansionEq}) and numerically, respectively. Domain $\mathcal{D}$ is defined as a rectangular cuboid of volume $2R \times \theta_{max} \times 2\pi$, where $r\in [0,2R]$, $\theta\in [0,\theta_{max}]$ and $\phi \in [0, 2\pi)$. The value of $\theta_{max}$ has been set to $0.8\pi/2$. The total points in the rectangular lattice of indices $\{(i,j,k)\}$ has been defined to be 8000. A plot of the $L^2$ relative error norm as a function of number of terms in Eq.~(\ref{EVectorFieldStairCaseVExpansionEq}) is shown in Fig.~\ref{EErrorFig}. We can observe an asymptotic convergence of the relative error for around $M=40$ terms of the expansion series. We can also observe that the highest convergence rate occurs for the $\phi$ component of the electric field. On the contrary, the radial component is the slowest component to converge. Hence, special care has to be taken with the analytical solution given by this term for few truncation terms.

\subsection{Finite Element Method approximation of the electrostatic problem}

Consider the spatial domain $\mathcal{D}$, with boundary $\partial\mathcal{D}$ and $\boldsymbol{n}$ the unit normal to the boundary.
The weak form of the Laplace's equation for the potential field is readily obtained by multiplying it by a test function $\Psi\in \mathcal{V}$, with $\mathcal{V}$ being a suitable function space that satisfies the Dirichlet boundary conditions (i.e. the non-uniform distribution for the electric potential over the planar region), and integrating it by parts. Hence, the weak problem is to find $\Phi\in \mathcal{V}$ such that
\[
\int_{\mathcal{D}} \boldsymbol{\nabla}\Phi\cdot\boldsymbol{\nabla}\Psi \, {\rm d} x = 
 \int_{\partial\mathcal{D}} (\boldsymbol{\nabla}\Phi\cdot \boldsymbol{n}) \Psi \, {\rm d} s \quad \text{for all } \Psi\in \mathcal{V}.
\]

\begin{figure}[H]
\centering%errorFEM.pdf
\includegraphics[width=0.45\textwidth]{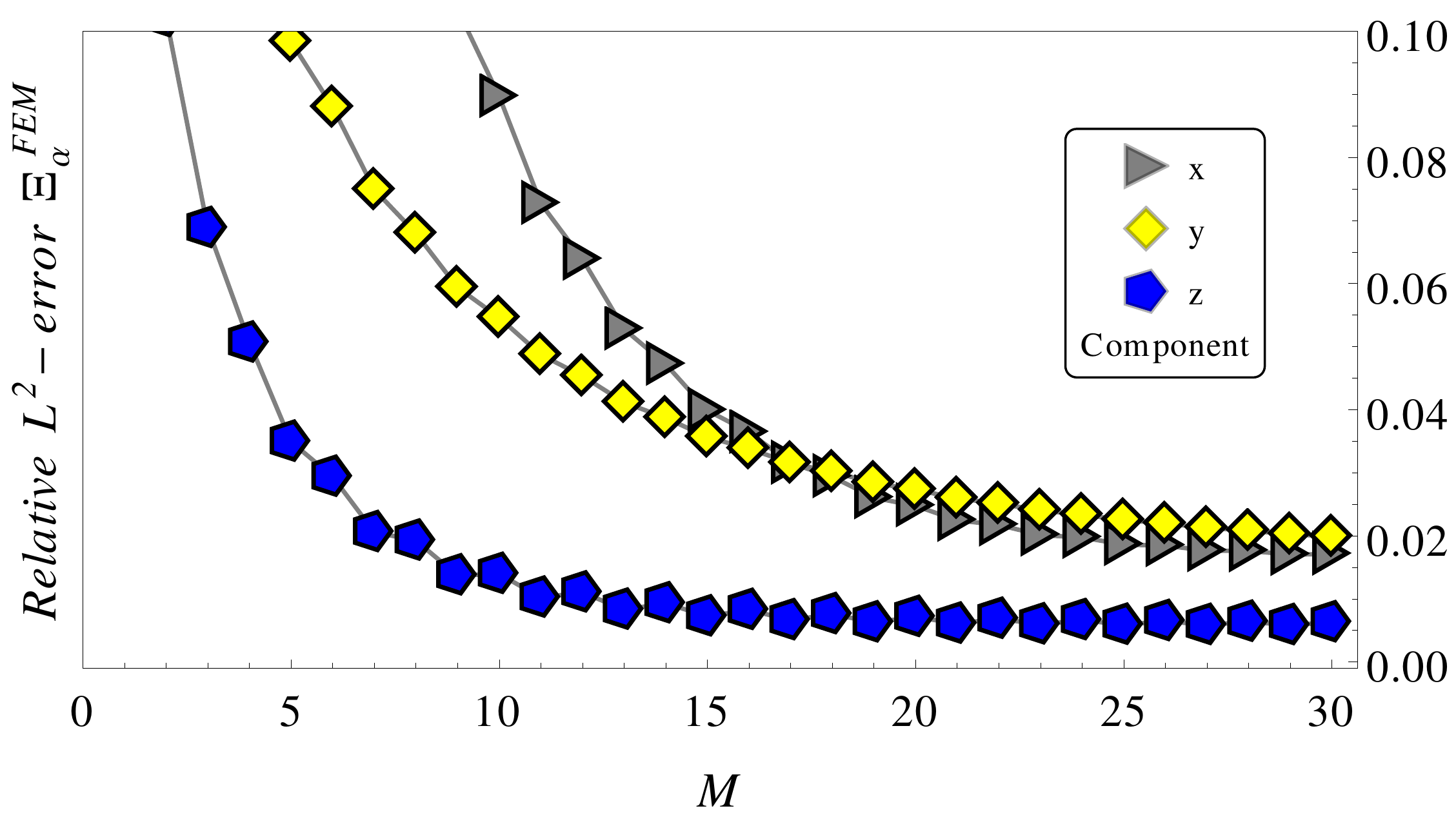}
    \caption{$L^2$ Relative error norm. Numerical values were computed via FEM. }
\label{EErrorFEMFig}
\end{figure}

We then formulate the discrete Galerkin approximation of the Laplace's problem.
Let us first denote by $\mathcal{T}_h =\left\{\mathcal{D}^e\right\}$ the finite element partition of the domain $\mathcal{D}$, with index $e$ ranging from 1 to the number of elements $n_{el}$ in the finite mesh.
The diameter of the element partition is denoted by $h$.
We define the finite test function spaces as made of continuous piecewise polynomial functions in space.
The Galerkin approximation of the weak Laplace's problem considers replacing $\mathcal{V}$ by the finite subspaces $\mathcal{V}_h\subset \mathcal{psi}$, where the subscript $h$ refers the discrete finite element space.
The problem to be solved now is to find $\Phi_h\in \mathcal{V}_h$ such that 
\begin{equation}
\int_{\mathcal{D}} \boldsymbol{\nabla}\Phi_h\cdot\boldsymbol{\nabla}\Psi_h \, {\rm d} x = 
 \int_{\partial\mathcal{D}} (\boldsymbol{\nabla}\Phi_h\cdot \boldsymbol{n}) \Psi_h \, {\rm d} s \quad \text{for all } \Psi_h\in \mathcal{V}_h. \label{FEM}    
\end{equation}{}

\begin{figure}[ht]  
  \begin{minipage}[b]{0.45\linewidth}
    
   \centering 
   \includegraphics[width=1\textwidth]{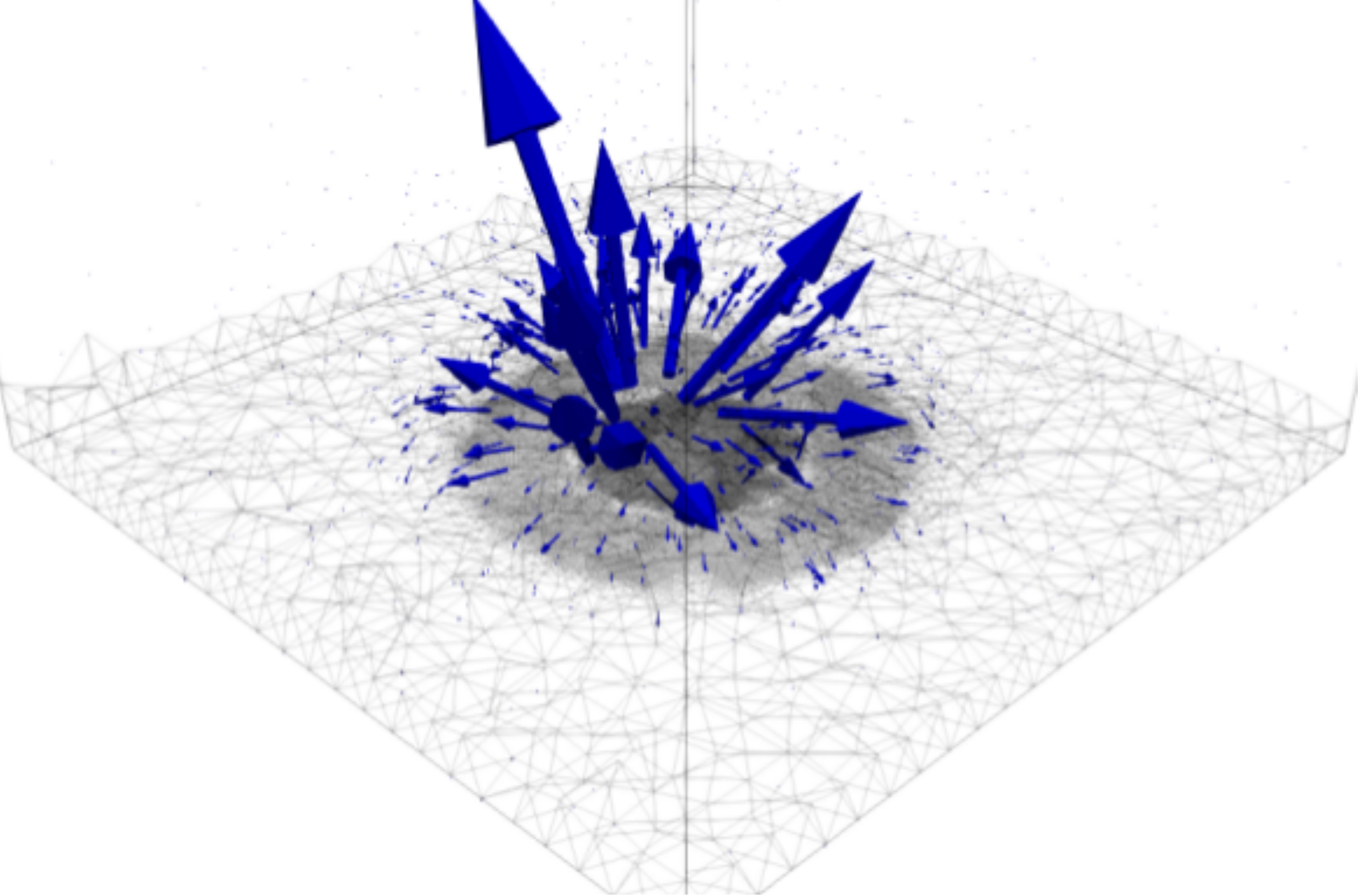}
   \caption*{(a)} %cropFig.pdf
    
  \end{minipage} 
  \begin{minipage}[b]{0.5\linewidth}
    
    \includegraphics[width=1.01\textwidth,trim = 3cm 1.5cm 3cm 9cm, clip=true]{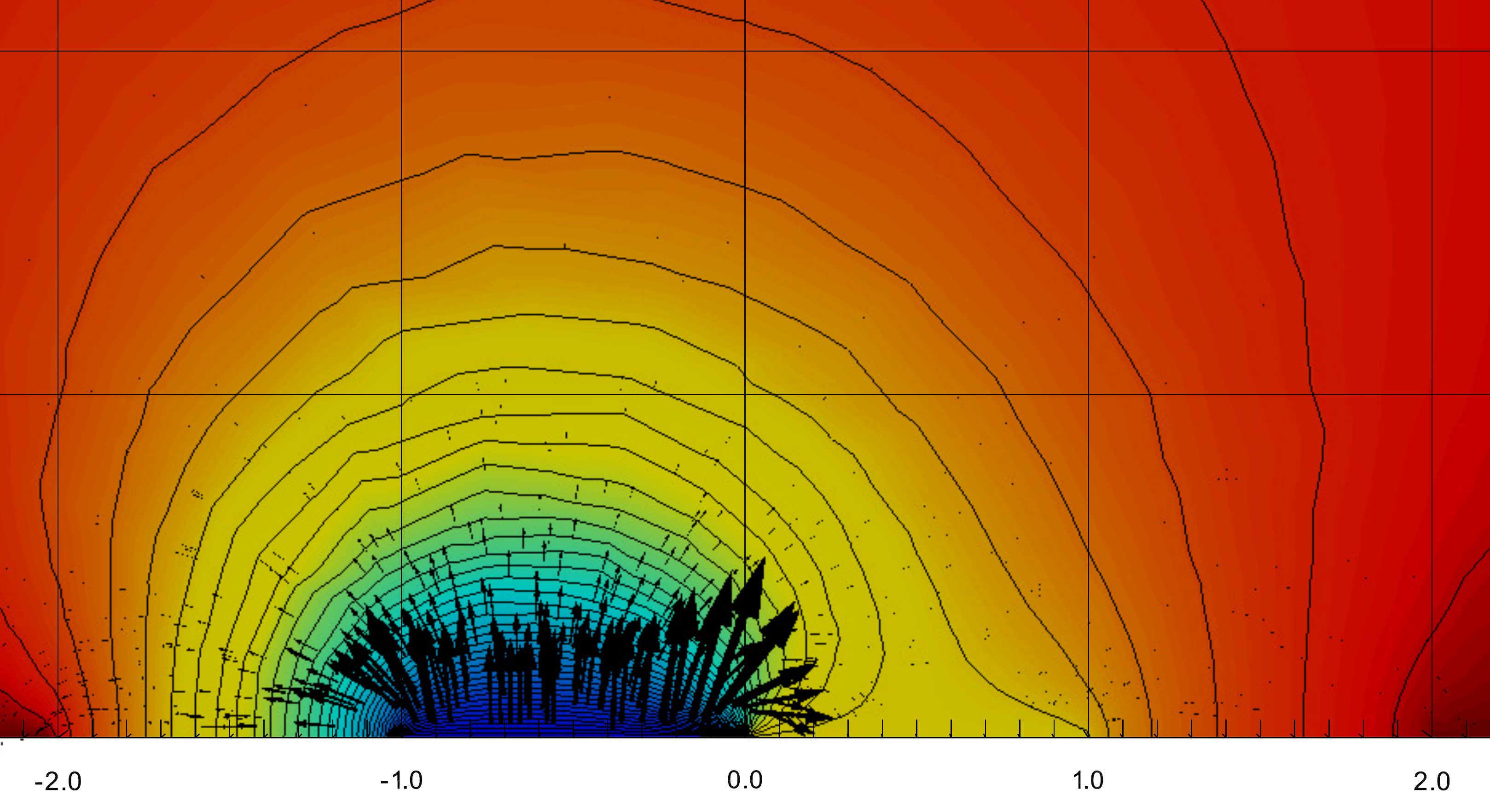}
    \caption*{(b) $\phi=0$} %FEMPhi0.eps
  
  \end{minipage} 
  \begin{minipage}[b]{0.5\linewidth}
    
    \includegraphics[width=1\textwidth,trim = 3cm 1.5cm 3cm 9cm, clip=true]{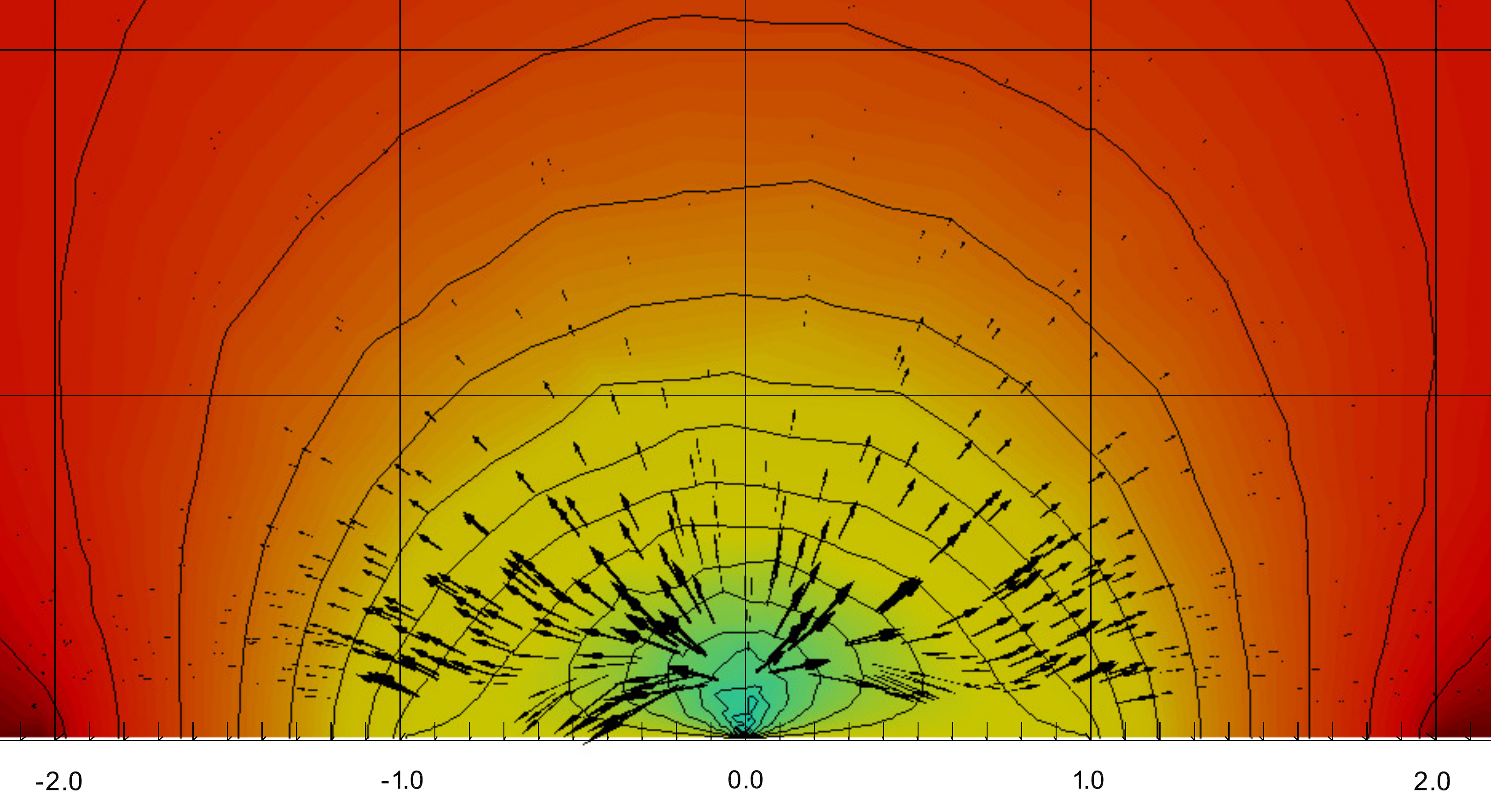}
    \caption*{(c) $\phi=\pi/2$}  %FEMPhi180.eps
  
  \end{minipage}
  \hfill
  \begin{minipage}[b]{0.5\linewidth}
    \includegraphics[width=1\textwidth,trim = 3cm 1.5cm 3cm 9cm, clip=true]{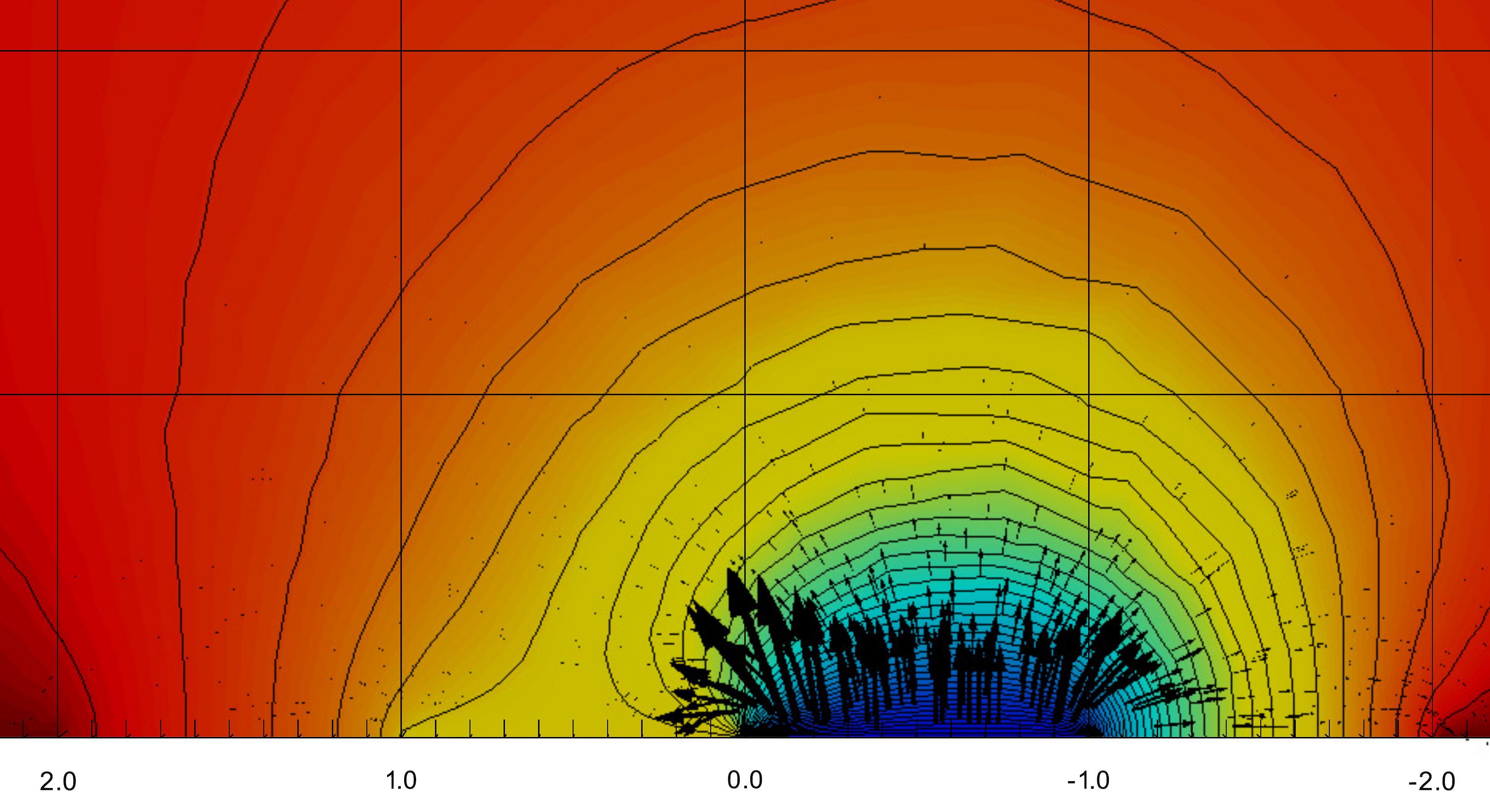}
    \caption*{(d) $\phi=3\pi/4$} %FEMPhi270.eps
    
  \end{minipage}
  
  \caption{ Electric field. (a) Numerical vector field computed with Eq.~(\ref{FEM}). Computations were performed for a Gaussian potential $\mathcal{V}$ (see Eq.~(\ref{gaussianContinousVEq})) using a $n_{el}=439014$ linear tetrahedron unstructured mesh with local refinement near the potential region. (b), (c) and (d) are projections of the $\boldsymbol{E}(\boldsymbol{r})$ on the plane $\phi=0, \pi/2$, and $3\pi/4$, respectively.  }
  
  \label{FEMFig}
  
\end{figure}

We obtained the results displayed in Fig.~\ref{FEMFig} for linear tetrahedron type of finite elements.
Specifically, we used an unstructured mesh composed by $n_{el}=439014$ to discretize the cuboid domain $\mathcal{D}:=[-4R,4R] \times [-4R,R] \times [0,8R]$, with local refinement near the planar circular region of radius $R$ that is centered on one of the boundaries. On Fig.~\ref{EErrorFEMFig} we show the $L^2$ relative error norm between the Eq.~(\ref{EVectorFieldStairCaseVExpansionEq}) and the numerical results obtained by the FEM. Solutions approach each other as $M$ is increased. Even when the current problem can be solved numerically, we stress the fact that truncated analytic series from Eq.~(\ref{EVectorFieldStairCaseVExpansionEq}) can be computationally less expensive than traditional numerical approaches if $\theta$ is not near to $\pi/2$. This is just the behaviour of analytic solution shown in  Fig.~\ref{EComparisonNBehaviourFig} where the truncated series tends faster to the exact result as $\theta$ is decreased from $\pi/2$.

\section*{Conclusion}
In this work we presented an approach to compute the electric field due to planar regions kept at a fixed but non-uniform potential $V(\phi)$. We used some analogies of this problem with magnetostatics to simplify the electrostatic problem. As we shown in Eq.~(\ref{electricFieldBiotSavartWithCorrectionEq}), the electric field can be found by evaluation of two one-dimensional integrals depending on $\phi$, where one of them is analogous to the well-known Biot-Savart law of magnetostatics. Using this approach, it is possible to find an exact series solution (see Eq.~(\ref{EVectorFieldStairCaseVExpansionEq})) of the problem when a region with circular or polygonal interconnected boundary is considered.  

Future work may be devoted to the extension of this methodology to the time-dependent surface electrode TDSE where the potential on $\mathcal{A}$ is not uniform and changing harmonically with time. In this case, we would like to check if this problem is analogous to the field radiated by a non-uniform harmonic current circulating on a closed-loop $c$. A motivation of that sequel is to being able to compute the radiation field on diverse geometries including fractal antennas.

\section*{Acknowledgments}
This work was supported by Vicerrector\'ia de investigaci\'on, Universidad ECCI. Robert Salazar also thanks Fundaci\'on Colfuturo and Departamento de Ciencias B\'asicas, Universidad ECCI.
%----------------------------------------------------------------------------------------
%	REFERENCE LIST
%----------------------------------------------------------------------------------------
\bibliographystyle{ieeetr} %alpha, apalike, ieeetr
\bibliography{bibliography.bib}

%\begin{thebibliography}{99} % Bibliography - this is %intentionally simple in this template

%\bibitem{andreotti1997studying} B. Andreotti, \emph{Studying Burgers' models to investigate the physical meaning of the alignments statistically observed in turbulence}, Phys. Fluids \textbf{9} : 3, March (1997)

%\bibitem{cohl1999compact} Cohl, Howard S., and Joel E. Tohline, \emph{A compact cylindrical Green's function expansion for the solution of potential problems}, The astrophysical journal \textbf{527} : 86 - 101 (1999) %DOI: https://doi.org/10.1086/308062

%\bibitem{abramowitz1965handbook} Abramowitz, Milton, and Irene A. Stegun. \emph{Handbook of Mathematical Functions With Formulas, Graphs, and Mathematical Tables.} (1964).

%\end{thebibliography}

%croos section karlie
%GOOD: http://www.eumetrain.org/satmanu/CMs/TrCyAt/print.htm 
%https://physics.stackexchange.com/questions/275799/why-is-the-eye-of-a-cyclone-a-forced-vortex
%http://www.chanthaburi.buu.ac.th/~wirote/met/tropical/textbook_2nd_edition/navmenu.php_tab_9_page_7.1.0.htm
%http://www.atmos.umd.edu/~dalin/andrew/part2.html
%https://nptel.ac.in/courses/119102007/2
%http://www.911omissionreport.com/steering_hurricanes.html
%https://www.youtube.com/watch?v=_brY_9ME8iE brooks
\end{document}